%% file: paperV3.tex
\documentclass[12pt]{article}
\pdfoutput=1
\usepackage[DIV13]{typearea}
\usepackage{indentfirst}
\usepackage{amsmath}
\usepackage{amsfonts}
\usepackage{mathrsfs}
\usepackage{amssymb}
\usepackage{mathtools}
\usepackage{bbm}
\usepackage{epsfig}
\usepackage{graphicx}
\usepackage{subfigure}
\usepackage{slashed}
\usepackage{multicol}
\usepackage[usenames,dvipsnames]{color}
\usepackage{cite}
\RequirePackage[colorlinks=true,urlcolor=blue,anchorcolor=blue,citecolor=blue,filecolor=blue,
               linkcolor=blue,menucolor=blue,linktocpage=true,pdfproducer=medialab]{hyperref}
\usepackage{cancel}
\usepackage{feynmp}
\usepackage{lscape}
\usepackage{multirow}
\usepackage{array}
\usepackage{pdfpages}
\usepackage{fancyhdr}
\usepackage{pifont}

\usepackage[left=.9in, right=.9in]{geometry}
\newcommand{\email}[1]{\href{mailto:#1}{\tt #1}}
\numberwithin{equation}{section}
\newcommand{\LL}{\mathscr{L}}

\def\cF{{\cal F}}
\def\cO{{\cal O}}
\def\cP{{\cal P}}

\def\cY{{\bf Y}}

\def\be{\begin{equation}}
\def\ee{\end{equation}}
\def\beq{\begin{equation}}
\def\eeq{\end{equation}}
\def\bc{\begin{center}}
\def\ec{\end{center}}
\def\bea{\begin{eqnarray}}
\def\eea{\end{eqnarray}}

\def\nn{\nonumber}

\renewcommand{\a}{\alpha}
\renewcommand{\b}{\beta}
\renewcommand{\d}{\delta}
\newcommand{\g}{\gamma}
\newcommand{\s}{\sigma}
\newcommand{\e}{\varepsilon}

\newcommand{\mean}[1]{\langle#1\rangle}
\newcommand{\derp}{\partial}

\newcommand{\hc}{\mathrm{h.c.}}

\newcommand{\diag}{{\rm{\bf diag}}}
\newcommand{\UH}{\mathbf{U}}
\newcommand{\MH}{\mathbf{M}}

\newcommand{\TL}{\mathbf{T}}
\newcommand{\VL}{\mathbf{V}}
\newcommand{\DL}{D}
\newcommand{\DLR}{\mathbf{D}}
\newcommand{\DLL}{\mathcal{D}}

\newcommand{\GF}{G_F}
\newcommand{\aem}{\alpha_\text{em}}

\newcommand{\tr}{{\rm Tr}}

\renewcommand{\to}{\rightarrow}

\newcommand{\Pu}{P_{\uparrow}}
\newcommand{\Pd}{P_{\downarrow}}

\newcommand{\BBu}{B^{\mu\nu}}
\newcommand{\BBd}{B_{\mu\nu}}
\newcommand{\WWu}{W^{\mu\nu}}
\newcommand{\WWd}{W_{\mu\nu}}

\newcommand{\st}{s_\theta}
\newcommand{\ct}{c_\theta}
\newcommand{\cdt}{c_{2\theta}}
\newcommand{\sdt}{s_{2\theta}}

\newcommand{\ie} {{\it i.e.}}
\newcommand{\Dfb}{\mbox{$\raisebox{2mm}{\boldmath ${}^\leftrightarrow$}\hspace{-4mm} \DL$}}

\newcommand{\blue}[1]{\color{blue} #1 \color{black}}

\begin{document}
\begin{titlepage}
\vspace*{-1cm}
\phantom{hep-ph/***} 
{\flushleft
{\blue{FTUAM-13-32}}
\hfill{\blue{IFT-UAM/CSIC-13-119}}\\
{\blue{YITP-SB-13-33}}
\hfill{\blue{DFPD-2013/TH/20}}\\}
\vskip 1cm
\begin{center}
\mathversion{bold}
{\LARGE\bf Disentangling a dynamical Higgs}\\
\mathversion{normal}
\vskip .3cm
\end{center}
\vskip 0.5  cm
\begin{center}
{\large I.~Brivio}~$^{a)}$,
{\large T.~Corbett}~$^{b)}$,
{\large O.\ J.\ P.\ \'Eboli}~$^{c)}$,
{\large M.B.~Gavela}~$^{a)}$,\\[2mm]
{\large J.\ Gonzalez--Fraile}~$^{d)}$,
{\large M.\ C.\ Gonzalez--Garcia}~$^{e,d,b)}$,
{\large L.~Merlo}~$^{a)}$,
{\large S.~Rigolin}~$^{f)}$
\\
\vskip .7cm
{\footnotesize
$^{a)}$~
Departamento de F\'isica Te\'orica and Instituto de F\'{\i}sica Te\'orica, IFT-UAM/CSIC,\\
Universidad Aut\'onoma de Madrid, Cantoblanco, 28049, Madrid, Spain\\
\vskip .1cm
$^{b)}$~
C.N.~Yang Institute for Theoretical Physics and Department of Physics and Astronomy, SUNY at Stony Brook, Stony Brook, NY 11794-3840, USA\\
\vskip .1cm
$^{c)}$~
  Instituto de F\'isica, Universidade de S\~ao Paulo, C.P. 66318, 05315-970, S\~ao Paulo SP, Brazil\\
\vskip .1cm
$^{d)}$~
Departament d'Estructura i Constituents de la Mat\`eria and ICC-UB, Universitat de Barcelona, 647 Diagonal, E-08028 Barcelona, Spain\\
\vskip .1cm
$^{e)}$~
Instituci\'o Catalana de Recerca i Estudis Avan\c{c}ats (ICREA)\\
\vskip .1cm
$^{f)}$~
Dipartimento di Fisica e Astronomia ``G.~Galilei'', Universit\`a di Padova and \\
INFN, Sezione di Padova, Via Marzolo~8, I-35131 Padua, Italy
\vskip .3cm
\begin{minipage}[l]{.9\textwidth}
\begin{center} 
\textit{E-mail:} 
\email{ilaria.brivio@uam.es},
\email{corbett.ts@gmail.com},
\email{eboli@fma.if.usp.br},
\email{belen.gavela@uam.es},
\email{fraile@ecm.ub.edu},
\email{concha@insti.physics.sunysb.edu},
\email{luca.merlo@uam.es},
\email{stefano.rigolin@pd.infn.it}
\end{center}
\end{minipage}
}
\end{center}
\vskip 0.5cm
\begin{abstract}
  The pattern of deviations from Standard Model predictions and
  couplings is different for theories of new physics based on a
  non-linear realization of the $SU(2)_L\times U(1)_Y$ gauge symmetry
  breaking and those assuming a linear realization.  We clarify this
  issue in a model-independent way via its effective Lagrangian
  formulation in the presence of a light Higgs particle, up to first
  order in the expansions: dimension-six operators for the linear
  expansion and four derivatives for the non-linear one.  Complete
  sets of gauge and gauge-Higgs operators are considered,
  implementing the renormalization procedure and deriving the Feynman
  rules for the non-linear expansion.  We establish the theoretical
  relation and the differences in physics impact between the two expansions.
  Promising discriminating signals include the
  decorrelation in the non-linear case of signals correlated in the
  linear one: some pure gauge versus gauge-Higgs couplings and also
  between couplings with the same number of Higgs legs. Furthermore,
  anomalous signals expected at first order in the non-linear
  realization may appear only at higher orders of the linear one, and
  vice versa. We analyze in detail the impact of both
  type of discriminating signals on LHC physics.
\end{abstract}
\end{titlepage}
\setcounter{footnote}{0}

\pdfbookmark[1]{Table of Contents}{tableofcontents}
\tableofcontents

%
%%%%%%%%%%%%%%%%%%%%%%%%%   1.  Introduction       %%%%%%%%%%%%%%%%%%%%%%%%
%

\section{Introduction}

The present ensemble of data does not show evidence for new exotic
resonances and points to a scenario compatible with the Standard Model
(SM) scalar boson (so-called ``Higgs'' for short)
\cite{Englert:1964et,Higgs:1964ia,Higgs:1964pj}. Either the SM is all
there is even at energies well above the TeV scale, which would
raise a number of questions about its theoretical consistency
(electroweak hierarchy problem, triviality, stability), or new physics
(NP) should still be expected around or not far from the TeV scale.

This putative NP could be either detected directly or studied
indirectly, analysing the modifications of the SM couplings. To this
aim, a rather model-independent approach is that of Lorentz and
gauge-invariant effective Lagrangians, which respect a given set of
symmetries including the low-energy established ones. These effective
Lagrangians respect symmetries in addition to $U(1)_\text{em}$ and
Lorentz invariance and as a consequence they relate and constrain
phenomenological couplings~\cite{Hagiwara:1986vm} based only on the
latter symmetries.

With a light Higgs observed, two main classes of effective Lagrangians
are pertinent, depending on how the electroweak (EW) symmetry breaking
is assumed to be realized: linearly for elementary Higgs particles or
non-linearly for ``dynamical" -composite- ones. It is important to
find signals which discriminate among those two categories and this
will be one of the main focuses of this paper.

In elementary Higgs scenarios, the effective Lagrangian provides a
basis for all possible Lorentz and $SU(3)_c\times SU(2)_L\times
U(1)_Y$ gauge invariant operators built out of SM fields. The latter
set includes a Higgs particle belonging to an $SU(2)_L$ doublet, and
the operators are weighted by inverse powers of the unknown
high-energy scale $\Lambda$ characteristic of NP: the leading
corrections to the SM Lagrangian have then canonical mass dimension
($d$) six~\cite{Buchmuller:1985jz,Grzadkowski:2010es}.  Many studies
of the effective Lagrangian for the linear expansion have been carried
out over the years, including its effects on Higgs production and decay
\cite{Hagiwara:1993qt,GonzalezGarcia:1999fq}, with a revival of
activity
~\cite{Corbett:2012dm,Corbett:2012ja}
after the Higgs discovery~\cite{Aad:2012tfa,Chatrchyan:2012ufa} (see
also
Refs.~\cite{Low:2012rj,Ellis:2012hz,Giardino:2012dp,Montull:2012ik,Espinosa:2012im,Carmi:2012in,Banerjee:2012xc,Bonnet:2012nm,Plehn:2012iz,Djouadi:2012rh,Batell:2012ca,Moreau:2012da,Cacciapaglia:2012wb,Azatov:2012qz,Masso:2012eq,Passarino:2012cb,Falkowski:2013dza,Giardino:2013bma,Ellis:2013lra,Djouadi:2013qya,Contino:2013kra,Dumont:2013wma,Elias-Miro:2013mua,Lopez-Val:2013yba,Jenkins:2013zja,Pomarol:2013zra,Banerjee:2013apa,Alloul:2013naa}
for studies of Higgs couplings in alternative and related frameworks).
Supersymmetric models are a typical example of the possible underlying
physics.

In dynamical Higgs scenarios, the Higgs particle is instead a
composite field which happens to be a pseudo-goldstone boson (GB) of a
global symmetry exact at scales $\Lambda_s$, corresponding to the masses
of the lightest strong resonances. The Higgs mass is protected by the
global symmetry, thus avoiding the electroweak hierarchy problem.
Explicit realizations include the revived and now popular models
usually dubbed ``composite Higgs"
scenarios~\cite{Kaplan:1983fs,Kaplan:1983sm,Banks:1984gj,Georgi:1984ef,Georgi:1984af,Dugan:1984hq,Agashe:2004rs,Contino:2006qr,Gripaios:2009pe,Marzocca:2012zn},
for various strong groups and symmetry breaking patterns \footnote{
  Also ``little Higgs"~\cite{ArkaniHamed:2001nc} (see
  Ref.~\cite{Schmaltz:2005ky} for a review) models and some
  higher-dimensional scenarios can be cast in the category of
  constructions in which the Higgs is a goldstone boson.}.  To the
extent that the light Higgs particle has a goldstone boson parenthood,
the effective Lagrangian is non-linear~\cite{PhysRev.177.2247} or ``chiral": a derivative
expansion as befits goldstone boson dynamics. The explicit breaking of
the strong group -necessary to allow a non-zero Higgs mass- introduces
chiral-symmetry breaking terms.  In this scenario, the characteristic
scale $f$ of the Goldstone bosons arising from the spontaneous
breaking of the global symmetry at the scale $\Lambda_s$ is
different\footnote{In the historical and simplest formulations of
  ``technicolor"~\cite{Susskind:1978ms,Dimopoulos:1979es,Dimopoulos:1981xc},
  the Higgs particle was completely removed from the low-energy
  spectrum, which only retained the three SM would-be-Goldstone bosons
  with a characteristic scale $f=v$.} from both the EW scale $v$
defined by the EW gauge boson mass, e.g. the $W$ mass $m_W=gv/2$, and the EW symmetry breaking (EWSB)
scale $\mean{h}$, and respects $\Lambda_s<4\pi f$. A model-dependent
function $g$ links the three scales, $v=g(f, \mean{h})$, and a
parameter measuring the degree of non-linearity of the Higgs dynamics
is usually introduced:
\begin{equation}
\xi\equiv (v/f)^2\,.
\end{equation}

The corresponding effective low-energy chiral Lagrangian is entirely
written in terms of the SM fermions and gauge bosons and of the
physical Higgs $h$. The longitudinal degrees of freedom of the EW
gauge bosons can be effectively described at low energies by a
dimensionless unitary matrix transforming as a bi-doublet of the
global symmetry:
\beq
\UH(x)=e^{i\sigma_a \pi^a(x)/v}\, , \qquad \qquad  \UH(x) \rightarrow L\, \UH(x) R^\dagger\,,
\eeq
where here the scale associated with the eaten GBs is $v$, and not $f$,
in order to provide canonically normalized kinetic terms, and $L$, $R$
denotes $SU(2)_{L,R}$ global transformations, respectively. Because of
EWSB, the $SU(2)_{L,R}$ symmetries are broken down to
the diagonal $SU(2)_C$, which in turn is explicitly broken by the
gauged $U(1)_Y$ and by the heterogeneity of the fermion masses.  On
the other hand, while insertions of the Higgs particle are weighted
down as $h/f$, as explained above, its couplings are now
(model-dependent) general functions. In all generality, the $SU(2)_L$
structure is absent in them and, as often pointed
out (e.g. Refs.~\cite{Grinstein:2007iv,Contino:2010mh}), the resulting effective Lagrangian can
describe many setups including that for a light SM singlet isoscalar.

To our knowledge, the first attempts to formulate a non-linear
effective Lagrangian in the presence of a ``non standard/singlet light
Higgs boson" go back to the
90's~\cite{Bagger:1993zf,Koulovassilopoulos:1993pw}, and later
works~\cite{Burgess:1999ha,Grinstein:2007iv}. More recently,
Ref.~\cite{Azatov:2012bz} introduced a relevant set of operators,
while Ref.~\cite{Alonso:2012px} derived a complete effective
Lagrangian basis for pure gauge and gauge-$h$ operators up to four
derivatives. Later on, Ref.~\cite{Buchalla:2013rka} added the pure
Higgs operator in Ref.~\cite{Contino:2011np}
as well as fermionic couplings, proposing a complete basis for all SM
fields up to four derivatives, and trading some of the operators in
Ref.~\cite{Alonso:2012px} by fermionic ones\footnote{The inferred
criticisms in Ref.~\cite{Buchalla:2013rka} to the results in
Ref.~\cite{Alonso:2012px} about missing and redundant operators are
incorrect: Ref.~\cite{Alonso:2012px} concentrated by definition in
pure gauge and gauge-$h$ couplings and those criticized as ``missing"
are not in this category; a similar comment applies to the redundancy
issue, explained by the choice mentioned above of trading some gauge
operators by fermionic ones in Ref.~\cite{Buchalla:2013rka}. Finally,
the $\xi$ weights and the truncations defined for the first time in
Ref.~\cite{Alonso:2012px} lead to rules for operator weights
consistent with those defined long ago in the Georgi-Manohar
counting~\cite{Manohar:1983md}, and more recently in
Ref.~\cite{Jenkins:2013sda}.}.

The effective linear and chiral Lagrangians with a light Higgs
 particle $h$ are intrinsically different, in particular from the
 point of view of the transformation properties under the $SU(2)_L$ symmetry. There is not a
 one-to-one correspondence of the leading corrections of both
 expansions, and one expansion is not the limit of the other unless
 specific constraints are imposed by hand -as illustrated below- or
 follow from  particular dynamics at high
 energies~\cite{SigmaInPreparation}. In the linear expansion, the
 physical Higgs $h$ participates in the scalar $SU(2)_L$ doublet
 $\Phi$; having canonical mass dimension one, this field appears
 weighted by powers of the cut-off $\Lambda$ in any non-renormalizable
 operator and, moreover, its presence in the Lagrangian must
 necessarily respect a pattern in powers of $(v+h)$. In the non-linear
 Lagrangian instead, the behavior of the $h$ particle does not abide
 any more to that of an $SU(2)_L$ doublet, but $h$ appears as a SM
 singlet.
%  and in fact, more generally,
%  the gauge $SU(2)_L$ symmetry is not explicit when realized
%  non-linearly
%  %. The
 %ordered series boils down to an expansion in derivatives of Lorentz
 %and $U(1)_{em}$ invariant couplings: 
  Less symmetry constraints means more possible invariant operators~\cite{Chanowitz:1987vj,Burgess:1992va,Burgess:1992gx} at a given
 order, and in summary:
\begin{itemize}
\item[-] In the non-linear realization, the chiral-symmetry breaking
interactions of $h$ are now generic/arbitrary functions
$\cF(h)$.
\item[-] Furthermore, a relative reshuffling of the order at which
couplings appear in each expansion takes
place~\cite{Alonso:2012jc,Alonso:2012px,Alonso:2012pz}. As a
consequence, a higher number of independent (uncorrelated) couplings
are present in the leading corrections for a non-linear Lagrangian. 
 \end{itemize}
Both effects increase the relative freedom of the purely
phenomenological Lorentz and $U(1)_{\rm{em}}$ couplings required at a
given order of the expansion, with respect to the linear
analysis. Decorrelations induced by the first point have been recently
stressed in Ref.~\cite{Isidori:2013cga} (analysing form factors for
Higgs decays), while those resulting from the second point above lead
to further discriminating signals and should be taken into account
as well. Both types of effects will be explored below.

In what respects the analysis of present LHC and electroweak data,
a first step in the direction of using a non-linear 
realization  was the substitution of the functional dependence on $(v+h)$
for a doublet Higgs in the linear expansion by a generic function
$\cF(h)$ for a generic SM scalar singlet $h$, mentioned in the first
point above. This has already led to a rich phenomenology \cite{Azatov:2012bz,Azatov:2012qz,Isidori:2013cla,Isidori:2013cga}.
Nevertheless, the scope of the decorrelations that a generic $\cF(h)$
induces between the pure gauge and the gauge-$h$ part of a given
operator is limited: whenever data set a strong constraint on the pure
gauge part of the coupling, that is on the global operator
coefficient, this constraint also affects  the gauge-$h$ part as it is
also proportional to the global coefficient; only in appealing to strong
and, in general, unnatural fine-tunings of the constants inside $\cF(h)$
could that constraint be overcome.

As for the second consequence mentioned above, the point is that if
higher orders in both expansions are considered, all possible Lorentz and
$U(1)_\text{em}$ couplings would appear in both towers (as it is
easily seen in the unitary gauge), but not necessarily at the same
leading or sub-leading order. One technical key to understand this
difference is the adimensionality of the field $\UH(x)$.  The induced
towering of the leading low-energy operators is different for the
linear and chiral regimes, a fact illustrated recently for the pure
gauge and gauge-$h$ effective non-linear
Lagrangian~\cite{Alonso:2012jc,Alonso:2012px,Alonso:2012pz}. More
recently, and conversely, an example was pointed
out~\cite{Buchalla:2013rka} of a $d=6$ operator of the linear
expansion whose equivalent coupling does not appear among the leading
derivative corrections in the non-linear expansion.

It will be shown below that, due to that reshuffling of the order at
which certain leading corrections appear, correlations that are
expected as leading corrections in one case may not hold in the other,
unless specific constraints are imposed by hand or follow from  high
energy dynamics. Moreover, interactions that are strongly suppressed
(subleading) in one regime may be leading order in the other.

In this paper we will first consider the basis of CP-even bosonic
operators for the general non-linear effective Lagrangian 
and analyse
in detail its complete and independent set of pure gauge and
gauge-Higgs operators, implementing the tree-level renormalization
procedure and deriving the corresponding Feynman rules. 
The similarities and differences with the couplings obtained in the
linear regime will be carefully determined, considering in particular the Hagiwara-Ishihara-Szalapski-Zeppenfeld (HISZ) basis~\cite{Hagiwara:1993ck,Hagiwara:1996kf}. Nevertheless, the physical results are checked to be independent of the specific linear basis used, as they should be. The comparison of the
effects in both realizations will be performed in the context of
complete bases of gauge and/or Higgs boson operators: all possible
independent (and thus non-redundant) such operators will be
contemplated for each expansion, and compared.
 For each non-linear
operator we will  identify  linear ones which lead to the same
gauge couplings, and it will be shown that up to $d=12$ linear operators  would be 
required to cover all the non-linear operators with at most four
derivatives.  We will then identify some of the most promising signals
to discriminate experimentally among both expansions in hypothetical
departures from the size and Lorentz structure of couplings predicted 
by the SM. This task is facilitated by the partial
use of results obtained earlier on the physics impact of the linear
regime on LHC physics from $d=6$ operators in
Refs.~\cite{Corbett:2012dm,Corbett:2012ja,Corbett:2013pja}, and from
previous analysis of $4$-point phenomenological couplings carried out
in
Refs.~\cite{Brunstein:1996fz,Belyaev:1998ih,Eboli:2000ad,Eboli:2003nq,Eboli:2006wa}. In
this paper we concentrate on the tree-level effects of operators, as a
necessary first step before loop effects are
considered~\cite{LoopPreparation}.
 
The structure of the paper can be easily inferred from the Table of Contents.

%
%%%%%%%%%%%%%%%%%%%%%%%%%   2.  The effective Lagrangian   %%%%%%%%%%%%
%

\section{The effective Lagrangian}
\label{EffectiveLagrangian}

We describe below the effective Lagrangian for a light dynamical
Higgs~\cite{Alonso:2012px} (see also Ref.~\cite{Buchalla:2013rka}),
restricted to the bosonic operators, except for the
Yukawa-like interactions, up to operators with four
derivatives\footnote{As usual, derivative is understood in the sense
of covariant derivative. That is, a gauge field and a momentum have
both chiral dimension one and their inclusion in non-renormalizable
operators is weighted down by the same high-scale
$\Lambda_s$. }. Furthermore, only CP-even operators will be taken into
account, under the assumption that $h$ is a CP-even scalar. 

The most up-to-date analysis to the Higgs results have established that 
the couplings of $h$ to the gauge bosons and the absolute value of the 
couplings to fermions are compatible with the SM ones. 
On the contrary, the sign of the couplings between $h$ and fermions is 
still to be measured, even if a
slight preference for a positive value is indicated in some two parameter
fits (see for example~\cite{Espinosa:2012im,Montull:2012ik,Azatov:2012qz}) 
which take into account one-loop induced EW corrections. 
It is then justified to write the effective Lagrangian as
a term $\LL_0$, which is in fact the SM Lagrangian except for the
mentioned sign (would the latter be confirmed positive, $\LL_0$ should
be exactly identified with the SM Lagrangian $\LL_0=\LL_{SM}$), and to
consider as corrections the possible departures from it due to the
unknown high-energy strong dynamics:
\beq
\LL_\text{chiral} = \LL_0+\Delta\LL\,.
\label{Lchiral}
\eeq 
This description is  data-driven  and, while being a
consistent chiral expansion up to four derivatives, the particular
division in Eq.~(\ref{Lchiral}) does not match that in number of
derivatives, usually adopted by chiral Lagrangian practitioners. For
instance, the usual custodial breaking term
$\tr(\TL\VL_\mu)\tr(\TL\VL^\mu)$ is a two derivative operator and is
often listed among the leading order set in the chiral expansion;
however, it is not present in the SM at tree level and thus here it belongs to $\Delta\LL$ by definition. Moreover,
data strongly constrain its coefficient so that it can
be always considered~\cite{Contino:2010mh} a subleading operator.

The first term in $\LL_\text{chiral}$ reads then 
\beq
\begin{split}
\LL_0 =& \frac{1}{2} (\derp_\mu h)(\derp^\mu h) -\dfrac{1}{4}\WWd^a W^{a\mu\nu}-\dfrac{1}{4}\BBd\BBu-\dfrac{1}{4} G^a_{\mu\nu}G^{a\mu\nu}- V (h)\\
 &-\dfrac{(v+h)^2}{4}\tr[\VL_\mu\VL^\mu]+ i\bar{Q}\slashed{D}Q+i\bar{L}\slashed{D}L\\
 &-\dfrac{v+ s_Y h}{\sqrt2}\left(\bar{Q}_L\UH \cY_Q Q_R+\hc\right)-\dfrac{v+ s_Y h}{\sqrt2}\left(\bar{L}_L\UH \cY_L L_R+\hc\right)\,,
\end{split}
\label{LLO}
\eeq 
where \mbox{$\VL_\mu\equiv \left(\DLR_\mu\UH\right)\UH^\dagger$} ($\TL\equiv\UH\sigma_3\UH^\dag$) is the vector (scalar) chiral field transforming in the adjoint of $SU(2)_L$. The covariant derivative reads 
\beq
\DLR_\mu \UH(x) \equiv \derp_\mu \UH(x) +igW_{\mu}(x)\UH(x) - 
                      \dfrac{ig'}{2} B_\mu(x) \UH(x)\sigma_3 \, , 
\eeq
with $W_\mu\equiv W_{\mu}^a(x)\sigma_a/2$ and $B_\mu$ denoting the
$SU(2)_L$ and $U(1)_Y$ gauge bosons, respectively. In Eq.~(\ref{LLO}),
the first line describes the $h$ and gauge boson kinetic terms, and
the effective scalar potential $V(h)$, accounting for the breaking of
the electroweak symmetry. The second line describes the $W$ and $Z$
masses and their interactions with $h$, as well as the kinetic terms
for GBs and fermions.  Finally, the third line corresponds to the
Yukawa-like interactions written in the fermionic mass eigenstate
basis, where $s_Y\equiv\pm$ encodes the experimental uncertainty on
the sign in the $h$-fermion couplings. A compact notation for the
right-handed fields has been adopted, gathering them into
doublets\footnote{The Cabibbo-Kobayashi-Maskawa mixing is understood
to be encoded in the definition of $Q_L$.} $Q_R$ and $L_R$. $\cY_Q$
and $\cY_L$ are two $6\times6$ block-diagonal matrices containing the
usual Yukawa couplings:
\beq
\cY_Q\equiv\diag\left(Y_U,\, Y_D\right)\,,\qquad\qquad
\cY_L\equiv  \diag\left(Y_\nu,\, Y_L\right)\,.
\eeq

$\Delta\LL$ in Eq.~(\ref{Lchiral}) includes all bosonic (that is, pure
gauge and gauge-$h$ operators plus pure Higgs ones) and Yukawa-like
operators that describe deviations from the SM picture due to the
strong interacting physics present at scales higher than the EW one,
in an expansion up to four derivatives~\cite{Alonso:2012px}:
\begin{align}
\Delta\LL= & \,\xi\left[c_B\cP_B(h) + c_W\cP_W(h)+c_G\cP_G(h) +c_C \cP_C(h) + c_T \cP_T(h)+c_H \cP_H(h)+c_{\Box H} \cP_{\Box H}(h)\right]\nn\\
   &+\xi \sum_{i=1}^{10} c_i\cP_i(h)+\xi^2\sum_{i=11}^{25} c_i\cP_i (h)+ \xi^4 c_{26}\cP_{26}(h)+ \Sigma_i \xi^{n_i} c^i_{HH} \cP^i_{HH}(h)
\label{DeltaL}
\end{align}
where $c_i$ are model-dependent constant coefficients, and the last
term account for all possible pure Higgs operators weighted by $\xi^{n_i}$ with $n_i\geq2$. The set of pure-gauge and gauge-$h$ operators are defined
by~\cite{Alonso:2012px}\footnote{The set of pure gauge and gauge-$h$ operators exactly matches that inRef.~\cite{Alonso:2012px}; nevertheless,  the labelling of some operators here and  their $\xi$-weights  are corrected with respect to those  in Ref.~\cite{Alonso:2012px}, see later.}:
\begin{description}
\item[Weighted by $\xi$:]
\beq
\hspace{-1cm}
\begin{aligned}
&\cP_C(h) = -\frac{v^2}{4}\tr(\VL^\mu \VL_\mu) \cF_{C}(h)\,\,
&&\cP_{4}(h) = ig' \BBd \tr(\TL\VL^\mu) \derp^\nu\cF_{4}(h)\\
&\cP_T(h) = \frac{v^2}{4} \tr(\TL\VL_\mu)\tr(\TL\VL^\mu) \cF_{T}(h)\,\,
&&\cP_{5}(h)= ig \tr(\WWd\VL^\mu) \derp^\nu\cF_{5}(h)\\
&\cP_{B}(h) =-\frac{g'^2}{4}\BBd \BBu \cF_B(h) \,\,
&&\cP_{6}(h) = (\tr(\VL_\mu\VL^\mu))^2 \cF_{6}(h)\\
&\cP_{W}(h) =-\frac{g^2}{4}\WWd^a W^{a\mu\nu} \cF_W(h) \,\,
&&\cP_{7}(h) = \tr(\VL_\mu\VL^\mu) \derp_\nu\derp^\nu\cF_{7}(h)  \\
&\cP_G(h) = -\frac{g_s^2}{4}G_{\mu\nu}^a G^{a\mu\nu}\cF_G(h)\,\,
&&\cP_{8}(h) = \tr(\VL_\mu\VL_\nu) \derp^\mu\cF_{8}(h)\derp^\nu\cF_{8}'(h) \\
&\cP_{1}(h) = gg' \BBd \tr(\TL \WWu) \cF_{1}(h)\,\,
&&\cP_{9}(h) = \tr((\DLL_\mu\VL^\mu)^2) \cF_{9}(h)\\
&\cP_{2}(h) = ig'\BBd \tr(\TL[\VL^\mu,\VL^\nu]) \cF_{2}(h)\,\,
&&\cP_{10}(h) = \tr(\VL_\nu\DLL_\mu\VL^\mu) \derp^\nu\cF_{10}(h)\\
&\cP_{3}(h)= ig \tr(\WWd [\VL^\mu,\VL^\nu]) \cF_{3}(h)
\end{aligned}
\label{Opxi}
\eeq
\item[Weighted by $\xi^2$:]
\beq
\hspace{-1cm}
\begin{aligned}
&\cP_{11}(h) = (\tr(\VL_\mu\VL_\nu))^2 \cF_{11}(h)\,\,
&&\cP_{19}(h) = \tr(\TL\DLL_\mu\VL^\mu)\tr(\TL\VL_\nu) \derp^\nu\cF_{19}(h)\\
&\cP_{12}(h) = g^2 (\tr(\TL\WWd))^2 \cF_{12}(h) \,\,
&&\cP_{20}(h) = \tr(\VL_\mu\VL^\mu) \derp_\nu\cF_{20}(h)\derp^\nu\cF_{20}'(h)   \\
&\cP_{13}(h) = ig \tr(\TL\WWd)\tr(\TL[\VL^\mu,\VL^\nu]) \cF_{13}(h) \,\,
&&\cP_{21}(h) = (\tr(\TL\VL_\mu))^2 \derp_\nu\cF_{21}(h)\derp^\nu\cF_{21}'(h)  \\
&\cP_{14}(h)  =g \e^{\mu\nu\rho\lambda} \tr(\TL\VL_\mu) \tr(\VL_\nu W_{\rho\lambda}) \cF_{14}(h)\,\,
&&\cP_{22}(h) = \tr(\TL\VL_\mu)\tr(\TL\VL_\nu) \derp^\mu\cF_{22}(h)\derp^\nu\cF_{22}'(h)\\
&\cP_{15}(h) = \tr(\TL\DLL_\mu\VL^\mu)  \tr(\TL\DLL_\nu\VL^\nu) \cF_{15}(h)  \,\,
&&\cP_{23}(h) = \tr(\VL_\mu\VL^\mu) (\tr(\TL\VL_\nu))^2 \cF_{23}(h)\\
& \cP_{16}(h) = \tr([\TL,\VL_\nu]\DLL_\mu\VL^\mu) \tr(\TL\VL^\nu) \cF_{16}(h)  \,\,
&&\cP_{24}(h) = \tr(\VL_\mu\VL_\nu)\tr(\TL\VL^\mu)\tr(\TL\VL^\nu) \cF_{24}(h)\\
&\cP_{17}(h) = ig \tr(\TL \WWd) \tr(\TL\VL^\mu) \derp^\nu\cF_{17}(h)\,\,
&&\cP_{25}(h)= (\tr(\TL\VL_\mu))^2 \derp_\nu\derp^\nu\cF_{25}(h)\\
&\cP_{18}(h) = \tr(\TL[\VL_\mu,\VL_\nu])\tr(\TL\VL^\mu) \derp^\nu\cF_{18}(h)\,\,
\end{aligned}
\label{Opxi2}
\eeq
\item[Weighted by $\xi^4$:]
\beq
\hspace{-1cm}
\cP_{26}(h) = (\tr(\TL\VL_\mu)\tr(\TL\VL_\nu))^2 \cF_{26}(h) \,.
\label{Opxi4}
\eeq
\end{description}
In Eq.~(\ref{Opxi2}), $\DLL_\mu$ denotes the covariant derivative on a field transforming 
in the adjoint representation of $SU(2)_L$, i.e. 
\begin{equation}
\DLL_\mu \VL_\nu \equiv \partial_\mu \VL_\nu +i g \left[ W_\mu, \VL_\nu \right] \,.
\end{equation}
Finally, the pure Higgs operators are:
\begin{description}
\item[Weighted by $\xi$:] this set includes two operators, one with two derivatives and one with four,
\beq
\cP_H(h) = \frac{1}{2}(\derp_\mu h)(\derp^\mu h) \cF_H(h)\,,
\qquad\qquad
\cP_{\square H}=\frac{1}{v^2}(\derp_\mu \derp^\mu h)^2\cF_{\square H}(h)\,.
\label{Opxih}
\eeq
In spite of not containing gauge interactions, they will be considered here as they affect the renormalization of SM parameters, and the propagator of the $h$ field, respectively.
\item[Weighted by $\xi^{\ge2}$:] this class consists of all possible pure Higgs operators with four derivatives weighted by $\xi^{\ge2}$, $\cP^i_{HH}(h)$. We refrain from listing them here, as pure-$h$ operators are beyond the scope of this work and therefore they will not be taken into account in the phenomenological sections below. An example of $\xi^2$-weighted operator would be~\cite{Goldberger:2007zk, Contino:2011np}
\beq
\cP_{DH}(h)={\frac{1}{v^4} \left((\derp_\mu h)(\derp^\mu h)\right)^2 \cF_{D H}(h)\,.}
\eeq
\end{description}

In another realm, note that $\cP_C(h)$, $\cP_T(h)$ and $\cP_H(h)$ are two-derivative operators and
would be considered among the leading terms in any formal analysis of the non-linear expansion (as explained after Eq.~(\ref{Lchiral})), a fact of no consequence below.
 
 The $\xi$ weights within $\Delta\LL$ do {\bf not} reflect an
 expansion in $\xi$, but a reparametrisation that facilitates the
 tracking of the lowest dimension at which a ``sibling'' of a given
 operator appears in the linear expansion. To guarantee the procedure, such an analysis requires to compare with a specific linear basis; complete linear bases are only available up to $d=6$ and here we use the completion of the original HISZ basis~\cite{Hagiwara:1993ck,Grzadkowski:2010es}, see Sect.~\ref{sec:Linear}.
  
 A sibling of a chiral
 operator $\cP_i(h)$ is defined as the operator of the linear
 expansion whose pure gauge interactions coincide with those described
 by $\cP_i(h)$. The canonical dimension $d$ of the sibling, that is
 the power of $\xi$, is thus an indicator of at which order in the
 linear expansion it is necessary and sufficient to go to account for
 those gauge interactions: operators weighted by $\xi^n$ require us to
 consider siblings of canonical dimension $d=4+2n$. It may happen that
 an operator in Eqs.~(\ref{Opxi})-(\ref{Opxih}) corresponds to a
 combination of linear operators with different canonical dimensions:
 the power of $\xi$ refers then to the lowest dimension of such
 operators that leads to the same phenomenological gauge
 couplings. The lowest dimensional siblings of the operators in
 Eqs.~(\ref{Opxi}) and (\ref{Opxih}) have $d=6$; those in
 Eqs.~(\ref{Opxi2}) have $d=8$; that of Eq.~(\ref{Opxi4}) has $d=12$. $\xi$ is not a physical
 quantity {\it per se} in the framework of the effective
 Lagrangian. If preferred by the reader, the $\xi$ weights can be
 reabsorbed in a redefinition of the coefficients $c_i$ and be
 altogether forgotten; nevertheless, they allow a fast connection with
 the analyses performed in the linear expansion, as illustrated later
 on.

In the Lagrangian above, Eq~(\ref{DeltaL}), we have chosen a definition of the operator coefficients which does not  make explicit the weights expected from Naive Dimensional Analysis (NDA)~\cite{Manohar:1983md,Jenkins:2013sda,Buchalla:2013eza}. While the NDA rules  are known not to apply to the gauge and scalar kinetic and mass terms, for the higher-order corrections they would imply suppressions by factors of the goldstone boson scale $f$ versus the high energy scale $\Lambda_s$. In particular, the coefficients of all operators in Eq. (2.6) except $P_C(h)$, as well as all operators in Eqs.~(\ref{Opxi2}), (\ref{Opxi4}) and (\ref{Opxih}), would be suppressed by the factor $f^2/\Lambda_s^2= 1/(16\pi^2)$. The coefficients can be easily redefined by the reader if wished.

The $\cF(h)$ functions encode the chiral interactions of the light
$h$, through the generic dependence on $(\mean{h}+h)$, and are model
dependent. Each function can be defined by $\cF(h)\equiv g_0(h,v) +
\xi g_1(h,v) + \xi^2 g_2(h,v) + \ldots$, where $g_i(h,v)$ are
model-dependent functions of $h$ and of $v$, once $\mean{h}$ is
expressed in terms of $\xi$ and $v$. Here we will assume that the
$\cF(h)$ functions are completely general polynomials of $\mean{h}$
and $h$ (not including derivatives of $h$). Notice that when using the
equations of motion (EOM) and integration by parts to relate
operators, $\cF(h)$ would be assumed to be redefined when convenient,
much as one customarily redefines the constant operator coefficients.

The insertions of the $h$ field, explicit or through generic
functions, deserve a separate comment: given their goldstonic origin,
they are expected to be suppressed by the goldstone boson scale as
$h/f$, as it has been already specified above.
This is encoded in the present formalism by the combination of the
$F_i(h)$ functions as defined in the text and the pertinent
$\xi$-weights which have been explicitly extracted from them, as they
constitute a useful tool to establish the relation with the linear
expansions. Consider an initial generic dependence on the $h$ field of
the form $(h + \mean{h})/f= \sqrt{\xi} (h+\mean{h})/v $: for instance
in the linear regime, in which $\mean{h}\sim v$, the $F_i(h)$
functions are defined in the text as leading to powers of $(1+h/v)$,
because the functional $\xi$-dependence has been made explicit in the
Lagrangian.\\

\subsubsection*{Connection to fermionic operators}

Several operators in the list in Eqs.~(\ref{Opxi})-(\ref{Opxi4}) are
independent only in the presence of massive fermions: these are
$\cP_{9}(h)$, $\cP_{10}(h)$, $\cP_{15}(h)$, $\cP_{16}(h)$, $\cP_{19}(h)$, one out of $\cP_{6}(h)$, $\cP_{7}(h)$ and $\cP_{20}(h)$, and one out of $\cP_{21}(h)$, $\cP_{23}(h)$ and $\cP_{25}(h)$. Indeed, $\cP_{9}(h)$, $\cP_{10}(h)$, $\cP_{15}(h)$, $\cP_{16}(h)$, and $\cP_{19}(h)$ contain the contraction $\DLL_\mu\,\VL^\mu$ that is
connected with the Yukawa couplings\cite{Alonso:2012px}, through the
manipulation of the gauge field EOM and the Dirac equations (see
App.~\ref{AppEOM} for details). When fermion masses are neglected,
these five operators can be written in terms of the other operators in
the basis (see Eq.~(\ref{PDVfermions})). Furthermore, using the light
$h$ EOM (see Eq.~(\ref{EOMh})), operator $\cP_{7}(h)$ ($\cP_{25}(h)$)
can be reduced to a combination of $\cP_{6}(h)$ and $\cP_{20}(h)$
($\cP_{21}(h)$ and $\cP_{23}(h)$), plus a term that can be absorbed in the
redefinition of the $h$-gauge boson couplings, plus a term containing
the Yukawa interactions (see App.~\ref{AppEOM} for details).
In summary, all those operators must be included to have a complete and
independent bosonic basis; nevertheless, in the numerical analysis in
Sect.~\ref{numerical} their effect will be disregarded as the impact
of fermion masses on data analysis will be negligible.

Other operators in the basis in Eqs.(\ref{Opxi})-(\ref{Opxih}) can be
traded by fermionic ones independently of the size of fermion masses,
applying the EOM for $ \DLL_\mu W^{\mu \nu}$ and $ \partial_\mu B^{\mu
\nu}$, see Eqs.~(\ref{EOMW}), (\ref{EOMB}) and (\ref{RelH}) in
App.~\ref{AppEOM}.  The complete list of fermionic operators that are
related to the pure gauge and gauge-$h$ basis in
Eqs.~(\ref{Opxi})-(\ref{Opxih}) can also be found there\footnote{For
completeness, the EOM of the gauge bosons, $h$ and $\UH(h)$, and the
Dirac equations as well as the full list of fermionic operators that
are related to the bosonic ones in Eqs.~(\ref{Opxi})-(\ref{Opxih}) are
presented in App.~\ref{AppEOM}. In this paper, we will only rely on
bosonic observables and therefore we will not consider any fermionic
operators other than those mentioned.}.  This trading procedure can
turn out to be very
useful~\cite{Giudice:2007fh,Corbett:2012ja,Elias-Miro:2013mua,Pomarol:2013zra,Jenkins:2013zja}
when analysing certain experimental data if deviations from the SM
values for the $h$-fermion couplings were found. A basis including all
possible fermionic couplings could be more useful in such a
hypothetical situation.  The bosonic basis defined above is instead
``blind"~\cite{DeRujula:1991se} to some deviations in fermionic
couplings. This should not come as a surprise: the choice of basis
should be optimized with respect to the experimental data under
analysis and the presence of blind directions is a common feature of
any basis. In this work we are focused in exploring directly the
experimental consequences of anomalous gauge and gauge-$h$ couplings
and Eqs.(\ref{Opxi})-(\ref{Opxih}) are the appropriate analysis tool.

\subsubsection*{Custodial symmetry}

In the list in Eqs.(\ref{Opxi})-(\ref{Opxih}), the operators
$\cP_{\Box H}(h)$, $\cP_T(h)$, $\cP_{1}(h)$, $\cP_{2}(h)$,
$\cP_{4}(h)$, $\cP_{9}(h)$, $\cP_{10}(h)$ and $\cP_{12-26}(h)$ are
custodial symmetry breaking, as either they: i) are related to fermion
masses; ii) are related to the hypercharge through $g'B_{\mu\nu}$;
iii) they contain the scalar chiral operator $\TL$ but no
$B_{\mu\nu}$.  Among these, only $\cP_T(h)$ and $\cP_1(h)$ are
strongly constrained from electroweak precision test, while the
phenomenological impact of the remaining operators has never been
studied and therefore they could lead to interesting effects.

If  instead by ``custodial breaking" operators one refers only to those in iii), a complete set of bosonic custodial preserving operators is given by the following eighteen operators:
\beq
\cP_{H}(h)\,,\quad
\cP_{\Box H}(h)\,,\quad
\cP_{C}(h)\,,\quad
\cP_{B}(h)\,,\quad
\cP_{W}(h)\,,\quad
\cP_{G}(h)\,,\quad
\cP_{1-11}(h)\,,\quad
\cP_{20}(h)\,.
\eeq
Furthermore, if fermion masses are neglected, this ensemble is further reduced to a set of fourteen independent operators, given by
\beq
\cP_{H}(h)\,,\quad
\cP_{C}(h)\,,\quad
\cP_{B}(h)\,,\quad
\cP_{W}(h)\,,\quad
\cP_{G}(h)\,,\quad
\cP_{1-5}(h)\,,\quad
\cP_{8}(h)\,,\quad
\cP_{11}(h)\,,
\eeq
plus any two among the following three operators:  
\beq
\cP_{6}(h)\,,\quad
\cP_{7}(h)\,,\quad
\cP_{20}(h)\,.
\eeq
Under the same assumptions (no beyond SM sources of custodial breaking and massless fermions), a subset of only twelve operators  has been previously proposed in Ref.~\cite{Azatov:2012bz}, as this reference in addition restricted  to operators that lead to cubic and quartic vertices of GBs and gauge bosons {\it and} including one or two Higgs bosons.

\vspace{0.5cm}

The Lagrangian in Eq.(\ref{Lchiral}) is very general and can be used
to describe an extended class of Higgs models, from the SM scenario
with a linear Higgs sector (for $\mean{h}=v$, $\xi=0$ and $s_Y=1$), to
the technicolor-like ansatz (for $f\sim v$ and omitting all terms in
$h$) and intermediate situations with a light scalar $h$ from
composite/holographic Higgs models
\cite{Dimopoulos:1981xc,Kaplan:1983fs,Kaplan:1983sm,Banks:1984gj,Georgi:1984ef,
Georgi:1984af,Dugan:1984hq,Agashe:2004rs,Contino:2006qr,Gripaios:2009pe}
(in general for $f\ne v$) up to dilaton-like scalar frameworks
\cite{Halyo:1991pc,Goldberger:2007zk,Vecchi:2010gj,Campbell:2011iw,Matsuzaki:2012mk,
Chacko:2012vm,Bellazzini:2012vz}
(for $f\sim v$), where the dilaton participates in the electroweak
symmetry breaking.

%
%%%%%%%%%%%%%%%%%%%%%%%%%   3. Comparison with the linear regime   %%%%%%%%%%%%
%

\section{Comparison with the linear regime}
\label{ComparisonLinear}

The chiral and linear approaches are essentially different from each
other, as explained in the introduction.  The reshuffling with respect
to the linear case of the order at which the leading operators appear
plus the generic dependence on $h$ imply that correlations among
observables present in one scenario may not hold in the other and,
moreover, interactions that are strongly suppressed in one case may be
leading corrections in the other. As the symmetry respected by the
non-linear Lagrangian is smaller, more freedom is generically expected
for the latter. In this section, for the sake of comparison we will
first present the effective Lagrangian in the linear regime, restricting to the HISZ basis~\cite{Hagiwara:1993ck,Hagiwara:1996kf}, and
discuss then the coincidences and differences expected in observable
predictions. The relation to another basis~\cite{Giudice:2007fh} can be found in App.~\ref{AppSILH}.

%%%
% 3.1 The effective Lagrangian in the linear regime
%%%

\subsection{The effective Lagrangian in the linear regime}
\label{sec:Linear}

Following the description pattern in Eq.~(\ref{Lchiral}), the
effective Lagrangian in the linear regime can be written accordingly
as
\beq
\LL_\text{linear} = \LL_{SM}+\Delta\LL_\text{linear}\,,
\label{Llinear}
\eeq
where the relation with the non-linear Lagrangian in Eq.~(\ref{LLO})
is given by $\LL_{SM}=\LL_{0}|_{s_Y=1}$, and $\Delta\LL_\text{linear}$
contains operators with canonical dimension $d>4$, weighted down by
suitable powers of the ultraviolet cut-off scale
$\Lambda$. Restricting to $CP$-even and baryon and lepton number
preserving operators, the leading $d=6$ corrections
\begin{equation}
\Delta \LL^{d=6}_\text{linear} = \sum_i \frac{f_i}{\Lambda^2} \cO_i\,, 
\label{DeltaLlinear}
\end{equation}
may be parametrized via a complete basis of 
operators~\cite{Buchmuller:1985jz,Grzadkowski:2010es}.  
Only a small subset of those modify the Higgs couplings 
to gauge bosons. Consider the HISZ basis~\cite{Hagiwara:1993ck,Hagiwara:1996kf}:
\begin{align}
&\cO_{GG} = \Phi^\dagger \Phi \,G^a_{\mu\nu} G^{a\mu\nu}\,,
&&\cO_{WW} = \Phi^{\dagger} \hat{W}_{\mu \nu} \hat{W}^{\mu \nu} \Phi\,,\nn \\
&\cO_{BB} = \Phi^{\dagger} \hat{B}_{\mu \nu} \hat{B}^{\mu \nu} \Phi\,,\nn 
&&\cO_{BW} =  \Phi^{\dagger} \hat{B}_{\mu \nu} \hat{W}^{\mu \nu} \Phi\,, \\
&\cO_W  = (\DL_{\mu} \Phi)^{\dagger} \hat{W}^{\mu \nu}  (\DL_{\nu} \Phi)\,, 
&&\cO_B  =  (\DL_{\mu} \Phi)^{\dagger} \hat{B}^{\mu \nu}  (\DL_{\nu} \Phi)\,,
\label{LinearOpsGauge}  \\
&\cO_{\Phi,1} =  \left (\DL_\mu \Phi \right)^\dagger \Phi\  \Phi^\dagger
\left (\DL^\mu \Phi \right )\,,
&&\cO_{\Phi,2} = \frac{1}{2} \partial^\mu\left ( \Phi^\dagger \Phi \right)
\partial_\mu\left ( \Phi^\dagger \Phi \right)\,,\nn \\
&\cO_{\Phi,4} = \left (\DL_\mu \Phi \right)^\dagger \left(\DL^\mu\Phi 
\right)\left(\Phi^\dagger\Phi \right )\,,\nn
\end{align}
where $\DL_\mu\Phi= \left(\partial_\mu+ \frac{i}{2} g' B_\mu 
+ \frac{i}{2}g\sigma_i W^i_\mu \right)\Phi $ and 
 $\hat{B}_{\mu \nu} \equiv \frac{i}{2} g'B_{\mu \nu}$ and 
$\hat{W}_{\mu\nu} \equiv \frac{i}{2} g\sigma_i W^i_{\mu\nu}$.
An additional operator is commonly added in phenomenological analysis,
\beq
\cO_{\Phi,3}=\dfrac{1}{3}\left(\Phi^\dag\Phi\right)^3\,,
\label{LinearOpsGaugePhi3}
\eeq
which is a pure Higgs operator. An equivalent basis of ten operators in the linear expansion is often
used nowadays instead of the previous set of ten linear operators: the
so-called SILH~\cite{Giudice:2007fh} Lagrangian, in which four of the
operators above are traded by combinations of them and/or by a
fermionic one via EOM (the exact relation with the SILH basis can be
found in App.~\ref{AppSILH}).

The pure Higgs interactions described by the $\xi$-weighted operator $\cP_{\square H}$ of the chiral expansion, Eq.~(\ref{Opxih}), are contained in the linear operator,
\beq
\cO_{\square \Phi}=\left(\DL_\mu \DL^\mu\Phi\right)^\dag\left(\DL_\nu \DL^\nu\Phi\right)\,.
\label{OBoxPhi}
\eeq 

Let us now explore the relation between the linear and non-linear analyses. 
Beyond the different $h$-dependence of the operators, that is 
(in the unitary gauge):
\beq
\Phi=\frac{1}{\sqrt{2}}\left(\begin{array}{c} 0 
\\ v+h(x)\end{array}\right)\qquad {\rm vs.}\qquad \cF(h)\,,
\eeq
it is interesting to explore the relation among the linear operators
in Eqs.~(\ref{LinearOpsGauge}) and those in the chiral expansion. A
striking distinct feature when comparing both basis is the different
number of independent couplings they span.  This is best illustrated
for instance truncating the non-linear expansion at order $\xi$ -which
may be specially relevant for small $\xi$- and comparing the result
with the $d=6$ linear basis that contributes to gauge-Higgs couplings: while the latter basis exhibits ten independent couplings, the former depends on sixteen. A more precise illustration follows when
taking momentarily $\cF_i(h)= \left(1+h/v\right)^n$, with $n=2$ in general, in all $\cP_i(h)$
under discussion, which would lead to:
\begin{align}
&\begin{aligned}
\cO_{BB} &=\,\,\frac{v^2}{2}\cP_B(h)\,,\qquad
&\cO_{WW} &
=\,\,\frac{v^2}{2}\cP_W(h)\,, \\
\cO_{GG} &
=\,\,-\frac{2v^2}{g_s^2} \cP_G(h)\,,\qquad
&\cO_{BW} &
=\,\,\frac{v^2}{8}\cP_1(h)\,, \\
\cO_{B} &
=\,\,\frac{v^2}{16}\cP_2(h)+\frac{v^2}{8}\cP_4(h)\,,\qquad
&\cO_{W} &
=\,\,\frac{v^2}{8}\cP_3(h)-\frac{v^2}{4}\cP_5(h)\,, \\
\cO_{\Phi,1} &
=\,\,\frac{v^2}{2}\cP_H(h)-\frac{v^2}{4}\cF(h)\cP_T(h)\,,\qquad
&\cO_{\Phi,2} &=\,\, v^2 \cP_H(h)\,, \\
\cO_{\Phi,4} &=\,\,\frac{v^2}{2}\cP_H(h)+\frac{v^2}{2}\cF(h) \cP_C(h)\,,
\end{aligned}
\label{LinearChiralCorrelations}\\
&\cO_{\Box\Phi} =\frac{v^2}{2}\cP_{\Box H}(h) + \frac{v^2}{8}\cP_{6}(h)+\frac{v^2}{4}\cP_{7}(h)-v^2\cP_{8}(h)-\frac{v^2}{4}\cP_{9}(h)-\frac{v^2}{2} \cP_{10}(h)\,.\nn
\end{align}
These relations show that five chiral operators, $\cP_B(h)$,
$\cP_W(h)$, $\cP_G(h)$, $\cP_1(h)$ and $\cP_H(h)$ are then in a
one-to-one correspondence with the linear operators $\cO_{BB}$,
$\cO_{WW}$, $\cO_{GG}$, $\cO_{BW}$ and $\cO_{\Phi,2}$,
respectively. Also the operator $\cP_T(h)$ ($\cP_C(h)$) corresponds to
a combination of the linear operators $\cO_{\Phi,1}$ and
$\cO_{\Phi,2}$ ($\cO_{\Phi,4}$ and $\cO_{\Phi,2}$). In contrast, it
follows from Eq.~(\ref{LinearChiralCorrelations}) above that:
\begin{itemize}
\item[-] Only a specific combination of the non-linear operators 
$\cP_2(h)$ and $\cP_4(h)$ corresponds to the linear operator $\cO_B$.
\item[-] Similarly, a specific combination of the non-linear operators 
$\cP_3(h)$ and $\cP_5(h)$ corresponds to the linear operator $\cO_W$.
\item[-] Only a specific combination of the non-linear operators $\cP_{\Box H}(h)$, $\cP_{6}(h)$, $\cP_{7}(h)$, $\cP_{8}(h)$, $\cP_{9}(h)$ and $\cP_{10}(h)$ corresponds to the linear operator 
$\cO_{\Box\Phi}$.
\end{itemize} 
It is necessary to go to the next order in the linear basis, $d=8$, to
identify the operators which break these correlations (see Eq.~(\ref{Siblingsxi2})). It can be
checked that, for example for the first two correlations, the linear $d=8$ operators
\beq
\left((\DL_\mu\Phi)^\dag\Phi\right)\hat{B}^{\mu \nu}\left(\Phi^\dag\DL_\nu\Phi\right)
\qquad\text{and}\qquad
\left((\DL_\mu\Phi)^\dag\Phi\right)\hat{W}^{\mu \nu}\left(\Phi^\dag\DL_\nu\Phi\right)
\eeq
correspond separately to $\cP_4(h)$ and $\cP_5(h)$, respectively. 

A comment is pertinent when considering the $\xi$ truncation.
In the $\xi\rightarrow 0$
limit, in which $\cF(h)\rightarrow\left(1+h/v\right)^2$,
if the underlying theory is expected to account for EWSB,  the ensemble of 
the non-linear Lagrangian should converge to a linear-like pattern. Nevertheless, the size 
of $\xi$ is not known in a model-independent way; starting an analysis by
formulating the problem (only) in the linear expansion is somehow
assuming an answer from the start: that $\xi$ is necessarily small in
any possible BSM construction. Furthermore, the non-linear Lagrangian accounts for more 
exotic singlet scalars, and that convergence is not granted in general.

The maximal set of CP-even independent operators involving gauge
and/or the Higgs boson in any $d=6$ linear basis is made out of 16
operators: the ten~\cite{Hagiwara:1993ck,Hagiwara:1996kf} in Eqs.~(\ref{LinearOpsGauge}) and (\ref{LinearOpsGaugePhi3}), plus the operator~\cite{Grzadkowski:2010es} $\cO_{\square \Phi}$ defined in Eq.~(\ref{OBoxPhi}), and another five which only modify the gauge boson couplings and do not involve
the Higgs field\footnote{The Operators $\cO_{DW}$, $\cO_{DB}$ and
$\cO_{DG}$ are usually traded by $\cO_{WWW}$ and $\cO_{GGG}$ plus
fermionic operators. As in this paper we focus on
 bosonic observables, such translation is not pertinent. 
Taken by themselves, the ensembles discussed constitute a
non-redundant and complete set of gauge and/or Higgs operators. In
$\cO_{DG}$, $ \DLL^\mu$ denotes the covariant derivative acting on a
field transforming in the adjoint of $SU(3)_C$.}~\cite{Hagiwara:1993ck,Hagiwara:1996kf}: 
\beq
\begin{aligned} 
\cO_{WWW}&=i\epsilon_{ijk}\hat{W}_\mu^{i\,\nu} \hat{W}^{j\,\rho}_\nu \hat{W}^{k\,\mu}_\rho\,,
&&&\qquad\cO_{GGG}&=if_{abc}G_\mu^{a\,\nu} G^{b\,\rho}_\nu G^{c\,\mu}_\rho\,,\\
\cO_{DW}&=\left(\DLL^\mu\,\hat{W}_{\mu\nu}\right)^i\left(\DLL_\rho \hat{W}^{\rho\nu}\right)^i\,,
&&&\qquad\cO_{DB}&=\left(\partial^\mu \hat{B}_{\mu\nu}\right)\left(\partial_\rho \hat{B}^{\rho\nu}\right)\,,\\
\cO_{DG}&=\left(\DLL^\mu\,G_{\mu\nu}\right)^a\left(\DLL_\rho G^{\rho\nu}\right)^a\,.
\end{aligned}
\label{PureGaugeLinearOps}
\eeq
The Lorentz structures contained in these five operators are {\it not} present in
the non-linear Lagrangian expanded up to four derivatives: they would
appear only at higher order in that expansion, i.e. six
derivatives. They are not the siblings of any of the chiral operators
discussed in this work, Eqs.~(\ref{Opxi})-(\ref{Opxih}). 

The rest of this paper will focus on how the present and future LHC
gauge and gauge-$h$ data, as well as other data, may generically shed
light on the (non-)linearity of the underlying physics.  In particular
exploiting the decorrelations implied by the discussion above as well
as via new anomalous discriminating signals.

%%%
% 3.2 On the decorrelations
%%%
Disregarding fine tunings, that is, assuming in general all dimensionless operator
coefficients of $\cal{O}$(1), the pattern of dominant signals expected
from each expansion varies because the nature of some leading
corrections is different, or because the expected relation between
some couplings varies. In the next subsections we analyze first how
some correlations among couplings expected in the linear regime are
broken in the non-linear one. Next, it is pointed out that some
couplings expected if the EWSB is linearly realized
are instead expected to appear only as higher order corrections in the
non-linear case. Conversely and finally, attention is paid to new
anomalous couplings expected as leading corrections in the non-linear
regime which appear only at $d\ge8$ of the linear expansion.

\subsection{Decorrelation of signals with respect to the linear analysis}
\label{sec:decor}
\label{decorr}

The parameter $\xi$ is a free parameter in the effective
 approach. Nevertheless, in concrete composite Higgs models
 electroweak corrections imply $\xi\lesssim
 0.2-0.4$~\cite{Grojean:2013qca} (more constraining bounds $\xi\lesssim
 0.1-0.2$ have been advocated in older analyses~\cite{Barducci:2013wjc,Elias-Miro:2013gya,Falkowski:2013dza}), and it is therefore interesting for
 the sake of comparison to consider the truncation of $\Delta\LL$
 which keeps only the terms weighted by $\xi$ and disregard first
 those weighted by higher $\xi$ powers.  We will thus analyze first
 only those operators in Eqs.~(\ref{Opxi}) and (\ref{Opxih}). We will
 refer to this truncation as $\Delta\LL^\xi$ and define
 $\LL^\xi_{chiral}\equiv \LL_0+\Delta\LL^\xi$.

All operators in $\Delta\LL^\xi$ have by definition lowest dimensional
linear siblings of $d=6$. We will thus compare first
$\LL^\xi_{chiral}$ with the $d=6$ linear
expansion~\cite{Buchmuller:1985jz,Grzadkowski:2010es,Giudice:2007fh}. For
low enough values of $\xi$, that is when the new physics scale
$\Lambda_s\gg v$, $\LL^\xi_{chiral}$ is expected to collapse into the
$d=6$ linear Lagrangian if it should account correctly for EW symmetry
breaking via an $SU(2)_L$ doublet scalar, but the non-linear Lagrangian
encodes more general scenarios (for instance that for
a SM singlet) as well.

The comparison of the effects in the non-linear versus the linear expansion is illuminating when done in the context of the maximal set of independent (and thus non-redundant) operators on the gauge-boson/Higgs sector for each expansion: comparing  complete bases of those characteristics. 
The number of independent bosonic operators that induce leading
deviations in gauge-$h$ couplings turns out to be then different for both expansions: ten $d=6$ operators in
the linear expansion, see Eq.~(\ref{LinearOpsGauge}) and Eq.~(\ref{OBoxPhi}), for sixteen
$\xi$-weighted operators\footnote{Note that the first operator in Eq.~(\ref{Opxih}) impacts on the gauge-$h$ couplings via the renormalization of the $h$ field.} in the chiral one, see Eq.~(\ref{Opxi}) and (\ref{Opxih}). For
illustration, further details are given here on one example pointed
out in Sect.~\ref{sec:Linear}: $\cP_2(h)$ and $\cP_4(h)$ versus the
$d=6$ operator $\cO_B$. From Eq.~(\ref{LinearChiralCorrelations}) it
followed that only the combinations $\cP_2(h)+2\cP_4(h)$ have a $d=6$ linear
equivalent (with $\cF_i(h)$ substituted by $(1+h/v)^2$).  In the
unitary gauge $\cP_2(h)$ and $\cP_4(h)$ read:
\begin{align}
\cP_2(h) &=2 ieg^2A_{\mu\nu}W^{-\mu} W^{+\nu}\cF_2(h) - 2 \dfrac{ie^2g
}{\cos\theta_W} Z_{\mu\nu}W^{-\mu} W^{+\nu}\cF_2(h)\,,
\label{P2unitary}\\
\cP_4(h) &= - \dfrac{eg}{\cos\theta_W} A_{\mu\nu}Z^\mu\derp^\nu
\cF_4(h) + \dfrac{e^2}{ \cos^2\theta_W}
Z_{\mu\nu}Z^\mu\derp^\nu\cF_4(h)\,,
\label{P4unitary}
\end{align}
with their coefficients $c_2$ and $c_4$ taking arbitrary
(model-dependent) values.  In contrast, their $d=6$ sibling $\cO_B$
results in the combination: \beq
\begin{split}
\cO_B =& \dfrac{ieg^2}{8} A_{\mu\nu}W^{-\mu} W^{+\nu}(v+h)^2
-\dfrac{ie^2g}{8\cos\theta_W} Z_{\mu\nu}W^{-\mu} W^{+\nu}(v+h)^2 \\
& -\dfrac{eg}{4\cos\theta_W} A_{\mu\nu}Z^\mu\derp^\nu
h(v+h)+\dfrac{e^2}{4\cos^2\theta_W} Z_{\mu\nu}Z^\mu\derp^\nu h(v+h)\,.
\end{split}
\label{OBunitary}
\eeq In consequence, the following interactions encoded in $\cO_B$
-and for the precise Lorentz structures shown above- get decorrelated
in a general non-linear analysis:
\begin{itemize}
\item[-] $\gamma-W-W$ from $\gamma-Z-h$, and $Z-W-W$ from $Z-Z-h$;
these are examples of interactions involving different number of
external $h$ legs.
\item[-] $\gamma-W-W-h$ from $\gamma-Z-h$, and $Z-W-W-h$ from $Z-Z-h$,
which are interactions involving the same number of external $h$ legs.
\end{itemize} 
While such decorrelations are expected among the leading SM deviations
in a generic non-linear approach, they require us to consider $d=8$
operators in scenarios with linearly realized EW symmetry
breaking. This statement is a physical effect, which means that it holds irrespective of the linear basis used, for instance it also holds in the bases in Refs.~\cite{Grojean:2013kd,Elias-Miro:2013gya}. The study of the
correlations/decorrelations described represents an interesting method
to investigate the intimate nature of the light Higgs $h$.

The argument developed above focused on just one operator, for
illustration.  A parallel analysis on correlations/decorrelations also applies in another case, i.e. the interactions described by $\cP_3(h)$ and $\cP_5(h)$ versus those in the $d=6$ linear operator $\cO_W$.
Obviously, in order to firmly establish the pattern of deviations expected, all possible operators at a given order of an expansion should be considered together, and this will be done in the phenomenological Sect.~\ref{pheno} below.

\subsection{Signals specific to the linear expansion}
\label{linearspecific}

The $d=6$ operators in Eq.~(\ref{PureGaugeLinearOps}) have no
equivalent among the dominant corrections of the non-linear expansion,
Eqs.~(\ref{Opxi})-(\ref{Opxih}), all $\xi$ weights considered.  This
fact results in an interesting method to test the nature of the
Higgs. Considering for example the operator $\cO_{WWW}$ in
Eq.~(\ref{PureGaugeLinearOps}), the couplings
\begin{align}
 \multirow{3}{*}
{\parbox{2cm}
{\input{Fdiagrams/WWA}}}
\qquad\qquad&
f_{WWW}\dfrac{3ieg^2}{4}\Big[g_{\rho\mu}\left((p_+\cdot p_-)
p_{A\nu}-(p_A\cdot p_-) p_{+\nu}\right) \nn \\ 
& +g_{\mu\nu}\left((p_A\cdot p_- )p_{+\rho}-(p_A\cdot p_+)
p_{-\rho}\right)
\nn\\ 
&+ g_{\rho\nu}\left((p_A\cdot p_+) p_{-\mu}-(p_+\cdot p_-)
p_{A\mu}\right)+p_{A\mu}p_{+\nu}p_{-\rho}-p_{A\nu}p_{+\rho}p_{-\mu}\Big]\,,\nn\\
\\ 
\multirow{3}{*}{\parbox{2cm}
{\input{Fdiagrams/WWZ}}}
\qquad\qquad&
f_{WWW}\dfrac{3ig^3\cos\theta_W}{4}\Big[g_{\rho\mu}\left((p_+\cdot
p_-) p_{Z\nu}-(p_Z\cdot p_-) p_{+\nu}\right) \nn\\ 
&+g_{\mu\nu}\left((p_Z\cdot p_- )p_{+\rho}-(p_Z\cdot p_+)
p_{-\rho}\right) \nn\\ 
&+g_{\rho\nu}\left((p_Z\cdot p_+)
p_{-\mu}-(p_+\cdot p_-)
p_{Z\mu}\right)+p_{Z\mu}p_{+\nu}p_{-\rho}-p_{Z\nu}p_{+\rho}p_{-\mu}\Big]\,,\nn
\end{align}
should be observable with a strength similar to that of other
couplings described by $d=6$ operators, if the EW breaking is linearly
realized by the underlying physics. On the contrary, for a subjacent
non-linear dynamics their strength is expected to be suppressed
(i.e. be of higher order)~\cite{Buchalla:2013rka} \footnote{This coupling is
usually referred to in the literature as $\lambda_V$
\cite{Hagiwara:1986vm}.}. A similar discussion holds for the other operators in Eq.~(\ref{PureGaugeLinearOps}).

%%%
% 3.3 New signals in the chiral Lagrangian
%%%

\subsection{New signals specific to the non-linear expansion}
\label{newsignals}

For large $\xi$, all chiral operators weighted by $\xi^n$ with
$n\ge2$, Eqs.~(\ref{Opxi2})-(\ref{Opxih}), are equally relevant to the
$\xi$-weighted ones in Eq.~(\ref{Opxi}), and therefore their siblings
require operators of dimension $d\ge8$.  Of special interest is
$\cP_{14}(h)$ which belongs to the former class, as some of the couplings encoded in
it are absent from the SM Lagrangian. This fact provides a viable
strategy to test the nature of the physical Higgs.
 
In App.~\ref{AppFR}, the Feynman rules for all couplings appearing in
the non-linear Lagrangian for gauge and gauge-$h$ operators can be
found. A special column indicates directly the non-standard structures
and it is easy to identify among those entries the couplings weighted
only by $\xi^n$ with $n\ge 2$. Here, we report explicitly only the example of the anomalous $Z-W-W$ and $\gamma-Z-W-W$ vertices, assuming for simplicity that the $\cF_{14}(h)$
function admits a polynomial expansion in $h/v$.
The operator $\cP_{14}(h)$ contains the couplings
\beq
\e^{\mu\nu\rho\lambda}\derp_\mu W^+_\nu W^-_\rho Z_\lambda 
\cF_{14}(h)\,,\qquad\qquad
\e^{\mu\nu\rho\lambda}Z_\mu A_\nu W^-_\rho W^+_\lambda  \cF_{14}(h)\,,
\label{P10Interac}
\eeq
which correspond to an anomalous $Z-W-W$ triple vertex and to an
anomalous $\gamma-Z-W-W$ quartic vertex, respectively. The
corresponding Feynman diagrams and rules read
\beq
\begin{aligned}
\parbox{3cm}{\input{Fdiagrams/WWZ}}&\qquad\qquad
-\xi^2\,\dfrac{g^3}{\cos\theta_W}\,\e^{\mu\nu\rho\lambda}
[p_{+\lambda}-p_{-\lambda}]\,,\\[5mm]
\parbox{3cm}{\input{Fdiagrams/WWZA}}&\qquad\qquad
-2\,\xi^2\,\dfrac{e g^3}{\cos\theta_W}\,\e^{\mu\nu\rho\lambda}\,.
\end{aligned}
\eeq
These couplings are  present neither in the SM nor in the $d=6$
linear Lagrangian and are anomalous couplings due to their Lorentz nature.
A signal of these type of interactions at colliders with
a strength comparable with that expected for the couplings in the
$d=6$ linear Lagrangian would be a clear hint of a strong dynamics in
the EWSB sector. More details are given in the
phenomenological sections below.

%
%%%%%%%%%%%%%%%%%%%%%%%%%   4.  Phenomenology       %%%%%%%%%%%%%%%%%%%%%
%

\section{Phenomenology}
\label{pheno}

Prior to developing the strategies suggested above to investigate the nature of the Higgs particle, the renormalization procedure is illustrated next. 

%%%
% 4.1 Renormalisation Procedure
%%%

\subsection{Renormalization Procedure}\label{sec:renormalization}

Five electroweak parameters of the SM-like Lagrangian $\LL_0$ are
relevant in our analysis, when neglecting fermion masses: $g_s$, $g$,
$g'$, $v$ and the $h$ self-coupling $\lambda$. The first four can be
optimally constrained by four observables that are extremely well
determined nowadays, while as a fifth one the Higgs mass $m_h$ can be used;
in summary:
\beq
\begin{aligned}
\a_s&&&\text{world average~\cite{Beringer:1900zz},}\\
G_F&&&\text{extracted from the muon decay rate~\cite{Beringer:1900zz},}\\
\aem&&&\text{extracted from Thomson scattering~\cite{Beringer:1900zz},}\\
m_Z&&&\text{extracted from the $Z$ lineshape at LEP I
~\cite{Beringer:1900zz},}\\
m_h&&&\text{now measured at LHC~\cite{Aad:2012tfa,Chatrchyan:2012ufa}.}
\end{aligned}
\label{inputs}
\eeq 
This ensemble of observables defines the so-called Z-scheme: they
will be kept as input parameters, and all predictions will be
expressed as functions of them.  Accordingly, whenever a dependence on
the parameters $g$, $g'$, $v$ (and $e$) or the weak mixing angle % Weinberg 
$\theta_W$ may
appear in the expressions below, it should be interpreted as
corresponding to the combinations of experimental inputs as follows:
\beq
\begin{aligned}
 e^2 &= 4 \pi \a_\text{em}\,, \qquad
 &\sin^2\theta_W &= \dfrac{1}{2}\left
(1-\sqrt{1-\dfrac{4\pi\a_\text{em}}{\sqrt{2}\GF m_Z^2}}\right)\,,\\
 v^2 &= \dfrac{1}{\sqrt{2}\GF}\,,\qquad
 &\Big(g &= \dfrac{e}{\sin\theta_W}\,,\qquad g' = \left.
\dfrac{e}{\cos\theta_W}\,\Big)\right|
_{\theta_W,\,e \text{ as above}}\,.
\end{aligned}
\label{param}
\eeq
The abbreviations $\st$ ($\sdt$) and $\ct$ ($\cdt$) will stand below
for $\sin\theta_W$ ($\sin2\theta_W$) and $\cos\theta_W$
($\cos2\theta_W$), respectively. Furthermore, for concreteness, we
assume a specific parametrization for the $\cF_i(h)$ functions:
\beq 
\cF_i(h)\equiv1+2\tilde a_i\dfrac{h}{v}+\tilde
b_i\dfrac{h^2}{v^2}+\ldots 
\eeq
where the dots stand for higher powers of $h/v$ that will not be
considered in what follows; to further simplify the notation $a_i$ and
$b_i$ will indicate below the products $a_i\equiv c_i\tilde a_i $ and
$b_i\equiv c_i\tilde b_i $, respectively, where $c_i$ are the global
operator coefficients.

Working in the unitary gauge to analyze the impact that the couplings
in $\Delta\LL$ in Eq.~(\ref{DeltaL}) have on $\LL_0$, it is
straightforward to show that $\cP_B(h)$, $\cP_W(h)$, $\cP_G(h)$,
$\cP_H(h)$, $\cP_1(h)$ and $\cP_{12}(h)$ introduce corrections to the SM kinetic
terms, and in consequence field redefinitions are necessary to obtain
canonical kinetic terms. Among these operators, $\cP_B(h)$, $\cP_W(h)$
and $\cP_G(h)$ can be considered innocuous operators with respect to
$\LL_0$, as the impact on the latter of $c_B$, $c_W$ and $c_G$ can be
totally eliminated from the Lagrangian by ineffectual field and
coupling constant redefinitions; they do have a physical impact though
on certain BSM couplings in $\Delta\LL$ involving external scalar
fields.

With canonical kinetic terms, it is then easy to identify the
 contribution of $\Delta\LL$ to the input parameters\footnote{The BSM
 corrections that enter into the definition of the input parameters
 will be generically denoted by the sign ``$\delta$", while the
 predicted measurable departures from SM expectations will be
 indicated below by ``$\Delta$".}:
\beq
\begin{aligned}
\dfrac{\delta \aem}{\aem}&\simeq 4e^2\,c_1\,\xi+4e^2\,c_{12}\xi^2 \,,
&&\qquad\qquad\dfrac{\delta \GF}{\GF}&\simeq 0\,,\\
\dfrac{\delta m_Z}{m_Z}&\simeq -c_T\,\xi-2e^2\,c_1\,\xi+2e^2\,
\cot^2\theta_W\,c_{12}\,\xi^2\,,
&&\qquad\qquad\dfrac{\delta m_h}{m_h}&\simeq 0\,,
\end{aligned}
\eeq
keeping only  terms linear in the coefficients $c_i$. 
  Expressing all other SM parameters in
$\LL_\text{chiral}$ in terms of the four input parameters leads to  the
predictions to be described next. 

\subsubsection*{$W$ mass}
The prediction for the $W$ mass  departs from the SM expectation by 
\beq
\begin{split}
\dfrac{\Delta m_W^2}{m_W^2}&=
\dfrac{4e^2}{\cdt} c_1\,\xi
+\dfrac{2\ct^2}{\cdt}c_T\,\xi
-\dfrac{4e^2}{\st^2}\,c_{12}\,\xi^2\\
&\equiv
\dfrac{e^2}{2\cdt}  f_{BW}\frac{v^2}{\Lambda^2}
-\frac{\ct^2}{2\cdt}f_{\Phi,1} \frac{v^2}{\Lambda^2}\,, 
\end{split}
\eeq
where the second line shows for comparison the corresponding
expression in the linear expansion at order $d=6$.

\subsubsection*{$S$ and $T$ parameters}
$\cP_{1}(h)$ and $\cP_T(h)$ generate tree-level contributions to the
oblique parameters $S$ and $T$~\cite{Peskin:1990zt}, which read
\begin{equation}
\aem\Delta S =  -8e^2 c_1\xi 
\qquad\qquad \hbox{ and } \qquad\qquad
\aem\Delta T =  2 c_T\xi\,.
\label{eq:STtree}
\end{equation}

\subsubsection*{Triple gauge--boson couplings}

The effective operators described in the non-linear Lagrangian,
Eqs.~(\ref{Opxi})-(\ref{Opxi4}), give rise to triple gauge--boson
couplings $\gamma W^+ W^-$ and $Z W^+W^-$. Following
Ref.~\cite{Hagiwara:1986vm}, the CP-even sector of the Lagrangian that
describes trilinear gauge boson vertices (TGV) can be parametrized as:
\begin{align}
\LL_{WWV} =& - \,i g_{WWV} \Bigg\{ 
g_1^V \Big( W^+_{\mu\nu} W^{- \, \mu} V^{\nu} - 
W^+_{\mu} V_{\nu} W^{- \, \mu\nu} \Big) 
   \,+\, \kappa_V W_\mu^+ W_\nu^- V^{\mu\nu}\, \label{eq:classical}\\ 
& -  ig_5^V \epsilon^{\mu\nu\rho\sigma}
\left(W_\mu^+\partial_\rho W^-_\nu-W_\nu^-\partial_\rho W^+_\mu\right)
V_\s \,+\,
 g_{6}^V \left(\derp_\mu W^{+\mu} W^{-\nu}-\derp_\mu W^{-\mu} W^{+\nu}\right)
V_\nu  \Bigg\}\,,\nn
\end{align}
where $V \equiv \{\gamma, Z\}$ and $g_{WW\gamma} \equiv e=g
\sin\theta_W$, $g_{WWZ} = g \cos\theta_W$ (see Eq.~(\ref{param}) for
their relation to observables). In this equation $W^\pm_{\mu\nu}$ and
$V_{\mu\nu}$ stand exclusively for the kinetic part of the gauge field
strengths. In contrast with the usual parameterization proposed in
Ref.~\cite{Hagiwara:1986vm}, the coefficient $\lambda_V$ (associated
with a linear $d=6$ operator) is omitted here as this coupling does not
receive contributions from the non-linear effective chiral Lagrangian
expanded up to four derivatives. 
Conversely, we have introduced the
coefficients $g_{6}^V$ associated to operators that contain the
contraction $\DLL_\mu\VL^\mu$; its $\partial_\mu\VL^\mu$ part vanishes
only for on-shell gauge bosons; in all generality $\DLL_\mu\VL^\mu$
insertions could only be disregarded\footnote{See for example
Ref.~\cite{Feruglio:1996bx} for a general discussion on possible
``off-shell'' vertices associated to $d=4$ and $d=6$ operators.} in
the present context when fermion masses are neglected, as explained in
Sect.~\ref{EffectiveLagrangian} and App.~\ref{AppEOM}.

Electromagnetic gauge invariance requires $g_{1}^{\gamma} =1$ and
 $g_5^\gamma=0$, and in consequence the TGV CP-even sector described
 in Eq.~(\ref{eq:classical}) depends in all generality on six
 dimensionless couplings $g_1^{Z}$, $g_5^Z$, $g_{6}^{\gamma,Z}$ and
 $\kappa_{\gamma,Z}$. Their SM values are $g_1^{Z}=\kappa_{\gamma}=
 \kappa_Z=1$ and $g_5^Z=g_{6}^{\gamma}=g_{6}^Z=0$.  Table \ref{tab:tgv}
 shows the departures from those SM values due to the effective
 couplings in Eq.~(\ref{DeltaL}); it illustrates the $\xi$ and
 $\xi^2$-weighted chiral operator contributions. For the sake of
 comparison, the corresponding expressions in terms of the
 coefficients of $d=6$ operators in the linear expansion are shown as
 well. A special case is that of the linear operator $\cO_{\Box\Phi}$, whose physical interpretation is not straightforward \cite{Jansen:1993jj,Jansen:1993ji,Grinstein:2007mp} and will be analyzed in detail in Ref.~\cite{BoxInPreparation};  the corresponding coefficient $f_{\Box\Phi}$ does not appear in Table \ref{tab:tgv} as contributing to the measurable couplings, while nevertheless the symbol $(*)$ recalls the theoretical link between some chiral operators and their sibling $\cO_{\Box\Phi}$. The analysis of Table~\ref{tab:tgv} leads as well to relations between measurable quantities, which are collected later on in
 Eq.~(\ref{old1}) and subsequent ones.

\begin{table}[h!]
\centering
\footnotesize
\begin{tabular}{|c||c||c|c||c|}
\hline
&&\multicolumn{1}{c}{}&&\\[-2mm]
& Coeff. & \multicolumn{2}{c||}{Chiral} & Linear \\[2mm]
& $\times e^2/\st^2$& $\times \xi$ & $\times \xi^2$ 
& $\times v^2/\Lambda^2$ \\[2mm]
\hline
\hline
&&&&\\[-3mm]
$\Delta\kappa_\gamma$ 
&$1$
&$-2c_1+2c_2+c_3$  
&$-4c_{12}+2c_{13}$
&$\frac{1}{8}(f_W+f_B- 2 f_{BW})$
\\[2mm]
$\Delta g_{6}^\gamma$
&$1$
&$-c_{9}$
&$-$
&$(*)$
\\[2mm]
\hline
&&&&\\[-3mm]
$\Delta g_1^Z$
&$\frac{1}{\ct^2}$
&$\frac{\sdt^2}{4e^2\cdt}c_T +\frac{2\st^2}{\cdt}c_1+ c_3$
&$-$
&
$ 
\frac{1}{8}f_W + \frac{\st^2}{4\cdt}f_{BW}-\frac{\sdt^2}{16e^2\cdt} f_{\Phi,1} 
$
\\[2mm]
$\Delta\kappa_Z$
&$1$
&$\frac{\st^2}{e^2\cdt}c_T+\frac{4\st^2}{\cdt}c_1
-\frac{2\st^2}{\ct^2}c_2+ c_3$
&$-4c_{12}+2c_{13}$
&$\frac{1}{8}f_W -\frac{\st^2}{8 \ct^2}f_B 
+\frac{\st^2}{2\cdt}f_{BW} -\frac{\st^2}{4e^2\cdt}f_{\Phi,1}$
\\[2mm]
$\Delta g_5^Z$
&$\frac{1}{\ct^2}$
&$-$
&$c_{14}$
&$-$
\\[2mm]
$\Delta g_{6}^Z$
&$\frac{1}{\ct^2}$
&$\st^2c_{9}$
&$-c_{16}$
&$(*)$
\\[2mm]
\hline
\end{tabular}
\caption{\em Effective couplings parametrizing the $V W^+ W^-$
vertices defined in Eq.~(\ref{eq:classical}).  The coefficients in the
second column are common to both the chiral and linear expansions. In
the third and fourth columns the specific contributions from the
operators in the chiral Lagrangian are shown.
For comparison, the last column exhibits the corresponding
contributions from the linear $d=6$ operators. The star $(*)$ in the last column indicates the link between the chiral operator $\cP_9(h)$ and its linear sibling $\cO_{\Box\Phi}$, without implying a physical impact of the latter on the  $V W^+ W^-$ observables, as explained in the text and in Ref.~\cite{BoxInPreparation}.}
\label{tab:tgv}
\end{table}

\boldmath
\subsubsection*{$h$ couplings to SM gauge-boson pairs}
\unboldmath
The effective operators described in Eqs.~(\ref{Opxi})-(\ref{Opxi4})
also give rise to interactions involving the Higgs and two gauge
bosons, to which we will refer as HVV couplings. The latter can be
phenomenologically parametrized as 
\begin{align}
\LL_\text{HVV}\equiv&
\phantom{+}g_{Hgg} \,G^a_{\mu\nu} G^{a\mu\nu} h+
g_{H \gamma \gamma} \, A_{\mu \nu} A^{\mu \nu} h+ 
g^{(1)}_{H Z \gamma} \, A_{\mu \nu} Z^{\mu} \partial^{\nu} h + 
g^{(2)}_{H Z \gamma} \, A_{\mu \nu} Z^{\mu \nu} h \nn
\\
&+g^{(1)}_{H Z Z}  \, Z_{\mu \nu} Z^{\mu} \partial^{\nu} h + 
g^{(2)}_{H Z Z}  \, Z_{\mu \nu} Z^{\mu \nu} h+
g^{(3)}_{H Z Z}  \, Z_\mu Z^\mu h + g^{(4)}_{H Z Z}  \, Z_\mu Z^\mu \Box h\nn
\\
&+ g^{(5)}_{H Z Z}  \, \derp_\mu Z^\mu Z_\nu \derp^\nu h+ g^{(6)}_{H Z Z}  \, \derp_\mu Z^\mu \derp_\nu Z^\nu  h\label{eq:lhvv}
\\
&+ g^{(1)}_{H W W}  \, \left (W^+_{\mu \nu} W^{- \, \mu} \partial^{\nu} h + \hc \right) +
g^{(2)}_{H W W}  \, W^+_{\mu \nu} W^{- \, \mu \nu} h+
g^{(3)}_{H W W}  \, W^+_{\mu} W^{- \, \mu} h\nn
\\
&+g^{(4)}_{H W W}  \, W^+_\mu W^{-\mu} \Box h+ g^{(5)}_{H W W}  \, 
+\left(\derp_\mu W^{+\mu} W^-_\nu \derp^\nu h+\hc\right)+ g^{(6)}_{H W W}  \, \derp_\mu W^{+\mu} \derp_\nu W^{-\nu}  h\,,\nn
\end{align}
where $V_{\mu \nu} = \partial_\mu V_\nu - \partial_\nu V_\mu$
with $V=\{A, Z, W, G\}$.  
Separating the contributions into SM ones plus corrections, 
\beq
g_i^{(j)}\simeq  g_i^{(j)SM} + \Delta g_i^{(j)}\,,
\label{gSM}
\eeq
it turns out that
\beq
\begin{aligned}
 g^{(3)SM}_{HZZ}=\frac{m_Z^2}{v}\,\,,\qquad\qquad
g^{(3)SM}_{HWW}=\frac{2m_Z^2\ct^2}{v}\,, 
\end{aligned}
\eeq
while the tree-level SM value for all other couplings in
Eq.~(\ref{eq:lhvv}) vanishes (the SM loop-induced value for $g_{Hgg}$,
$g_{H\gamma\gamma}$ and $g^{(2)}_{HZ\gamma}$ will be taken into
account in our numerical analysis, though); 
in these expressions, $v$ is as defined in
Eq.~(\ref{param}). Table \ref{tab:hvv} shows the expressions for the
corrections $\Delta g_{Hgg}$, $\Delta g_{H \gamma \gamma}$, $\Delta
g^{(1,2)}_{H Z \gamma}$, $\Delta g^{(1,2,3,4,5,6)}_{H W W}$, and
$\Delta g^{(1,2,3,4,5,6)}_{H Z Z}$ 
induced at tree-level by the effective non-linear couplings under discussion. 
In writing Eq.~(\ref{eq:lhvv}) we have introduced the coefficients $\Delta g^{(4,5,6)}_{H Z Z}$ and $\Delta g^{(4,5,6)}_{H W W}$: $\Delta g^{(4)}_{H VV}$ become redundant for on-shell $h$; $\Delta g^{(5,6)}_{H VV}$ vanish for on-shell $W_\mu$ and $Z_\mu$ or massless fermions.
Notice also that the
leading chiral corrections include operators weighted by $\xi$ powers
up to $\xi^2$. For the sake of comparison, the corresponding
expressions in terms of the coefficients of the linear $d=6$ operators
in Eq.~(\ref{LinearChiralCorrelations}) are also shown\footnote{Alternatively the coefficient of 
$\Delta g^{(3)}_{H WW}$ can be defined
in terms of the {\sl measured} value of $M_W$ as $M_W^2/e^2$. 
In this case  the entries
in columns 3--5 read $-4 c_H+4(2 a_C-c_C)$, $-32\frac{e^2}{\st^2}$,
and $-2f_{\Phi,1}+4 f_{\Phi,4}-4f_{\Phi,2}$ respectively.}.

\begin{table}[h!]
\hspace{-1.2cm}
\footnotesize
\begin{tabular}{|c||c||c|c||c|}
\hline
&&\multicolumn{1}{c}{}&&\\[-2mm] 
& Coeff. & \multicolumn{2}{c||}{Chiral} & Linear \\[2mm]
& $\times e^2/4v$& $\times \xi$& $\times \xi^2$& $\times v^2/\Lambda^2$ \\[2mm]
\hline
\hline
&&&&\\[-3mm]
$\Delta g_{Hgg}$ & $\frac{g^2_s}{e^2}$ & $-2a_G$ & $-$  &$-4f_{GG}$ \\[2mm]
$\Delta g_{H \gamma \gamma}$ & $1$  & $-2(a_B+a_W)+ 8a_1$ & $8 a_{12}$ &$ -(f_{BB}+f_{WW})+f_{BW}$  \\[2mm]
$\Delta g^{(1)}_{H Z \gamma}$ & $\frac{1}{\sdt}$  & $-8 (a_5+2 a_4)$ & $-16 a_{17}$  & $2(f_W-f_B) $\\[2mm]
$\Delta g^{(2)}_{H Z \gamma}$ & $\frac{\ct}{\st}$  & $4\frac{\st^2}{\ct^2}a_B-4a_W+8\frac{\cdt}{\ct^2}a_1$ & $16 a_{12}$ & $2\frac{\st^2}{\ct^2} f_{BB} -2f_{WW}+\frac{\cdt}{\ct^2} f_{BW}$ \\ [2mm]
$\Delta g^{(1)}_{H Z Z}$ & $\frac{1}{\ct^2}$  & $-4\frac{\ct^2}{\st^2}a_5+8 a_4$& $-8 \frac{\ct^2}{\st^2} a_{17}$  & $ \frac{\ct^2}{\st^2} f_W+f_B$ \\[2mm]
$\Delta g^{(2)}_{H Z Z}$ & $-\frac{\ct^2}{\st^2}$ &$2\frac{\st^4}{\ct^4} a_B+2 a_W+8\frac{\st^2}{\ct^2} a_1$ &$-8a_{12}$ & $\frac{\st^4}{\ct^4}f_{BB}+f_{WW}+\frac{\st^2}{\ct^2} f_{BW}$ \\[2mm]
$\Delta g^{(3)}_{H Z Z}$ & $\frac{m_Z^2}{e^2}$ & $-2c_H+2(2a_C-c_C)-8(a_T-c_T)$ & $-$ &$f_{\Phi,1}+ 2 f_{\Phi,4}-2 f_{\Phi,2}$ \\[2mm]
$\Delta g^{(4)}_{H Z Z}$ & $-\frac{1}{\sdt^2}$ & $16 a_{7}$ & $32 a_{25}$& (*) \\[2mm] 
$\Delta g^{(5)}_{H Z Z}$ & $-\frac{1}{\sdt^2}$ & $16a_{10}$ & $32a_{19}$ 
&(*)\\[2mm] 
$\Delta g^{(6)}_{H Z Z}$ & $-\frac{1}{\sdt^2}$ & $16a_{9}$ & $32a_{15}$ &(*)\\[2mm] 
$\Delta g^{(1)}_{H WW}$ & $\frac{1}{\st^2}$  & $-4 a_5$ & $-$  & $f_W$ \\[2mm]
$\Delta g^{(2)}_{H WW}$ & $\frac{1}{\st^2}$ & $-4 a_W$ & $-$  & $-2f_{WW}$ \\[2mm]
$\Delta g^{(3)}_{H WW}$ & $\frac{m_Z^2\ct^2}{e^2}$ & $-4c_H+4(2a_C-c_C)+\frac{32e^2}{\cdt} c_1
+\frac{16\ct^2}{\cdt}c_T$  & $-\frac{32e^2}{\st^2}\,c_{12}$  & 
$\frac{-2(3\ct^2-\st^2)}{\cdt}f_{\Phi,1}+ 4 f_{\Phi,4}-4 f_{\Phi,2}+\frac{4  e^2}{\cdt}  f_{BW}$\\[2mm]
$\Delta g^{(4)}_{H W W}$ & $-\frac{1}{\st^2}$ & $8a_{7}$ & $-$ &(*)\\[2mm] 
$\Delta g^{(5)}_{H W W}$ & $-\frac{1}{\st^2}$ & $4a_{10}$ & $-$ &(*) \\[2mm] 
$\Delta g^{(6)}_{H W W}$ & $-\frac{1}{\st^2}$ & $8a_{9}$ & $-$ &(*)\\[2mm] 
\hline
\end{tabular}
\caption{\em The trilinear Higgs-gauge bosons couplings defined in
Eq.~(\ref{eq:lhvv}). The coefficients in the second column are common
to both the chiral and linear expansions. The contributions from the
operators weighted by $\xi$ and $\xi^{\ge2}$ are
listed in the third and fourth columns, respectively.  For comparison,
the last column exhibits the corresponding expressions for the linear
expansion at order $d=6$. 
The star $(*)$ in the last column indicates the link between the chiral operators $\cP_7(h)$, $\cP_9(h)$ and $\cP_{10}(h)$, and their linear sibling $\cO_{\Box\Phi}$, without implying a physical impact of the latter on the observables considered, as explained in the text and in Ref.~\cite{BoxInPreparation}.}
\label{tab:hvv}
\end{table}

Notice that the bosonic operators $\cP_{H}(h)$ and $\cP_{C}(h)$ induce
universal shifts to the SM-like couplings of the Higgs to weak gauge
bosons. Similarly $\cP_{H}(h)$, induces universal shifts to the Yukawa
couplings to fermions, see Eq.~({\color{blue} FR.32}) in Appendix~\ref{AppFR}. It is straightforward to
identify the link between the coefficients of these operators and the
parameters $a$ and $c$ defined in
Refs.~\cite{Azatov:2012bz,Espinosa:2012im,Azatov:2012qz} assuming
custodial invariance, which reads\footnote{Supplementary terms are
present when taking into account the custodial breaking couplings
considered in this paper.}
\beq
a= 1-\dfrac{\xi c_H}{2}+\dfrac{\xi(2a_C-c_C)}{2}\,,\qquad\qquad
c=s_Y\left(1-\dfrac{\xi c_H}{2}\right)\,.
\eeq

\subsubsection*{Quartic gauge--boson couplings}

The quartic gauge boson couplings also receive contributions from the operators in Eqs.~(\ref{Opxi})-(\ref{Opxi4}). 
The corresponding effective Lagrangian reads
\begin{align}
\LL_{4X}\, \equiv \, g^2 & \Bigg\{ \, g^{(1)}_{ZZ} (Z_\mu Z^\mu)^2 
\,+\, g^{(1)}_{WW}\, W^+_\mu W^{+\mu} W^-_\nu W^{-\nu} \,-\, 
g^{(2)}_{WW} (W^+_\mu W^{-\mu})^2 \, \nn \\
&\hspace{0.cm} \,+ \, 
g^{(3)}_{VV'} W^{+\mu} W^{-\nu}\left( V_\mu V'_\nu + V'_\mu V_\nu \right) 
\,-\, 
g^{(4)}_{VV'} W^+_\nu W^{-\nu} V^\mu V'_\mu \, \nn \\  
&\hspace{0.cm} \,+\, 
ig^{(5)}_{VV'} \e^{\mu\nu\rho\s} W^+_\mu W^-_\nu V_\rho V'_\s \Bigg\}\,,
\label{eq:4v}
\end{align}
where $VV'=\{\gamma\gamma, \gamma Z, ZZ\}$. Notice that all these
couplings are $C$ and $P$ even, except for $g^{(5)}_{VV'}$ that is
$CP$ even but both $C$ and $P$ odd. 
Some of these couplings are nonvanishing at tree-level in the SM:
\beq
\begin{aligned}
g^{(1)SM}_{WW}&=\frac{1}{2}\,,\qquad
&g^{(2)SM}_{WW}&=\frac{1}{2}\,,\qquad
&g^{(3)SM}_{ZZ}&=\frac{\ct^2}{2}\,,\qquad
&g^{(3)SM}_{\g\g}&=\frac{\st^2}{2}\,,\\
g^{(3)SM}_{Z\g}&=\frac{\sdt}{2}\,,\qquad
&g^{(4)SM}_{ZZ}&=\ct^2\,,\qquad
&g^{(4)SM}_{\g\g}&=\st^2\,,\qquad
&g^{(4)SM}_{Z\g}&=\sdt\,,
\end{aligned}
\eeq
where the notation defined in Eq.~(\ref{gSM}) has been used and the
 expression for the weak mixing angle can bee found in
 Eq.~(\ref{param}).  Table \ref{tab:4V} shows the contributions to the
 effective quartic couplings from the chiral operators in
 Eqs.~(\ref{Opxi})-(\ref{Opxi4}) and from the linear operator in
 Eq.~(\ref{LinearOpsGauge}). 

\begin{table}[ht!]
\hspace*{-2cm}
\footnotesize
\centering
\renewcommand{\arraystretch}{2}
\begin{tabular}{|*2{>{$}c<{$}||}*2{>{$}c<{$}|}{|>{$}c<{$}|}}
\hline
& \text{Coeff. }& \multicolumn{2}{c||}{Chiral}  &  \text{Linear} \\[-1mm]
&\times e^2/4\st^2& \multicolumn{1}{c|}{$\times \xi$} & 
\multicolumn{1}{c||}{$\times \xi^2$} & \times v^2/\Lambda^2 \\
\hline
\hline
\Delta g^{(1)}_{WW}&
1&  \frac{\sdt^2}{e^2\cdt}c_T+\frac{8\st^2}{\cdt}c_1 +4 c_3 &
 2c_{11}-16c_{12} +8 c_{13}
 &\frac{f_W}{2}+\frac{\st^2}{\cdt}f_{BW}-\frac{\sdt^2}{4\cdt e^2}f_{\Phi1}\\

\Delta g^{(2)}_{WW}&
1&  \frac{\sdt^2}{e^2\cdt}c_T+\frac{8\st^2}{\cdt}c_1 +4 c_3 -4 c_{6}& -2c_{11} -16 c_{12} +8 c_{13}
&\frac{f_W}{2} +\frac{\st^2}{\cdt}f_{BW}-\frac{\sdt^2}{4\cdt e^2}f_{\Phi1}+(*)\\\hline 

\Delta g^{(1)}_{ZZ}&
\frac{1}{\ct^4}& c_{6} &  c_{11} +2 c_{23} +2 c_{24} 
+ 4 c_{26} \xi^2 & (*)\\
\Delta g^{(3)}_{ZZ}&
\frac{1}{\ct^2} &  \frac{\sdt^2\ct^2}{e^2\cdt}c_T
+\frac{2\sdt^2}{\cdt}c_1 +4\ct^2 c_3-2\st^4 c_{9}&
2c_{11}+4\st^2c_{16} + 2 c_{24}
& \frac{f_W\ct^2}{2} +\frac{\sdt^2}{4\cdt}f_{BW}-\frac{\sdt^2\ct^2}{4e^2\cdt}f_{\Phi1} +(*)\\ %
\Delta g^{(4)}_{ZZ}&
\frac{1}{\ct^2}& \frac{2\sdt^2\ct^2}{e^2\cdt}c_T
+\frac{4\sdt^2}{\cdt}c_1 +8\ct^2 c_3 - 4c_{6} & - 4c_{23} 
& f_W\ct^2+2\frac{\sdt^2}{4\cdt}f_{BW}-\frac{\sdt^2\ct^2}{2e^2\cdt}f_{\Phi1}+(*)\\\hline %
\Delta g^{(3)}_{\g\g}&
\st^2 & - 2c_{9} & -  
& (*) \\\hline %
\Delta g^{(3)}_{\g Z}& 
\frac{\st}{\ct}& \frac{\sdt^2}{e^2\cdt}c_T+\frac{8\st^2}{ \cdt}c_1 + 4c_3 + 4\st^2 c_{9}&
  - 4c_{16} & \frac{f_W}{2} +\frac{\st^2}{\cdt}f_{BW}-\frac{\sdt^2}{4\cdt e^2}f_{\Phi1}+(*)\\ %
\Delta g^{(4)}_{\g Z}&
\frac{\st}{\ct} &  \frac{2\sdt^2}{e^2\cdt}c_T
+\frac{16\st^2}{ \cdt}c_1 +8 c_3 & - & f_W+2\frac{\st^2}{\cdt}f_{BW}-\frac{\sdt^2}{2\cdt e^2}f_{\Phi1} \\
\Delta g^{(5)}_{\g Z}&
\frac{\st}{\ct}& -  &  8c_{14}   & -\\\hline
\end{tabular}
\caption{\em Effective couplings parametrizing the vertices of four gauge 
bosons defined in Eq.~(\ref{eq:4v}). The contributions from the
operators weighted by $\xi$ and $\xi^{\ge2}$ are
listed in the third and fourth columns, respectively.  For comparison,
the last column exhibits the corresponding expressions for the linear
expansion at order $d=6$. 
The star $(*)$ in the last column indicates the link between the chiral operators $\cP_6(h)$ and $\cP_9(h)$, and their linear sibling $\cO_{\Box\Phi}$, without implying a physical impact of the latter on the observables considered, as explained in the text and in Ref.~\cite{BoxInPreparation}.}
\label{tab:4V}
\end{table}

\subsubsection*{(De)correlation formulae}

Some operators of the non-linear Lagrangian in
Sect.~\ref{EffectiveLagrangian} participate in more than one of the
couplings in Tables \ref{tab:tgv} and \ref{tab:hvv}. This fact leads
to interesting series of relations that relate different couplings.
First, simple relations on the TGV sector results:
\begin{align}
\Delta\kappa_Z+\frac{s_\theta^2}{c_\theta^2}\Delta \kappa_\gamma
-\Delta g_1^Z&=\dfrac{16e^2}{\st^2}(2c_{12}-c_{13})\xi^2\,,\label{old1}\\
\Delta g_{6}^\gamma+\dfrac{c_\theta^2}{s_\theta^2}\Delta g_{6}^Z&=-\dfrac{e^2}{s_\theta^4}\,c_{16}\,\xi^2\,,
\end{align}
while other examples of relations involving HVV couplings are: 
\begin{align}
g^{(1)}_{HWW}-\ct^2\, g^{(1)}_{HZZ}-\ct\st\, g^{(1)}_{HZ\gamma}
&=\dfrac{2e^2}{v\st^2}a_{17}\xi^2\,,\label{new1}\\
2\ct^2\,g^{(2)}_{HZZ}+2\st\ct\,g^{(2)}_{HZ\gamma}+
2\st^2\,g_{H\gamma\gamma}-g^{(2)}_{HWW}&
=\dfrac{4e^2}{v\st^2}a_{12}\xi^2\,,\label{new2}\\
\Delta g^{(4)}_{HZZ}-\dfrac{1}{2c_\theta^2}\Delta g^{(4)}_{HWW}&
=-\dfrac{8e^2}{v\sdt^2}\,a_{25}\,\xi^2\,,\\
\Delta g^{(5)}_{HZZ}-\dfrac{1}{c_\theta^2}\Delta g^{(5)}_{HWW}&
=-\dfrac{8e^2}{v\sdt^2}\,a_{19}\,\xi^2\,,\\
\Delta g^{(6)}_{HZZ}-\dfrac{1}{2c_\theta^2}\Delta g^{(6)}_{HWW}&
=-\dfrac{8e^2}{v\sdt^2}\,a_{15}\,\xi^2\,.
\end{align}
The non-vanishing terms on the right-hand side of
 Eqs.~(\ref{old1})-(\ref{new2}) stem from $\xi^2$-weighted terms in
 the non-linear Lagrangian. It is interesting to note that they would
 vanish in the following cases: i) the $d=6$ linear limit\footnote{Eq.~(\ref{old1})
 with vanishing right-hand side was already
 known~\cite{Hagiwara:1993ck,Hagiwara:1995vp} to hold in the linear
 regime at order $d=6$.}; ii) in the $\xi-$truncated non-linear
 Lagrangian; iii) in the custodial preserving limit. The first two relations with a vanishing right-hand side where already found in Ref.~\cite{Contino:2013kra}. Any hypothetical deviation from zero in the data
 combinations indicated by the left-hand side of those equations would
 thus be consistent with either $d=8$ corrections of the linear
 expansion or a non-linear realisation of the underlying dynamics.
    
Furthermore, we found an interesting correlation which only holds in the linear
regime at order $d=6$, mixes TGV and HVV couplings, and stems from
comparing Tables \ref{tab:tgv} and \ref{tab:hvv}:
\beq
\Delta\kappa_Z-\Delta
g_1^Z=\dfrac{v\st}{2\ct}\left[\left(\ct^2-\st^2\right)\left(g^{(1)}_{HZ\gamma}
+2g^{(2)}_{HZ\gamma}\right)+2\st\ct\left(2g_{H\g\g}-g^{(1)}_{HZZ}
-2g^{(2)}_{HZZ}\right)\right]\,.
\label{CorrelationMixed}
\eeq
This relation does not hold in the non-linear regime, not even when
only $\xi-$weighted operators are considered.  Its verification from
experimental data would  be an excellent test of BSM physics in
which the EWSB is linearly realized and dominated by
$d=6$ corrections.

The above general (de)correlations are a few examples among many
~\cite{SigmaInPreparation}.

 When in addition the strong
experimental constraints on the $S$ and $T$ parameters are applied,
disregarding thus $c_T$ and $c_1$ (equivalently, $f_{\phi 1}$ and
$f_{BW}$ for the linear case), supplementary constraints follow, {\em e.g.}:
\beq
\begin{aligned}
\dfrac{2}{m^2_Z} g^{(3)}_{HZZ}-\dfrac{1}{m^2_W+\delta m^2_W}g^{(3)}_{HWW}
&=\dfrac{16e^2}{v\,\st^2}a_{12}\xi^2\,,\\
2 g_{H\g\g}+\dfrac{\ct}{\st}g^{(2)}_{HZ\gamma}-g^{(2)}_{HWW}
&=-\dfrac{4e^2}{v\,\st^2}a_{12}\xi^2\,,\\
2 g^{(2)}_{HZZ}+\dfrac{\st}{\ct}g^{(2)}_{HZ\gamma}-g^{(2)}_{HWW}
&=-\dfrac{4e^2}{v\,\st^2}a_{12}\xi^2\,,\\
\dfrac{-2\st^2}{\ct^2-\st^2} g_{H\g\g}
+\dfrac{2\ct^2}{\ct^2-\st^2}g^{(2)}_{HZZ}-g^{(2)}_{HWW}
&=-\dfrac{4e^2}{v\,\st^2}a_{12}\xi^2\,.
\end{aligned}
\eeq
where again the non-zero entries on the right-hand sides vanish in
either the $d=6$ linear or the $\xi$-truncated non-linear limits.

\subsubsection*{Counting of degrees of freedom for the HVV Lagrangian}

Given the present interest in the gauge-$h$ sector, we analyze here the number of degrees of freedom involved in the HVV Lagrangian, Eq.~(\ref{eq:lhvv}), for on-shell and off-shell gauge and Higgs bosons, with massive and massless fermions.

This can be schematically resumed as follows: for the massive fermion case, 
\[
\begin{array}{llllll}
\text{phen. couplings:}	& 16 		& \xrightarrow{\text{i)}}& 12\,\, (\Delta g^{5,6}_{HVV}=0) 
												& \xrightarrow{\text{ii)}}& 10\,\, (\Delta g^{4}_{HVV}\text{ redundant})\\
\text{op. coefficients:}	& 17 		&\xrightarrow{\text{i)}}& 13\,\, (\cP_{11},\cP_{12},\cP_{16},\cP_{17}\text{ irrelevant} )
												&\xrightarrow{\text{ii)}}& 11\,\, (\cP_{7},\cP_{25}\,\text{ redundant})
\end{array}
\]
where the first line refers to the phenomenological couplings appearing in Eq.~(\ref{eq:lhvv}), while the second one to the operator coefficients of the non-linear basis in Eq.~(\ref{DeltaL}). Moreover, i) denotes the limit of on-shell gauge bosons, i.e. $\derp^\mu Z_\mu=0$ and $\derp^\mu W^{\pm}_\mu=0$, while ii) refers to the limit of, in addition, on-shell $h$. In brackets we indicate the couplings and the operator coefficients that are irrelevant for redundant under the conditions i) or ii). 
 
If fermion masses are set to zero, the conditions $\derp^\mu Z_\mu=0$ and $\derp^\mu W^{\pm}_\mu=0$ hold also for off-shell gauge bosons, and therefore the counting starts with 12 phenomenological couplings and 13 operator coefficients.

This analysis for the number of operator coefficients refers to the full non-linear Lagrangian in Eq.~(\ref{DeltaL}), which includes the custodial breaking operators.

\vspace{1 cm}
Up to this point, as well as in Apps.~\ref{AppEOM}, \ref{AppSiblings}
and \ref{AppFR} for the EOM, $d=6$ siblings and Feynman rules,
respectively, all non-linear pure gauge and gauge-$h$ operators of the
chiral Lagrangian Eq.~(\ref{Lchiral}) have been taken into account.
The next subsection describes the results of the numerical analysis,
and there instead the value of fermion masses on external legs will be
neglected. This means that operators $\cP_{9}(h)$, $\cP_{10}(h)$, $\cP_{15}(h)$, $\cP_{16}(h)$, and $\cP_{19-21}(h)$ become redundant then, and will not be analyzed.

%%%
% 4.2 Numerical Analysis
%%%

%%%%%%%%%%%%%%%%%%%%%%%%%%%%%%%%%%%%%%%%%%%%%%%%%%%%%%%
\boldmath
\subsection{Present bounds on operators  weighted by $\xi$}
\label{numerical}
\unboldmath
\begin{figure}[htb!]
\centering
\includegraphics[width=0.6\textwidth,height=0.35\textheight]{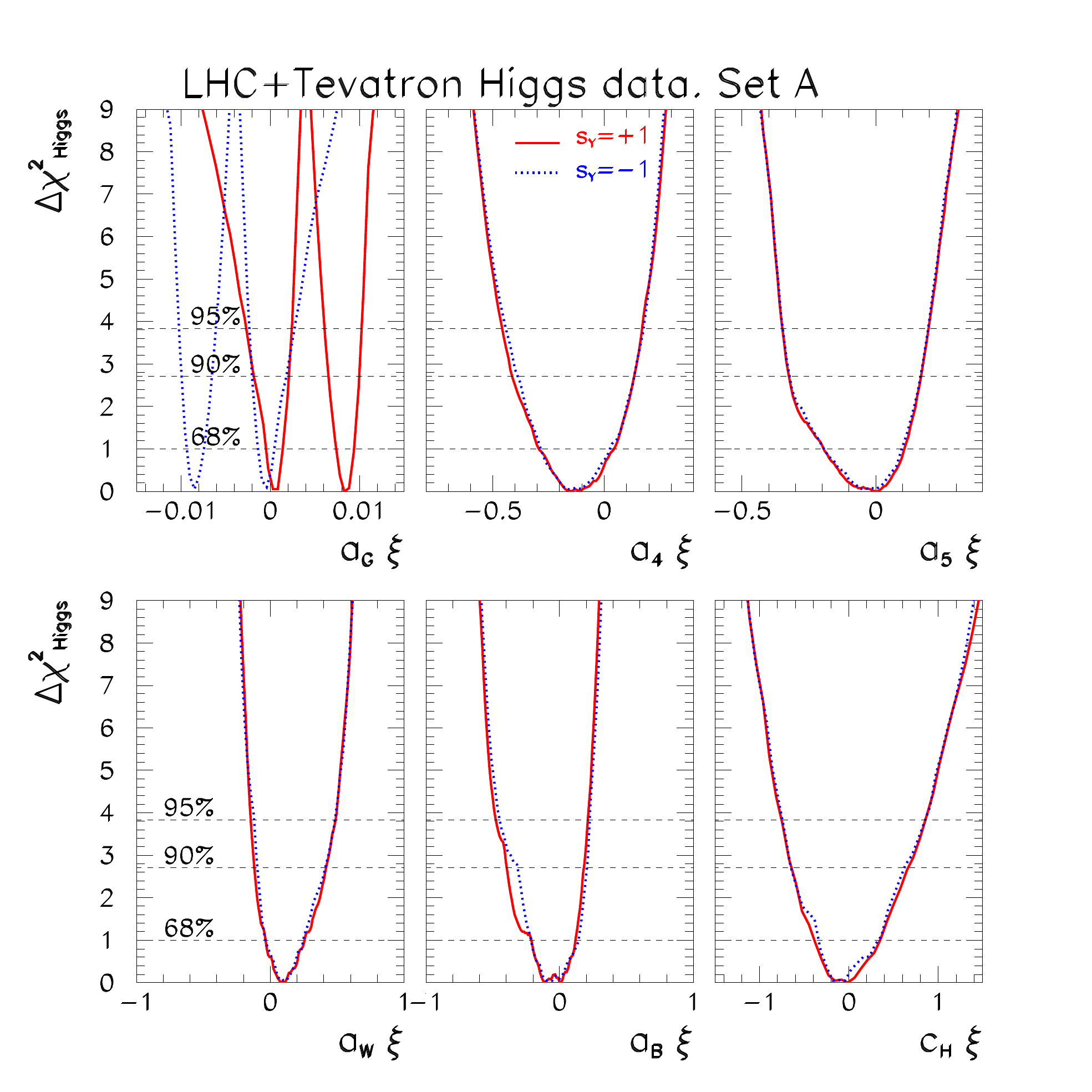}
\includegraphics[width=0.6\textwidth,,height=0.35\textheight]{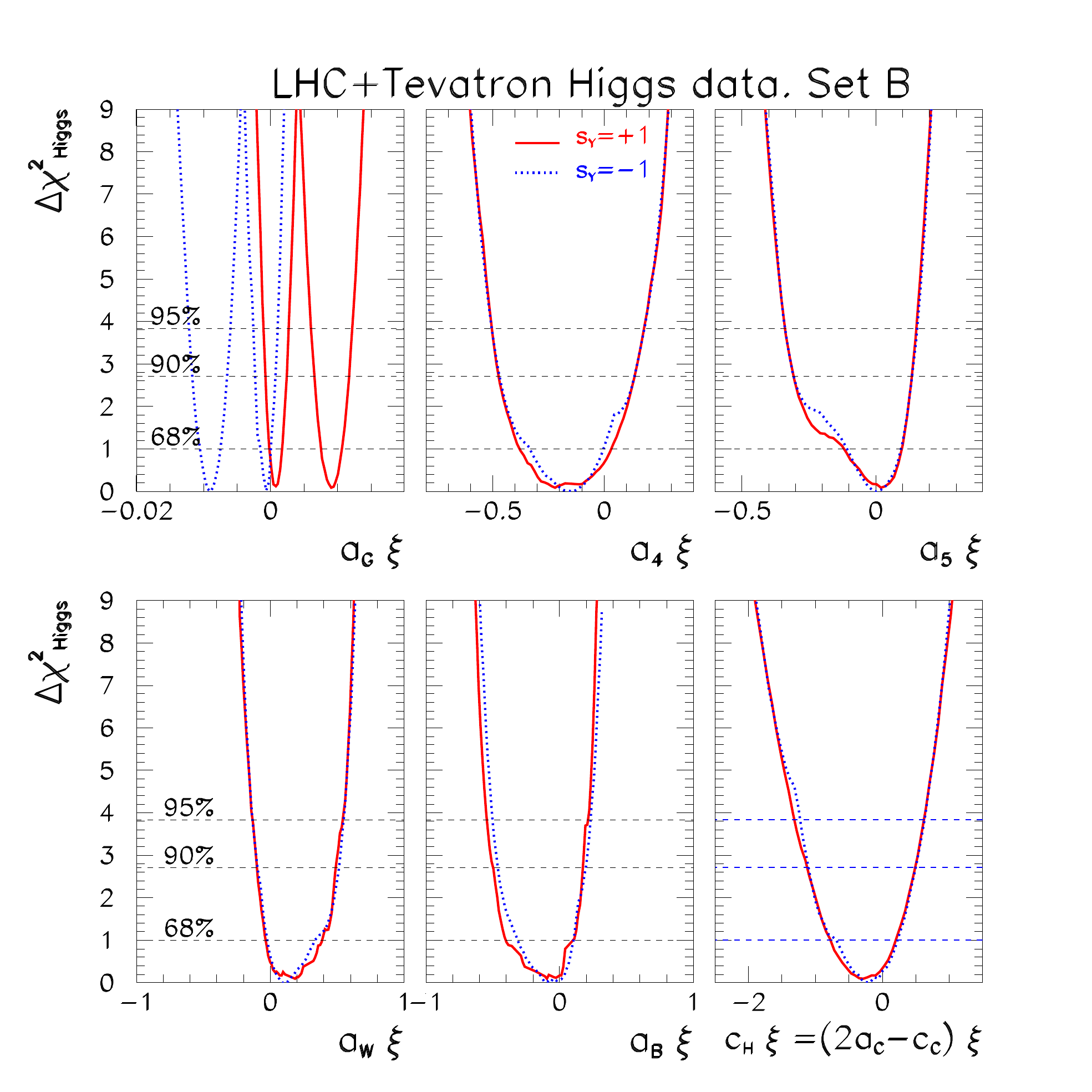}
\caption{\em $\Delta\chi^2_{\rm Higgs}$ dependence on the coefficients 
of the seven bosonic operators in Eq.~(\ref{eq:Hbase}) from the
analysis of all Higgs collider (ATLAS, CMS and Tevatron) data. In
each panel, we marginalized over the five undisplayed variables. 
The six upper (lower) panels corresponds to analysis with set {\bf A}
({\bf B}). In each panel the red solid (blue dotted) line stands for
the analysis with the discrete parameter $s_Y=+(-)$.}
\label{fig:1dima_6bos}
\end{figure}
At present, the most precise determination of $S$, $T$, $U$ from a
global fit to electroweak precision data (EWPD) yields the following
values and correlation matrix~\cite{Beringer:1900zz}
\begin{eqnarray}
\Delta S=0.00\pm 0.10 & \Delta T=0.02\pm 0.11 & \Delta U=0.03\pm 0.09\\
& & \nonumber \\
&\rho=\left(\begin{array}{ccc}
1 & 0.89 & -0.55 \\
0.89 & 1 &-0.8 \\
-0.55 & -0.8 &1 
\end{array}\right) \; . 
\label{eq:STUexp} 
\end{eqnarray}
Operators $\cP_{1}(h)$ 
and
$\cP_T(h)$ contribute at tree-level to these observables, 
see Eq.~(\ref{eq:STtree}) and consequently they are severely constrained. 
The corresponding 95\% CL allowed ranges for their coefficients  read
\begin{equation}
  -4.7\times 10^{-3} \leq \xi c_1\leq 4\times 10^{-3} 
\;\;\; \hbox{ and } \;\;\;
-2\times 10^{-3} \leq \xi c_T \leq 1.7 \times 10^{-3}  \;.
\label{eq:c1ctbounds}
\end{equation}
These constraints render the contribution of $\cP_{1}(h)$ and
$\cP_T(h)$ to the gauge-boson self-couplings and to the present Higgs
data too small to give any observable effect. Consequently we will not
include them in the following discussion.

As for the $\xi$-weighted TGV contributions from $\cP_{2}(h)$ and
$\cP_3(h)$, their impact  on  the coefficients $\Delta\kappa_\gamma$, $\Delta
g_1^Z$ and $\Delta\kappa_Z$ was described in Table~\ref{tab:tgv}.
At present, the most precise determination of TGV in this scenario
results from the two--dimensional analysis in Ref.~\cite{LEPEWWG} which
was performed in terms of $\Delta\kappa_\gamma$ and $\Delta g_1^Z$
with $\Delta\kappa_Z$ determined by the relation %in Table~\ref{tab:tgv}
Eq.~(\ref{old1}) with the right-handed side set to zero:
\begin{equation}
\kappa_\gamma=0.984^{+0.049}_{-0.049} \;\;\; \hbox{ and } \;\;\; 
g_1^Z=1.004^{+0.024}_{-0.025}\,,
\label{eq:tgvdata}
\end{equation}
with a correlation factor $\rho=0.11$. 
In Table \ref{tab:bounds} we list the corresponding 90\%
CL allowed ranges on the coefficients $c_2$ and $c_3$ 
from the analysis of the TGV data. 

Now, let us focus on the constraints on $\xi-$weighted operators
stemming from the presently available Higgs data on HVV couplings.
There are seven bosonic operators in this category
\footnote{ In present Higgs data analysis, the Higgs state
is on-shell and in this case  $\Delta g^{(4)}_{HVV}$ 
can be recasted as a $m_H^2$ correction to 
 $\Delta g^{(3)}_{HVV}$. Thus the contribution from $c_7$, 
i.e. the coefficient of  $\cP_{7}(h)$ to the Higgs observables, 
can be reabsorved in a redefinition of $2 a_C-c_C$.}
\beq
\cP_G(h)\,, \;\cP_4(h)\,,\; \cP_5(h)\,,\;
\cP_B(h)\,, \;\cP_W(h)\,,\; \cP_H(h)\,,\;  \cP_C(h).
\label{eq:Hbase}
\eeq  
To perform a
seven-parameter fit to the present Higgs data is technically beyond 
the scope of this paper and we will consider sets of ``only" six of 
them simultaneously. 
We are presenting below two such analysis. Leaving out a different
coupling in each set. In the first one, {\bf A}, 
we will neglect $\cP_C(h)$ 
and in the second one, {\bf B}, we will link its contribution to that of 
$\cP_H(h)$, so the 6 parameters in each set read: 
\begin{eqnarray}
&&\mathrm{{\bf Set \,\,A }:} \qquad
a_G\;,\;a_4\;, \;a_5\;,a_B\;,\;a_W\;,\;c_H\;,\; 2a_C-c_C=0\; 
 \;, 
\label{eq:HbaseA} \\
&&\mathrm{{\bf Set \,\,B }:} \qquad
a_G\;,\;a_4\;, \;a_5\;,a_B\;,\;a_W\;,\;c_H=2a_C-c_C  \;.
\label{eq:HbaseB}
\end{eqnarray}
For both sets we will explore the sensitivity of the results to
the sign of the $h$-fermion couplings by performing analysis with both 
values  of the discrete parameter $s_Y=\pm$.

As mentioned above, $\cP_H(h)$ and $\cP_C(h)$ induce a universal shift
to the SM-like HVV couplings involving electroweak gauge bosons, see
Eq.~({\color{blue} FR.15}) and ({\color{blue} FR.17}), 
while $\cP_H(h)$ also induces a universal
shift to the Yukawa Higgs-fermion couplings, 
see Eq.~({\color{blue} FR.32}).  In
consequence, the two sets above correspond to the case in which the
shift of the Yukawa Higgs-fermion couplings is totally unrelated to
the modification of the HVV couplings involving electroweak bosons
({\bf set B}), and to the case in which the shift of SM-like HVV
couplings involving electroweak bosons and to the Yukawa Higgs-fermion
couplings are the same ({\bf set A}). In both sets we keep all other
five operators which induce modifications of the HVV couplings with
different Lorentz structures than those of the SM as well as tree-level
contributions to the loop-induced vertices $h\gamma\gamma$, $h\gamma
Z$ and $hgg$.

Notice also that a combination of  $\cP_H(h)$ and  $\cP_C(h)$  can be traded 
via the EOM (see third line in Eq.~(\ref{RelH})) by that of fermion-Higgs
couplings $\cP_{f,\alpha\beta}(h)$ plus that of other operators
already present in the six-dimensional gauge-$h$ set analyzed. So our
choice allows us to stay close to the spirit of this work (past and
future data confronting directly the gauge and gauge-$h$ sector),
while performing a powerful six-dimensional exploration of possible
correlations.

Technically, in order to obtain the present constraints on the
coefficients of the bosonic operators listed in Eqs.~(\ref{eq:HbaseA})
and (\ref{eq:HbaseB}) we perform a chi--square test using the
available data on the signal strengths ($\mu$). We took into account
data from Tevatron D0 and CDF Collaborations and from LHC, CMS, and
ATLAS Collaborations at 7 TeV and 8 TeV for final states
$\gamma\gamma$, $W^+W^-$, $ZZ$, $Z\gamma$, $b\bar b$, and
$\tau\bar\tau$ \cite{Tuchming:2013wja,atlastau,atlasb,atlaszz,atlasww,
  atlasgamgam,atlasgamgamnew,CMStau,CMSb,CMSb2,CMSzz,CMSww,CMSgamgam,
  Chatrchyan:2013vaa}. For CMS and ATLAS data, the included results on
$W^+W^-$, $ZZ$ and $Z\gamma$ correspond to leptonic final states,
while for $\gamma\gamma$ all the different categories are included
which in total accounts for 56 data points.  We refer the reader to
Refs.~\cite{Corbett:2012dm,Corbett:2013pja} for details of the Higgs
data analysis.

The results of the analysis are presented in Fig.~\ref{fig:1dima_6bos}
which displays the chi--square ($\Delta \chi^2_{\rm Higgs}$)
dependence from the analysis of the Higgs data on the six bosonic
couplings for the two sets {\bf A} and {\bf B} of operators and for
the two values of the discrete parameter $s_Y=\pm$.  In each
panel $\Delta \chi^2_{\rm Higgs}$ is shown after marginalizing over
the other five parameters.  As seen in this figure, there are no
substantial difference between both sets in the determination of the
five common parameters with only slight differences in $a_G$ (more
below). The quality of the fit is equally good for both sets (
$|\chi^2_{min,\rm A}-\chi^2_{min,\rm B}|<0.5$).  The SM lays at
$\chi^2_{SM}=68.1$ within the 4\% CL region in the six dimensional
parameter space of either set.

In Fig.~\ref{fig:1dima_6bos}, for each set, the two curves of $\Delta
\chi^2_{\rm Higgs}$ for $s_Y=\pm$ are defined with respect to the
same $\chi^2_{min}$ corresponding to the minimum value of the two
signs. However, as seen in the figure, the difference is
inappreciable.  In other words, we find that in both six-parameter
analysis the quality of the description of the data is equally good
for both signs of the $h$-fermion couplings. Quantitatively for either
set $|\chi^2_{min,+}-\chi^2_{min,-}|$ is compatible with zero within
numerical accuracy.  If all the anomalous couplings are set to zero
the quality of the fit is dramatically different for both signs with
$\chi^2_{-}-\chi^2_{+}=26$. This arises from the different sign of the
interference between the $W$- and top-loop contributions to
$h\gamma\gamma$ which is negative for the SM value $s_Y=+$ and
positive for $s_Y=-$ which increases
$BR_{-}(h\rightarrow\gamma\gamma)/BR_{SM}(h\rightarrow\gamma\gamma)\sim
2.5$, a value strongly disfavoured by data. However, once the effect
of the 6 bosonic operators is included -- in particular that of
$\cP_B(h)$ and $\cP_W(h)$ which give a tree-level contribution to the
$h\gamma\gamma$ vertex -- we find that both signs of the $h$-fermion
couplings are equally probable.

In the figure we also see that in all cases $\Delta\chi^2_{\rm Higgs}$
as a function of $a_G$ exhibits two degenerate minima. They are due to
the interference between SM and anomalous contributions possessing
exactly the same momentum dependence. Around the secondary minimum the
anomalous contribution is approximately twice the one due to the
top-loop but with an opposite sign. The gluon fusion Higgs production
cross section is too depleted for $a_G$ values between the minima,
giving rise to the intermediate barrier. Obviously the allowed values
of $a_G$ around both minima are different for $s_Y=+$ and $s_Y=-$ as a
consequence of the different relative sign of the $a_G$ and the
top-loop contributions to the $hgg$ vertex. In the convention chosen
for the chiral Lagrangian, the relative sign of both contributions is
negative (positive) for $s_Y=+$, ($s_Y=-$) so that the non-zero
minimum occurs for $a_G$ around $0.01$ ($-0.01$). The precise value of
the $a_G$ coupling at the minima is slightly different for the
analysis with set ${\bf A}$ and ${\bf B}$ due to the effect of the
coefficient $c_H$ near the minima, which shifts the contribution of
the top-loop by a slightly different quantity in both analysis.

\begin{table}
\centering
\begin{tabular}{|c|l|l|}
\cline{2-3}
\multicolumn{1}{c|}{}&\multicolumn{2}{c|}{}\\[-3mm]
\multicolumn{1}{c|}{} &
\multicolumn{2}{c|}{90\% CL allowed range}
\\[1mm]
\cline{2-3} 
\multicolumn{1}{c}{}&\multicolumn{1}{|c}{}&\multicolumn{1}{|c|}{}\\[-3mm]
\multicolumn{1}{c|}{} &
\multicolumn{1}{c|}{Set A} & 
\multicolumn{1}{c|}{Set B} \\ [1mm]
\hline
\multicolumn{1}{|c|}{}&\multicolumn{1}{|c}{}&\multicolumn{1}{|c|}{}\\[-4mm]
\multicolumn{1}{|c|}{$a_G\xi (\cdot 10^{-3})$} & {\scriptsize $s_Y=+1$}: 
$[-1.8,2.1]\cup[6.5, 10]$ & {\scriptsize $s_Y=+1$}: $[-0.78, 2.4]\cup[6.5, 12]$
\\[2mm]
\multicolumn{1}{|c|}{} & {\scriptsize $s_Y=-1$}: $[-9.9,
-6.5]\cup[-2.1, 1.8]$
& {\scriptsize $s_Y=-1$}: $[-12, -6.5]\cup[-2.3, 0.75]$ \\[1mm]
\hline
\multicolumn{1}{|c|}{}&\multicolumn{2}{c|}{}\\[-4mm]
$a_4\xi$ & \multicolumn{2}{|c|}{$[-0.47, 0.14]$}
\\[1mm]
\hline
\multicolumn{1}{|c|}{}&\multicolumn{2}{c|}{}\\[-4mm]
$a_5\xi$ & \multicolumn{2}{|c|}{$[-0.33, 0.17]$}
\\[1mm]
\hline
\multicolumn{1}{|c|}{}&\multicolumn{2}{c|}{}\\[-4mm]
$a_W\xi$ & \multicolumn{2}{|c|}{$[-0.12, 0.51]$}
\\[1mm]
\hline
\multicolumn{1}{|c|}{}&\multicolumn{2}{c|}{}\\[-4mm]
$a_B\xi$ & \multicolumn{2}{|c|}{$[-0.50, 0.21]$}
\\[1mm]
\hline
\multicolumn{1}{|c|}{}&\multicolumn{1}{|c|}{}&\multicolumn{1}{|c|}{}\\[-4mm]
\multicolumn{1}{|c|}{$c_H\xi$} & 
\multicolumn{1}{|c|}{$[-0.66, 0.66]$} & 
\multicolumn{1}{|c|}{$[-1.1, 0.49]$}
\\[1mm]
\hline
\hline
\multicolumn{1}{|c|}{}&\multicolumn{2}{c|}{}\\[-4mm]
$c_2\xi$ & \multicolumn{2}{|c|}{$[-0.12, 0.076]$}
\\[1mm]
\hline
\multicolumn{1}{|c|}{}&\multicolumn{2}{c|}{}\\[-4mm]
$c_3\xi$ & \multicolumn{2}{|c|}{$[-0.064, 0.079]$}
\\[1mm]
\hline
\end{tabular}
\caption{\em 90\% CL allowed ranges of the coefficients 
  of the operators contributing to Higgs data
  ($a_G$, $a_4$, $a_5$, $a_W$, $a_B$, and $c_H$) and to TGV ($c_2$ and
  $c_3$). For the coefficients $a_4$, $a_5$, $a_W$, and $a_B$, for which the
  range is almost the same for analysis with both sets 
  and both values of $s_Y$  
  we show the inclusive range of the four analysis.
  For $c_H$ the allowed range is the same for both signs of $s_Y$.}
\label{tab:bounds}
\end{table}
%%%%%%%%%%%%%%%
\begin{figure}[ht!]
\centering
\includegraphics[width=0.48\textwidth]{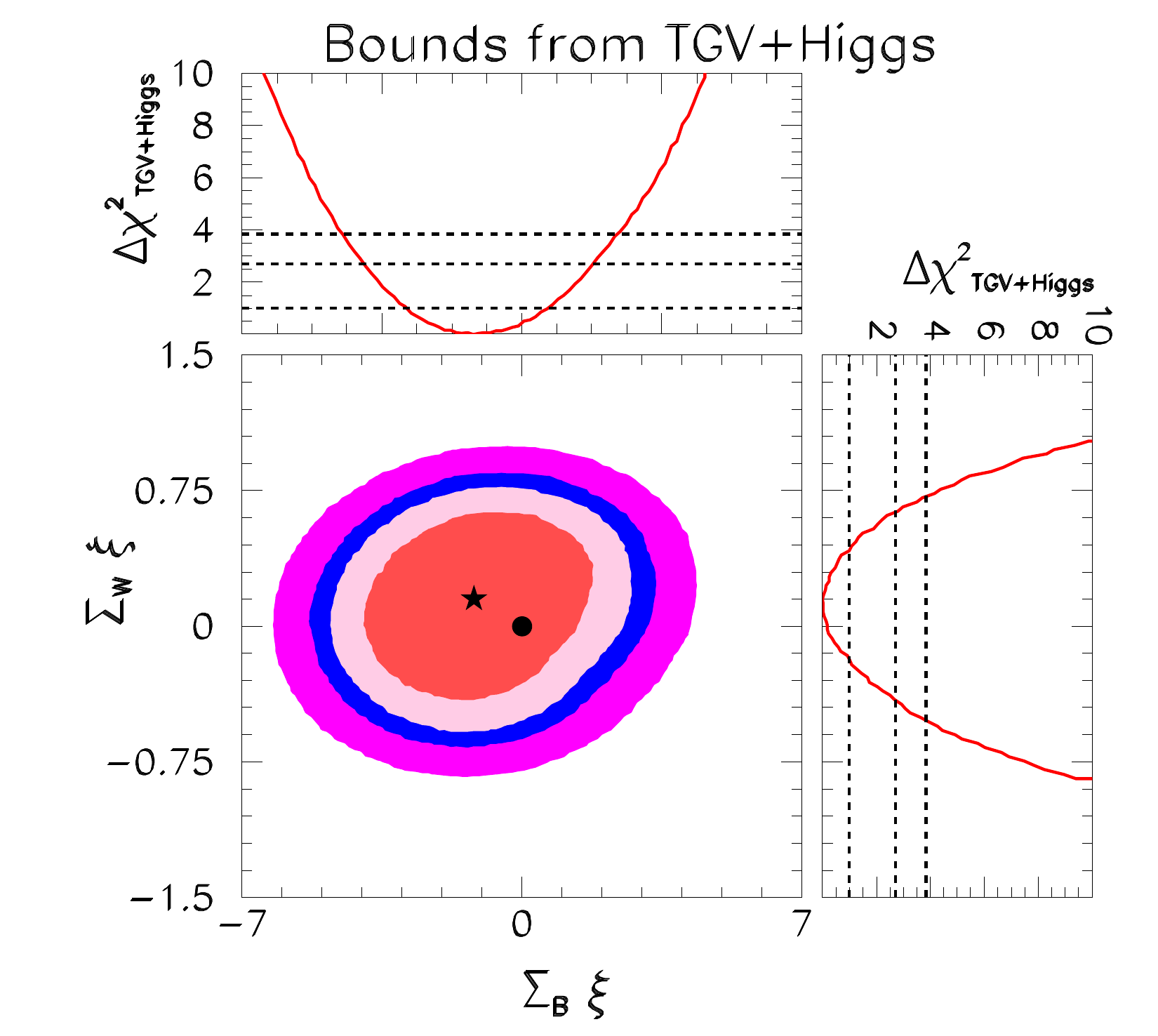}
\includegraphics[width=0.48\textwidth]{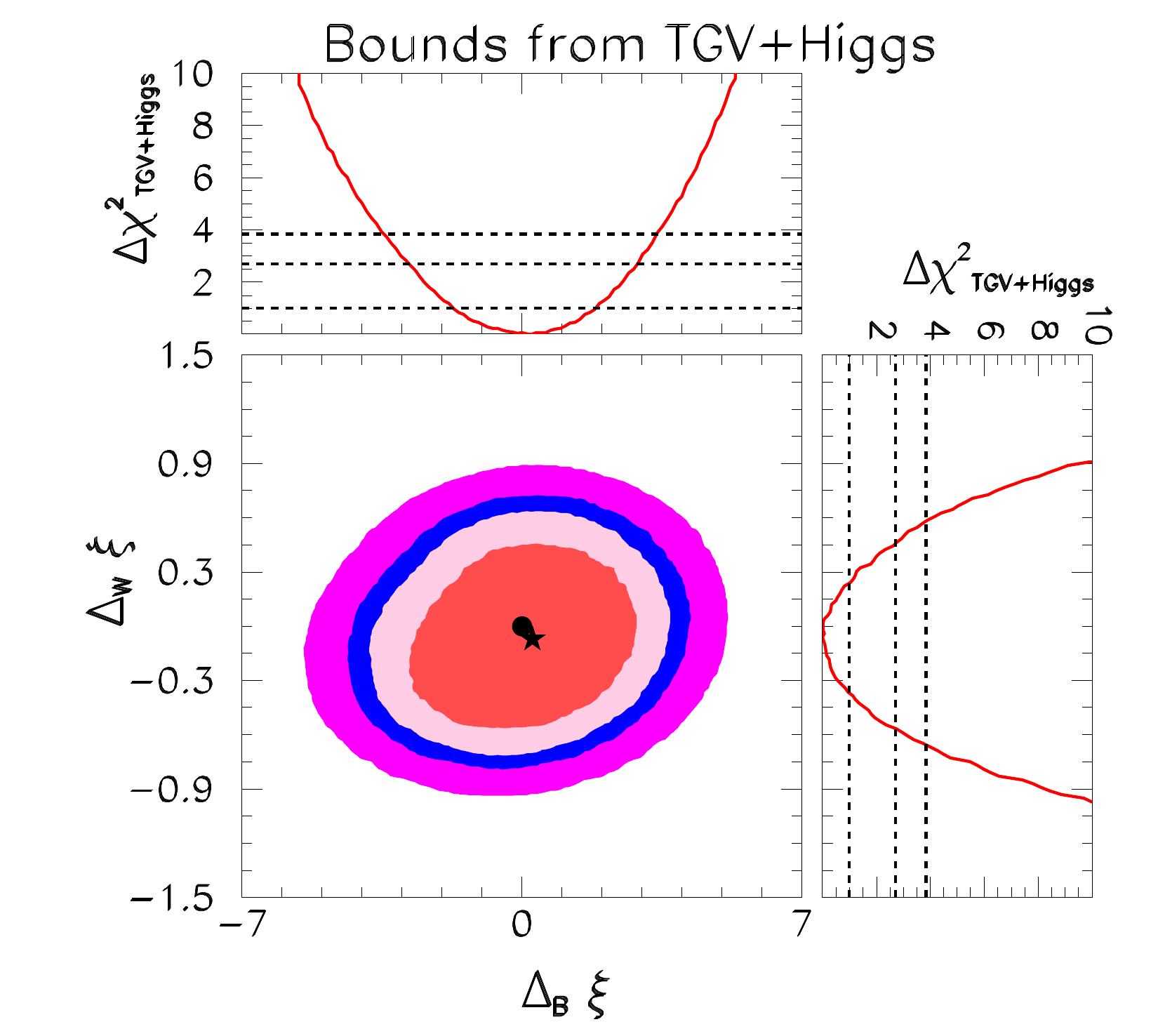}
\caption{\em
{\bf Left}:A BSM sensor irrespective of the type of expansion:
  constraints from  TGV and Higgs data on the combinations
  $\Sigma_B=4(2c_2+a_4)$ and $\Sigma_W=2(2c_3-a_5)$, which converge to
  $f_B$ and $f_W$ in the linear $d=6$ limit. 
The dot  at $(0,0)$ signals the SM expectation.
{\bf Right}:A non-linear versus linear discriminator: constraints on the
combinations $\Delta_B=4(2c_2-a_4)$ and $\Delta_W=2(2c_3+a_5)$, which
would take zero values in the linear (order $d=6$) limit (as well
as in the SM), indicated by the dot at $(0,0)$.  
For both figures the lower left panels
  shows the 2-dimensional allowed regions at 68\%, 90\%, 95\%, and
  99\% CL  after marginalization with respect to the other 
  six parameters 
($a_G$, $a_W$, $a_B$, $c_H$, $\Delta_B$, and $\Delta_W$)
and ($a_G$, $a_W$, $a_B$, $c_H$, $\Sigma_B$, and $\Sigma_W$) respectively.
The star corresponds to the best fit point of the analysis. 
The upper left and lower right panels give the
corresponding 1-dimensional projections over each of the two combinations. 
}
\label{fig:nl1}
\end{figure}

Fig.~\ref{fig:1dima_6bos} also shows that in all cases the curves for
$a_4$ and $a_5$ are  almost ``mirror symmetric''. This is due to
the strong anticorrelation between those two coefficients, because
they are the dominant contributions to the Higgs branching ratio into
two photons, which is proportional to $a_4+a_5$. In Table
\ref{tab:bounds} we list the corresponding 90$\%$ CL allowed
ranges for the six coefficients, for the different variants of the
analysis. With the expected uncertainties attainable in the Higgs
signal strengths in CMS and ATLAS at 14 TeV with an integrated
luminosity of 300 fb$^{-1}$~\cite{CMS:2013xfa,ATLAS:2013hta}, we
estimate that the sensitivity to those couplings can improve by a
factor $\cO(3-5)$ with a similar analysis.

We finish by stressing that in the context of
$\xi$--weighted operators in the chiral expansion the results from
TGV analysis and those from the HVV analysis apply to two
independent sets of operators as discussed in Sec.~\ref{sec:decor}.
This is unlike the case of the linear expansion for which $2 c_2=a_4$
and $2 c_3=-a_5$, which establishes an interesting complementarity in the
experimental searches for new signals in TGV and HVV couplings
in the linear regime \cite{Corbett:2013pja}. 
Conversely,  in the event of some anomalous observation in
either of these two sectors, the presence of this (de)correlation 
would allow for direct testing of the nature of the Higgs
boson. 
This is illustrated in Fig.~\ref{fig:nl1},
where the results of the combined analysis of the TGV and HVV 
data are projected into
combinations of the coefficients of the operators $\cP_2(h)$,
$\cP_3(h)$, $\cP_4(h)$ and $\cP_5(h)$: 
\beq
\begin{aligned}
\Sigma_B\equiv 4(2c_2+a_4)\,, \qquad\qquad \Sigma_W\equiv 2(2c_3-a_5)\,,\\
\Delta_B\equiv 4(2c_2-a_4)\,, \qquad\qquad \Delta_W\equiv 2(2c_3+a_5)\,,
\end{aligned}
\eeq defined such that at order $d=6$ of the linear regime
$\Sigma_B=c_B$, $\Sigma_W=c_W$, while $\Delta_B=\Delta_W=0$. With
these variables, the $(0,0)$ coordinate corresponds to the SM in
Fig.~\ref{fig:nl1} left panel, while in Fig.~\ref{fig:nl1} right panel
it corresponds to the linear regime (at order $d=6$). Would future
data point to a departure from $(0,0)$ in the variables of the first
figure it would indicate BSM physics irrespective of the linear or
non-linear character of the underlying dynamics; while such a
departure in the second figure would be consistent with a non-linear
realization of EWSB.  For concreteness the figures are
shown for the $s_Y=+$ analysis with set {\bf A}, but very similar
results hold for the other variants of the analysis.

%%%%%%%%%%%%%%%%%%%%%%%%%%%%%%%%%%%%%%%%%%%%%%%%%%%%%%%%%%%%%%%%%%%%%%
\boldmath
\subsection{$\xi^2$-weighted couplings: LHC potential to study $g_5^Z$}
\label{sec:numg5}
\unboldmath
One interesting property of the $\xi^2$-chiral Lagrangian is the presence of 
operator $\cP_{14}(h)$ that generates a non-vanishing $g_5^Z$ TGV, which
is a $C$ and $P$ odd, but $CP$ even operator; see
Eq.~(\ref{eq:classical}).  Presently, the best direct limits on this
anomalous coupling come from the study of $W^+W^-$ pairs and single
$W$ production at LEP II
energies~\cite{Abbiendi:2003mk,Achard:2004ji,Schael:2004tq}. Moreover,
the strongest bounds on $g_5^Z$ originate from its impact on
radiative corrections to  $Z$
physics~\cite{Eboli:1994jh,Dawson:1994qh,Eboli:1998hb}; see
Table~\ref{tab:exdata} for the available direct and indirect limits on
$g_5^Z$.

We can use the relation in Table~\ref{tab:tgv} to translate the
existing bounds on $g_5^Z$ into limits on $\cP_{14}(h)$. The
corresponding limits can be seen in the last column of
Table~\ref{tab:exdata}. We note here that these limits were obtained
assuming only a non-vanishing $g_5^Z$ while the rest of anomalous TGV
were set to their corresponding SM value.

%%%%%%%%%%%%%%%
\begin{table}[b!]
\begin{tabular}{c|c|c|c|}
\cline{2-4} 
\multicolumn{1}{c}{}&\multicolumn{1}{|c|}{}&\multicolumn{2}{c|}{}\\[-3mm]
\multicolumn{1}{c|}{} &
Measurement ($\pm 68\%$ CL region) & \multicolumn{2}{|c|}{$95\%$ CL region} \\[1mm]
  \hline
  \multicolumn{1}{|c|}{}&&&\\[-3mm]
  \multicolumn{1}{|c|}{Experiment} & $g_5^Z$ 
  & $g_5^Z$ & $c_{14}\xi^2$
  \\[1mm]
  \hline
  \multicolumn{1}{|c|}{}&&&\\[-3mm]
  \multicolumn{1}{|c|}{OPAL~\cite{Abbiendi:2003mk}} & $-0.04^{+0.13}_{-0.12}$ 
  & $[-0.28, 0.21 ]$ & $[-0.16, 0.12 ]$
  \\[1mm]
  \hline
  \multicolumn{1}{|c|}{}&&&\\[-3mm]
  \multicolumn{1}{|c|}{L3~\cite{Achard:2004ji}} & $0.00^{+0.13}_{-0.13}$ 
  & $[-0.21, 0.20 ]$ & $[-0.12, 0.11 ]$
  \\[1mm]
  \hline
  \multicolumn{1}{|c|}{}&&&\\[-3mm]
  \multicolumn{1}{|c|}{ALEPH~\cite{Schael:2004tq}} & $-0.064^{+0.13}_{-0.13}$ 
  & $[-0.317, 0.19 ]$ & $[-0.18, 0.11 ]$
  \\[1mm]
  \hline
  \multicolumn{4}{c}{}
  \\
  \hline
   \multicolumn{1}{|c}{}&&&\\[-3mm]
  \multicolumn{2}{|c|}{$90\%$ CL region  from indirect bounds~\cite{Eboli:1994jh,Dawson:1994qh,Eboli:1998hb} }
  & $g_5^Z$: $[-0.08,0.04]$ & $c_{14}\xi^2$: $[-0.04,0.02]$
  \\[1mm]
  \hline
\end{tabular}
\caption{\em Existing direct measurements of $g_5^Z$ coming from LEP
  analyses~\cite{Abbiendi:2003mk,Achard:2004ji,Schael:2004tq} as well
  as the strongest constraints from the existing indirect bounds on
  $g_5^Z$ in the
  literature~\cite{Eboli:1994jh,Dawson:1994qh,Eboli:1998hb}. In the
  last column we show the translated bounds on $c_{14}\xi^2$. These
  bounds were obtained assuming only $g_5^Z$ different from zero while
  the rest of anomalous
TGV were set to the SM values.}
\label{tab:exdata}
\end{table}
%%%%%%%%%%%%%%%

At present, the LHC collaborations have presented some data analyses
of anomalous TGV ~\cite{Aad:2012twa,Aad:2012oea,Lombardo:2013daa,
Chatrchyan:2013fya,Chatrchyan:2012bd} but in none of them have they
included the effects of $g_5^Z$. A preliminary study on the potential
of LHC 7 to constrain this coupling was presented in
Ref.~\cite{Eboli:2010qd} where it was shown that the LHC 7 with a very
modest luminosity had the potential of probing $g_5^Z$ at the level of
the present indirect bounds.  In Ref.~\cite{Eboli:2010qd} it was also
discussed the use of some kinematic distributions to characterize the
presence of a non-vanishing $g_5^Z$. So far the LHC has already
collected almost $25$ times more data than the luminosity considered
in this preliminary study which we update here. Furthermore, in this
update we take advantage of a more realistic background evaluation, by
using the results of the experimental LHC analysis on other anomalous
TGV couplings ~\cite{Aad:2012twa}.

At the LHC, the anomalous coupling $g_5^Z$ contributes to $WW$ and
$WZ$ pair production, with the strongest limits originating from the
last reaction~\cite{Eboli:2010qd}.  Hence, the present study is
focused on the $WZ$ production channel, where we consider only the
leptonic decays of the gauge bosons for a better background
suppression, \ie, we analyze the reaction
\begin{equation}
 pp\rightarrow \ell^{\prime\pm}\ell^+\ell^-E_T^{miss}\,,
\label{eq:3lmiss}
\end{equation}
where $\ell^{(\prime)}=e$ or $\mu$. The main background for the
$g_5^Z$ analysis is the irreducible SM production of $WZ$ pairs.
There are further reducible backgrounds like $W$ or $Z$ production
with jets, $ZZ$ production followed by the leptonic decay of the $Z$'s
with one charged lepton escaping detection and $t\bar{t}$ pair
production.

We simulated the signal and the SM irreducible background using an
implementation of the anomalous operator $g_5^Z$ in
FeynRules~\cite{Christensen:2008py} interfaced with MadGraph
5~\cite{Alwall:2011uj} for event generation. We account for the
different detection efficiencies by rescaling our simulation to the
one done by ATLAS~\cite{Aad:2012twa} for the study of $\Delta
\kappa_Z$, $g_1^Z$ and $\lambda_Z$. However, we also cross checked the
results using a setup where the signal simulation is based on the same
FeynRules~\cite{Christensen:2008py} and MadGraph5~\cite{Alwall:2011uj}
implementation, interfaced then with PYTHIA~\cite{Sjostrand:2006za}
for parton shower and hadronization and with PGS 4~\cite{pgs} for
detector simulation.  Finally, the reducible backgrounds for the 7 TeV
analysis were obtained from the simulations presented in the ATLAS
search~\cite{Aad:2012twa}, and they were properly rescaled for the 8
TeV and 14 TeV runs.

In order to make our simulations more realistic, we closely follow the
TGV analysis performed by ATLAS~\cite{Aad:2012twa}.  Thus, the
kinematic study of the $WZ$ production starts with the usual detection
and isolation cuts on the final state leptons.  Muons are considered
if their transverse momentum with respect to 
the collision axis $z$, $p_T\equiv\sqrt{p_x^2+p_y^2}$, and pseudorapidity 
$\eta\equiv\frac{1}{2}\ln\frac{|\vec p| +p_z}{|\vec p| -p_z}$, satisfy  
\begin{eqnarray}
 p_T^\ell>15\mbox{  GeV  ,   } |\eta^\mu|<2.5\mbox{   .}
\end{eqnarray}
Electrons must comply with the same transverse momentum requirement
than that applied to muons; however, the electron pseudo-rapidity cut is
\begin{eqnarray}
 |\eta^e|<1.37 \mbox{      or      } 1.52< |\eta^e|<2.47\, .
\end{eqnarray}
To guarantee the isolation of muons (electrons),  we required that the
scalar sum of the $p_T$ of the particles within 
$\Delta R\equiv\sqrt{\Delta\eta^2+\Delta\phi^2}=0.3$ of the
muon (electron), excluding the muon (electron) track, is smaller than
15$\%$ (13\%) of the charged lepton $p_T$. In the case where the final
state contains both muons and electrons,  a further isolation
requirement has been imposed:
\begin{eqnarray}
 \Delta R_{e\mu} > 0.1\mbox{             .}
\end{eqnarray}

It was also required that at least two leptons with the same flavour and
opposite charge are present in the event and that their invariant mass
is compatible with the $Z$ mass, \ie
\begin{eqnarray}
 M_{\ell^+\ell^-}\in \left[M_Z-10,\ \ M_Z+10\right] \mbox{   GeV.}
\end{eqnarray}
A further constraint imposed is that a third lepton is present which passes the above
detection requirements and whose transverse momentum satisfies
\begin{eqnarray}
 p_T^\ell>20\mbox{  GeV .}
\end{eqnarray}
Moreover, with the purpose of suppressing most of the
$Z+\mbox{jets}$ and other diboson production background, we required
\begin{eqnarray}
 E_T^{\mbox{miss}}>25\mbox{  GeV  and      } M_T^W>20\mbox{  GeV  ,}
\end{eqnarray}
where $ E_T^{\mbox{miss}}$ is the missing transverse energy and the
transverse mass is defined as
\begin{eqnarray}
 M_T^W=\sqrt{2p_T^\ell E_T^{\mbox{miss}}\left(1-\cos(\Delta\phi)\right)}\,,
\end{eqnarray}
with $p_T^\ell$ being the transverse momentum of the third lepton, and
where $\Delta\phi$ is the azimuthal angle between the missing
transverse momentum and the third lepton.  Finally, it was required  that
at least one electron or one muon has a transverse momentum complying
with
\begin{eqnarray}
p_T^{e(\mu)}>25\ \  (20)\mbox{       GeV.}
\end{eqnarray}

Our Monte Carlo simulations have been tuned to the ATLAS
ones~\cite{Aad:2012twa}, so as to incorporate more realistic detection
efficiencies. Initially,  a global $k$-factor was introduced to account
for the higher order corrections to the process in Eq.~(\ref{eq:3lmiss}) by
comparing our leading order prediction to the NLO one used in the
ATLAS search~\cite{Aad:2012twa}, leading to $k\sim 1.7$.  Next, we
compared our results after cuts with the ones quoted by ATLAS in Table
1 of Ref.~\cite{Aad:2012twa}.  We tuned our simulation by applying a
correction factor per flavour channel ($eee$, $ee\mu$, $e\mu\mu$ and
$\mu\mu\mu$) that is equivalent to introducing a detection efficiency of
$\epsilon^e=0.8$ for electrons and $\epsilon^\mu=0.95$ for muons.  
These efficiencies have been employed in our simulations for signal and
backgrounds.

After applying all the above cuts and efficiencies, the cross section
for the process (\ref{eq:3lmiss}) in the presence of a non-vanishing
$g_5^Z$ can be written as \footnote{ We assumed in this study that all
  anomalous TGV vanish except for $g_5^Z$.}
\begin{equation}
  \sigma\ =\ \sigma_{\mbox {bck}}\ + \sigma_{SM}\ +\ \sigma_{\mbox{int}}\ g_5^Z\ 
        +\ \sigma_{\mbox{ano}}\ \big(g_5^Z\big)^2\,,
\end{equation}
where $\sigma_{SM}$ denotes the SM contribution to $W^\pm Z$
production, $\sigma_{\mbox{int}}$ stands for the interference between
this SM process and the anomalous $g_5^Z$ contribution and
$\sigma_{\mbox{ano}}$ is the pure anomalous contribution.
Furthermore, $\sigma_{\mbox{bck}}$ corresponds to all background
sources except for the SM EW $W^\pm Z$ production.  We present in
Table~\ref{tab:cs} the values of $\sigma_{SM}$, $\sigma_{\mbox{int}}$
and $\sigma_{\mbox{ano}}$ for center--of--mass energies of 7, 8 and 14
TeV, as well as the cross section for the reducible backgrounds.

%%%%%%%%%%%%%%%
\begin{table}[htb!]
\begin{center}
\begin{tabular}{|c|c|c|c|c|}
\hline 
&&&&\\[-2mm]
COM Energy  & $\sigma_{\mbox{bck}}$ (fb) & $\sigma_{SM}$ (fb) & $\sigma_{\mbox{int}}$ (fb) & $\sigma_{\mbox{ano}}$ (fb)
\\[1mm]
\hline
&&&&\\[-2mm]
7 TeV & 14.3 & 47.7 & 6.5 & 304
\\[1mm]
\hline
&&&&\\[-2mm]
8 TeV & 16.8 & 55.3 & 6.6 & 363
\\[1mm]
\hline
&&&&\\[-2mm]
14 TeV & 29.0 & 97.0 & 9.1 & 707
\\[1mm]
\hline
\end{tabular}
\caption{ {\em Values of the cross section predictions for the process
  $pp\rightarrow \ell^{\prime\pm}\ell^+\ell^-E_T^{miss}$ after
  applying all the cuts described in the text. $\sigma_{SM}$ is the SM
  contribution coming from EW $W^\pm Z$ production,}
  $\sigma_{\mbox{int}}$ {\em is the interference between this SM process
  and the anomalous $g_5^Z$ contribution,} $\sigma_{\mbox{ano}}$ {\em is the
  pure anomalous contribution and } $\sigma_{\mbox{bck}}$ {\em corresponds to
  all background sources except for the SM EW $W^\pm Z$ production.}}
\label{tab:cs}
\end{center}
\end{table}
%%%%%%%%%%%%%%%

\begin{figure}[htb!]
\centering
 \includegraphics[width=0.4\textwidth]{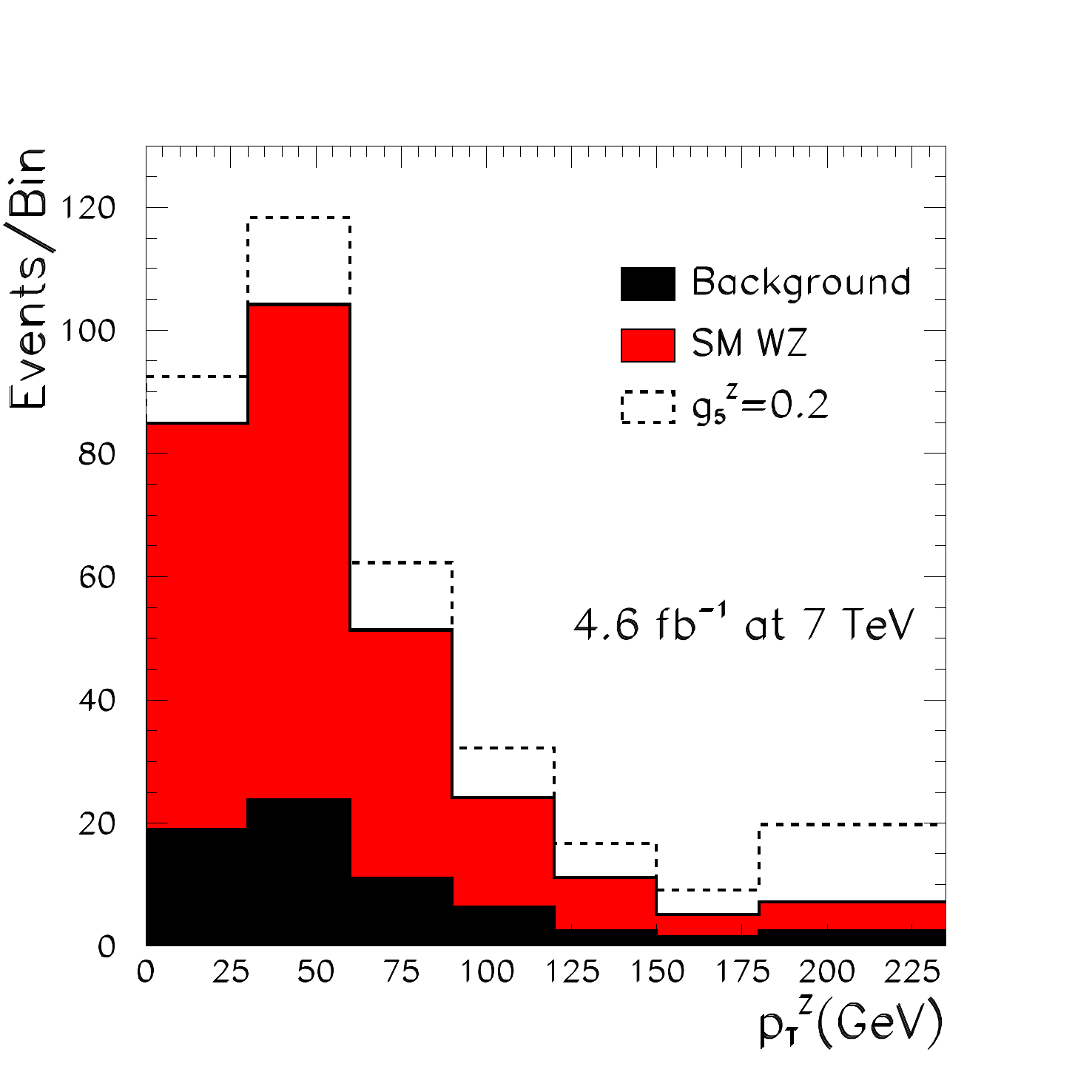}
  \includegraphics[width=0.42\textwidth]{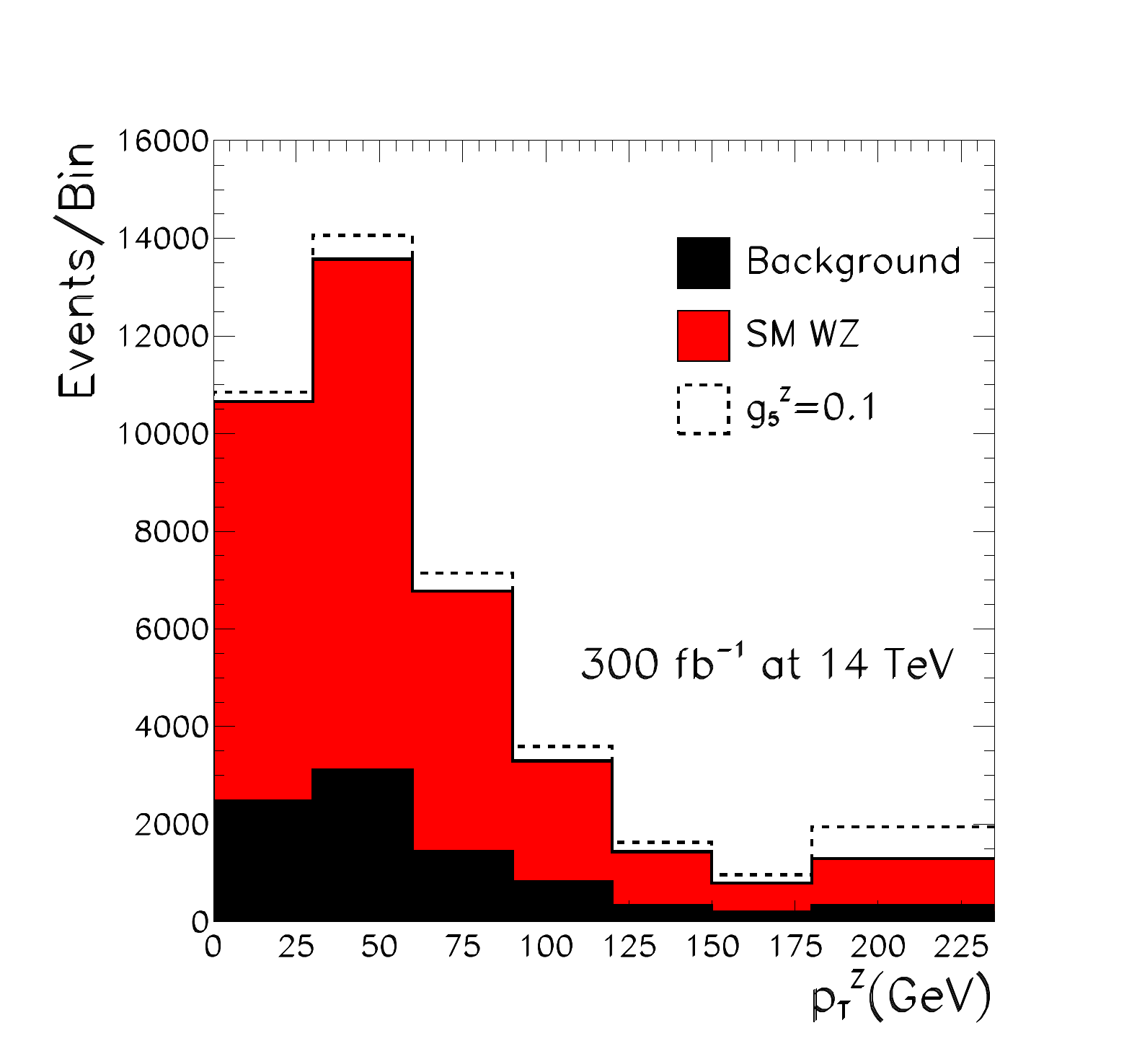}
  \caption{\em The left (right) panel displays the number of expected
    events as a function of the $Z$ transverse momentum  for a
    center--of--mass energy of 7 (14) TeV, assuming an integrated
    luminosity of $4.64$ ($300$) fb$^{-1}$. The black histogram
    corresponds to the sum of all background sources except for the SM
    electroweak $pp\rightarrow W^\pm Z$ process, while the red
    histogram corresponds to the sum of all SM backgrounds, and the
    dashed distribution corresponds to the addition of the anomalous
    signal for $g_5^Z=0.2$ ($g_5^Z=0.1$). The last bin contains all
    the events with $p_T^Z>180$ GeV.}
\label{fig:g5}
\end{figure}

In order to quantify the expected limits on $g_5^Z$,  advantage has been taken in this analysis 
of the fact that anomalous TGVs enhance the cross sections at high
energies. Ref.~\cite{Eboli:2010qd} shows that the variables
$M_{WZ}^{\mbox{rec}}$ (the reconstructed $W-Z$ invariant mass), $p_T^{\ell\ \mbox{max}}$ and $p_T^Z$ are able
to trace well this energy dependence, leading to similar sensitivities
to the anomalous TGV. Here, we chose $p_T^Z$ to study $g_5^Z$ because
this variable is strongly correlated with the subprocess
center--of--mass energy ($\hat{s}$), and, furthermore, it can be
directly reconstructed with good precision from the measured lepton
momenta. The left (right) panel of Figure~\ref{fig:g5} depicts the
number of expected events with respect to the $Z$ transverse momentum
for the 7 (14) TeV run and an integrated luminosity of $4.64$ ($300$)
fb$^{-1}$.  As illustrated by this figure, the existence of an
anomalous $g_5^Z$ contribution enhances the tail of the $p_T^Z$
spectrum, signaling the existence of new physics.

%%%%%%%%%%%%%%%%%%%%%%%%%%%%%%%%%%%%%%%%%%%%%%%%%%%%%
%%%%%%%%%%%%%%%%%%%%%%%%%%%%%%%%%%%%%%%%%%%%%%%%%%%%%

Two procedures have been used to estimate the LHC potential to probe
anomalous $g_5^Z$ couplings. In the first approach, we performed a
simple event counting analysis assuming that the number of observed
events correspond to the SM prediction ($g_5^Z=0$) and we look for the
values of $g_5^Z$ which are inside the $68\%$ and $95\%$ CL allowed
regions. As suggested by Ref.~\cite{Eboli:2010qd}, the
following additional cut was applied in this analysis to enhance the sensitivity to
$g_5^Z$: 
\begin{eqnarray}
 p_T^Z>90\mbox{       GeV.}
\end{eqnarray}
On a second analysis,  a simple $\chi^2$  was built based on the contents
of the different bins of the $p_T^Z$ distribution, in order to obtain
more stringent bounds. The binning used is shown in Fig.~\ref{fig:g5}.
Once again, it was assumed that the observed $p_T^Z$ spectrum corresponds
to the SM expectations and we sought for the values of $g_5^Z$ that
are inside the $68\%$ and $95\%$ allowed regions. The results of both
analyses are presented in Table~\ref{tab:results}.

%%%%%%%%%%%%%%%
\begin{table}[htb!]
\hspace{-1.2cm}
{\footnotesize
\begin{tabular}{c|c|c|c|c|}
\cline{2-5}
&\multicolumn{2}{|c|}{}&\multicolumn{2}{|c|}{}\\[-3mm]
& \multicolumn{2}{|c|}{$68\%$ CL range} & \multicolumn{2}{|c|}{$95\%$ 
CL range}
\\[1mm]
\hline
\multicolumn{1}{|c|}{}&&&&\\[-3mm]
\multicolumn{1}{|c|}{Data sets used} & Counting $p_T^Z>90$ GeV 
& $p_T^Z$ binned analysis & Counting $p_T^Z>90$ GeV & $p_T^Z$ binned analysis
 \\[1mm]
\hline
\multicolumn{1}{|c|}{}&&&&\\[-3mm]
\multicolumn{1}{|c|}{7+8 TeV} & $(-0.066,\ 0.058)$ 
& $(-0.057,\ 0.050)$ & $(-0.091,\ 0.083)$ & $(-0.080,\ 0.072)$
\\
\multicolumn{1}{|c|}{(4.64+19.6 fb$^{-1}$)} & &&&\\[1mm]
\hline
\multicolumn{1}{|c|}{}&&&&\\[-3mm]
\multicolumn{1}{|c|}{7+8+14 TeV} 
& $(-0.030,\ 0.022)$ & $(-0.024,\ 0.019)$ & $(-0.040,\ 0.032)$ 
& $(-0.033,\ 0.028)$
\\
\multicolumn{1}{|c|}{(4.64+19.6+300 fb$^{-1}$)}&&&&\\[1mm]
\hline
\end{tabular}
\caption{\em Expected sensitivity on $g_5^Z$ at the LHC for the two different 
  procedures described in the text.}
\label{tab:results}
}
\end{table}
%%%%%%%%%%%%%%%

We present in the first row of Table~\ref{tab:results} the expected
LHC limits for the combination of the 7 TeV and 8 TeV existing data
sets, where we considered an integrated luminosity of $4.64$ fb$^{-1}$
for the 7 TeV run and $19.6$ fb$^{-1}$ for the 8 TeV one. Therefore,
the attainable precision on $g_5^Z$ at the LHC 7 and 8 TeV runs is
already higher than the present direct bounds stemming from LEP and it
is also approaching the present indirect limits.  Finally, the last
row of Table~\ref{tab:results} displays the expected precision on
$g_5^Z$ when the 14 TeV run with an integrated luminosity of $300$
fb$^{-1}$ is included in the combination. Here, once more, it was
assumed that the observed number of events is the SM expected one.
The LHC precision on $g_5^Z$ will approach the per cent level, clearly
improving the present both direct and indirect bounds.

%%%%%%%%%%%%%%%%%%%%%%%%%%%%%%%%%%%%%%%%%%%%%%%%%%%%%%%%%%%%%%%%%%%%%%
\boldmath
\subsection{Anomalous quartic couplings}
\label{sec:numquartic}
\unboldmath
As shown in Sect.~\ref{newsignals}, in the chiral expansion several
operators weighted by $\xi$ or higher powers contribute to quartic
gauge boson vertices without inducing any modification to TGVs. 
Therefore, their coefficients are much less
constrained at present, and one can expect still larger deviations on
future studies of quartic vertices at LHC for
large values of $\xi$.
This is unlike in the
linear expansion, in which the modifications of quartic gauge
couplings that do not induce changes to TGVs appear only when the $d=8$
operators are considered~\cite{Eboli:2006wa}. For instance, the linear
operators similar to $\cP_{6}(h)$ and $\cP_{11}(h)$ are ${\cal L}_{S,0}$ 
and ${\cal L}_{S,1}$ in Ref.~\cite{Eboli:2006wa}.

Of the five operators giving rise to purely quartic gauge boson
vertices ($\cP_{6}(h)$, $\cP_{11}(h)$, $\cP_{23}(h)$, $\cP_{24}(h)$,
$\cP_{26}(h)$), none modifies quartic vertices including photons while
all generate the anomalous quartic vertex $ZZZZ$ that is not present
in the SM. Moreover, all these operators but $\cP_{26}(h)$ modify the
$ZZW^+W^-$ vertex, while only $\cP_{6}(h)$ and $\cP_{11}(h)$ also induce
anomalous contributions to $W^+W^-W^+W^-$.

Presently, the most stringent bounds on the coefficients of these
operators are indirect, from their one--loop contribution to the EWPD derived
in Ref.~\cite{Brunstein:1996fz} where it was shown that the five
operators correct  $\alpha\Delta T$ while render $\alpha\Delta
S=\alpha\Delta U=0$.  In Table~\ref{tab:QGCbounds} we give the updated
indirect bounds using the determination of the oblique parameters in
Eq.~(\ref{eq:STUexp}).

At the LHC these anomalous quartic
couplings can be directly tested in the production of three vector bosons
($VVV$) or in vector boson fusion (VBF) production of two gauge
bosons~\cite{Eboli:2000ad}. At lower center--of--mass energies the
best limits originate from the $VVV$ processes, while the VBF channel
dominates for the 14 TeV run~\cite{Eboli:2000ad,Eboli:2003nq,
Eboli:2006wa,Belyaev:1998ih,Degrande:2013yda}.

At the LHC with 14 TeV center-of-mass energy, the couplings $c_{6}$ and $c_{11}$
can be constrained by combining their impact on the VBF channels
\begin{equation}
  p p \to jj W^+ W^-   \;\;\;\hbox{and}\;\;\;
    p p \to jj (W^+ W^+ + W^-W^-) \,,
\end{equation}
where $j$ stands for a tagging jet and the final state $W$'s decay
into electron or muon plus neutrino. It was shown in
Ref.~\cite{Eboli:2006wa} that the attainable 99\% CL limits on these
couplings are
\begin{equation}
-12 \times 10^{-3} < c_{6} \, \xi< 10 \times 10^{-3} \;\;\;\;,\;\;\;\;
-7.7 \times 10^{-3} < c_{11}\, \xi^2 < 14 \times 10^{-3} 
\end{equation}
for an integrated luminosity of 100 fb$^{-1}$. Notice that the
addition of the channel $pp \to jjZZ$ does not improve significantly
the above limits~\cite{Belyaev:1998ih}.

\begin{table}
\begin{center}
\begin{tabular}{|c|c|}
\hline
&\\[-3mm]
coupling & 90\% CL allowed region
\\[1mm]
\hline
&\\[-3mm]
$c_{6} \,\xi$  & $[-0.23, 0.26]$
\\[1mm]
\hline
&\\[-3mm]
$c_{11}\, \xi^2$  & $[-0.094, 0.10]$
\\[1mm]
\hline
&\\[-3mm]
$c_{23}\, \xi^2$  & $[-0.092, 0.10]$
\\[1mm]
\hline
&\\[-3mm]
$c_{24}\, \xi^2$  & $[-0.012, 0.013]$
\\[1mm]
\hline
&\\[-3mm]
$c_{26}\, \xi^4$  & $[-0.0061, 0.0068]$
\\[1mm]
\hline
\end{tabular}
\caption{\em 90\% CL limits on the anomalous quartic couplings from their 
one-loop contribution to the EWPD. The bounds were obtained assuming only 
one operator different from zero at a time and for a
  cutoff scale $\Lambda_s = 2$ TeV.}
\label{tab:QGCbounds}
\end{center}
\end{table}
%%%%%%%%%%%%%%%

%
%%%%%%%%%%%%%%%%%%%%%%%%%   .  Conclusions       %%%%%%%%%%%%%%%%%%%%%%%%
%

\section{Conclusions}
\label{Conclusions}

In this paper we have made a comparative study of the 
departures from the Standard Model predictions in  theories
based on  linear  and non-linear realizations of  
$SU(2)_L\times U(1)_Y$ gauge symmetry breaking. To address
this question in a model-independent way, we have considered
effective Lagrangians containing either a light fundamental Higgs 
in the linear realization or a light 
dynamical Higgs in the non-linear one. 
We have exploited the fact that these two expansions are intrinsically
 different from the point of view of the presence or absence,
 respectively, of a global $SU(2)_L$ symmetry in the effective
 Lagrangian, with the light Higgs scalar behaving as a singlet in the
 chiral case.  Less symmetry means more possible invariant operators
 at a given order, and the result is that the non-linear realization
 for a light dynamical Higgs particle is expected to exhibit a larger
 number of independent couplings than linear ones.  This has been
 explored here concentrating on the CP-even operators involving pure
 gauge and gauge-$h$ couplings.  First, in
 Sec.~\ref{EffectiveLagrangian} we have presented the maximal set of
 independent (and thus non-redundant) operators of that type contained
 in the effective chiral Lagrangian for a light dynamical Higgs, up to
 operators with four derivatives. In Sec.~\ref{sec:Linear} the
 analogous complete basis of independent operators up to dimension six
 in the linear expansion is presented. Comparing both sets of operators, we have established the relations and differences between the chiral and the linear bases.

In particular, in Secs.~\ref{decorr} and \ref{newsignals} we have
identified two sources of discriminating signatures. For small values
of the $\xi$ parameter the counting of operators is not the same in
both sets, being larger by six for the chiral expansion. This implies
that, even keeping only operators weighted by $\xi$, the expected
deviations from the SM predictions in the Higgs couplings to gauge
bosons and that of the triple gauge boson self-couplings are
independent in the chiral expansion, unlike in the linear expansion at
dimension six; one interesting set of (de)correlated couplings is explored in details as indicators of a non-linear character.
Conversely, when considering operators weighted by $\xi^n$ with $n\geq
2$ in the chiral expansion, we find anomalous signals which appear
only at dimension eight of the linear Lagrangian; they may thus be
detected with larger (leading) strength for a non-linear realization
of EWSB than for a linear one, for sizeable values of $\xi$. 

In order to quantify the observability of the above effects we have
implemented the renormalization procedure as described in
Sec.~\ref{sec:renormalization} and derived the corresponding Feynman
rules for the non-linear expansion (which we present in the detail in
Appendix~\ref{AppFR}, for the complete set of independent operators
under discussion).  Neglecting external fermion masses only in the
numerical analysis, the results of our simulations for some of the
discriminating signatures at LHC are presented in
Secs.~\ref{numerical}--\ref{sec:numquartic}. To our knowledge, this is
the first six-parameter analysis in the context of the non-linear
expansion, focusing on the $\xi$-weighted pure gauge and gauge-$h$
effective couplings. In particular we have derived the present bounds
on the coefficients of the latter from the analysis of electroweak
precision physics, triple gauge boson coupling studies and Higgs data.
The results are summarized in Fig.~\ref{fig:1dima_6bos} and
Table~\ref{tab:bounds} and the corresponding level of decorrelation
between the triple gauge couplings and Higgs effects is illustrated in
Fig.~\ref{fig:nl1}: the presently allowed values for the parameters
$c_i\xi$ and $a_i\xi$ turn out to be of order 1, with only few
exceptions bounded to the per cent level. With the expected
uncertainties attainable in CMS and ATLAS at 14 TeV, that sensitivity
can be improved by a factor $\cO(3-5)$. Furthermore, our study of the
present sensitivity to the C and P odd operator in the analysis
of $WWZ$ vertex, with the accumulated luminosity of LHC7+8 and with
LHC14 in the future, show that per cent precision on the coupling of
the operator $\cP_{14}(h)$ is foreseeable. Similar precision should be
attainable for the coefficients of the operators leading to generic
quartic gauge couplings $\cP_{6}(h)$ and $\cP_{11}(h)$.

%%%%%%%%%%%%%%%%%%%%%%%%%%%%%%%%%%%%%%%%%%%%%%%%%%%%%%%%%%%%
% Acknowledgements
%%%%%%%%%%%%%%%%%%%%%%%%%%%%%%%%%%%%%%%%%%%%%%%%%%%%%%%%%%%%
\section*{Acknowledgements}
We acknowledge illuminating conversations with Rodrigo Alonso, Gino
Isidori, Aneesh Manohar, Michael Trott, and Juan Yepes.  We also
acknowledge partial support of the European Union network FP7 ITN
INVISIBLES (Marie Curie Actions, PITN-GA-2011-289442), of CiCYT
through the project FPA2009-09017, of CAM through the project HEPHACOS
P-ESP-00346, of the European Union FP7 ITN UNILHC (Marie Curie
Actions, PITN-GA-2009-237920), of MICINN through the grant
BES-2010-037869, of the Spanish MINECO’s “Centro de Excelencia Severo
Ochoa” Programme under grant SEV-2012-0249, and of the Italian
Ministero dell'Uni\-ver\-si\-t\`a e della Ricerca Scientifica through
the COFIN program (PRIN 2008) and the contract MRTN-CT-2006-035505.
The work of I.B. is supported by an ESR contract of the European Union
network FP7 ITN INVISIBLES mentioned above. The work of L.M. is
supported by the Juan de la Cierva programme (JCI-2011-09244). The
work of O.J.P.E. is supported in part by Conselho Nacional de
Desenvolvimento Cient\'{\i}fico e Tecnol\'ogico (CNPq) and by
Funda\c{c}\~ao de Amparo \`a Pesquisa do Estado de S\~ao Paulo
(FAPESP), M.C.G-G and T.C are supported by USA-NSF grant PHY-09-6739,
M.C.G-G is also supported by CUR Generalitat de Catalunya grant
2009SGR502 and together with J.G-F by MICINN FPA2010-20807 and
consolider-ingenio 2010 program CSD-2008-0037. J.G-F is further
supported by ME FPU grant AP2009-2546. I.B., J.G-F., M.C.G-G., B.G.,
L.M, and S.R. acknowledge CERN TH department and J.G-F. also
acknowledges ITP Heidelberg for hospitality during part of this work.

%%%%%%%%%%%%%%%%%%%%%%%%%%%%%%%%%%%%%%%%%%%%%%%%%%%%%%%%%%%%
% Appendix
%%%%%%%%%%%%%%%%%%%%%%%%%%%%%%%%%%%%%%%%%%%%%%%%%%%%%%%%%%%%
\appendix \small

%
%%%%%%%%%%%%%%%%%%%%%%%%%   Appendix A       %%%%%%%%%%%%%%%%%%%%%%%%
%
\section{EOM and fermion operators}
\label{AppEOM}

The EOM can be extracted from the $\LL_0$ part of the chiral
Lagrangian, Eq.~(\ref{LLO}); as we will work at first order in
$\Delta\LL$ they read\footnote{With alternative choices for the
  separation $\LL_0$ versus $\Delta\LL$ the EOM are correspondingly
  modified~\cite{Alonso:2012px,Alonso:2012pz,Buchalla:2013rka}: this
  is of no relevance to the focus of this paper, which explores the
  tree-level impact of effective operators.}:
\begin{align}
&(\DLL^\mu\WWd)^a =  \dfrac{g}{2}\bar{Q}_L\s^a \g_\nu Q_L 
+ \dfrac{g}{2}\bar{L}_L\s^a \g_\nu L_L+
\dfrac{igv^2}{4}\tr[\VL_\nu \s^a]\left(1+\dfrac{h}{v}\right)^2 \label{EOMW}\\
&\derp^\mu \BBd =  -\dfrac{ig'v^2}{4}\tr[\TL\VL_\mu] 
\left(1+\dfrac{h}{v}\right)^2+
g'\sum_{i=L,R}\left( \bar{Q}_i\bold{h}_i\g_\nu Q_i
+\dfrac{1}{6}\bar{L}_L\g_\nu L_L\right) \label{EOMB}\\
&\square h = -\dfrac{\delta V(h)}{\delta h}-\dfrac{v+h}{2}\tr[\VL_\mu\VL^\mu]-
\dfrac{s_Y}{\sqrt2}\left(\bar{Q}_L \UH \cY_Q Q_{R}+\bar{L}_L \UH \cY_L L_{R}
+\hc\right) \label{EOMh}\\
&\Big[ \DLR_\mu\Big(\dfrac{(v+h)^2}{2\sqrt2} 
\UH^\dag \DLR^\mu \UH \Big)\Big]_{ij}= 
\begin{cases}
-\left(v+s_Y h\right)\left[(\bar{Q}_R\cY_Q^\dag)_j (\UH^\dag Q_{L})_i  
+(\bar{L}_R\cY_L^\dag)_j (\UH^\dag L_{L})_i\right]\label{EOMU}\\
\hspace{6.5cm}\text{for}\quad i\neq j\\
0\hspace{6.3cm}\text{for}\quad i=j
\end{cases}\\
&i\slashed{D} Q_L= \dfrac{v+s_Y h}{\sqrt{2}} \UH \cY_Q  Q_R \qquad\qquad\qquad 
 i\slashed{D} Q_R = \dfrac{ v+s_Y h}{\sqrt{2}} \cY_Q^\dag \UH^\dag Q_L \label{EOMQ}\\
&i\slashed{D} L_L = \dfrac{v+s_Y h}{\sqrt{2}} \UH \cY_L  L_R \qquad\qquad\qquad
 i\slashed{D} L_R = \dfrac{ v+s_Y h}{\sqrt{2}} \cY_L^\dag \UH^\dag L_L\,,
\label{EOML}
\end{align}
where $\bold{h}_{L,R}$ are the $2\times2$ matrices of hypercharge for the left- and right-handed quarks. 

By using these EOM, it is possible to identify relations between some
bosonic operators listed in Eqs.(\ref{Opxi})-(\ref{Opxi4}) and
specific fermion operators. This allows us to trade those bosonic
operators by the corresponding fermionic ones: this procedure can turn
out to be very useful when analysing specific experimental data. For
instance, if deviations from the SM values of the $h$-fermion
couplings were found, then the following three operators,
\beq
\begin{aligned}
&\cP_{U,\a\b}(h)=-\frac{v}{\sqrt{2}}\bar{Q}_{L\a}\UH
\left(\cF_U(h)\Pu Q_{R}\right)_\b  +\hc\,,\\
&\cP_{D,\a\b}(h)=-\frac{v}{\sqrt{2}}\bar{Q}_{L\a}\UH
\left(\cF_D(h)\Pd Q_{R}\right)_\b  +\hc\,,\\
&\cP_{E,\a\b}(h)=-\frac{v}{\sqrt{2}}\bar{L}_{L\a}\UH
\left(\cF_E(h)\Pd L_{R}\right)_\b +\hc\,,
\label{Opfer}
\end{aligned}
\eeq
would be a good choice for an operator basis. In the previous equations 
the two projectors 
\begin{equation}
\Pu=\begin{pmatrix}1& \\ & 0 \end{pmatrix}
\hspace*{3cm}
\Pd=\begin{pmatrix}0& \\ & 1 \end{pmatrix},
\end{equation} 
have been introduced.

On the contrary, without including the operators in
Eqs.~(\ref{Opfer}), the bosonic basis defined in Eqs.~(\ref{Opxi})-
(\ref{Opxih}) is blind to these directions.  The fermionic operators
that arise applying the EOM to bosonic operators in the basis above is
presented in the following list:
\begin{description}
\item[Weighted by $\xi$:]
\beq
\begin{aligned}
\cP_{U,\a\b}(h) &=-\frac{v}{\sqrt{2}}\bar{Q}_{L\a}\UH
\left(\cF_U(h)\Pu Q_{R}\right)_\b  +\hc\\
\cP_{D,\a\b}(h) &=-\frac{v}{\sqrt{2}}\bar{Q}_{L\a}\UH
\left(\cF_D(h)\Pd Q_{R}\right)_\b  +\hc\\
\cP_{E,\a\b}(h) &=-\frac{v}{\sqrt{2}}\bar{L}_{L\a}\UH
\left(\cF_E(h)\Pd L_{R}\right)_\b +\hc\\
\cP_{1Q,\a\b}(h) &=\frac{\a}{2}\bar{Q}_{L\a} \g^\mu
\{\TL,\VL_\mu\}\left(\cF_{1Q}(h) Q_L\right)_\b \\
\cP_{1L,\a\b}(h) &=\frac{\a}{2}\bar{L}_{L\a} \g^\mu
\{\TL,\VL_\mu\}\left(\cF_{1L}(h)L_L\right)_\b \\
\cP_{1U,\a\b}(h) &=\frac{\a}{2}\bar{Q}_{R\a}\g^\mu
\left\{\s^3,\tilde{\VL}_\mu\right\} \left(\cF_{1U}(h)\Pu Q_R\right)_\b  \\ 
\cP_{1D,\a\b}(h) &=\frac{\a}{2}\bar{Q}_{R\a}\g^\mu
\left\{\s^3,\tilde{\VL}_\mu\right\} \left(\cF_{1D}(h)\Pd Q_R\right)_\b  \\ 
\cP_{1N,\a\b}(h) &=\frac{\a}{2}\bar{L}_{R\a}\g^\mu
\left\{\s^3,\tilde{\VL}_\mu\right\} \left(\cF_{1N}(h)\Pu L_R\right)_\b  \\ 
\cP_{1E,\a\b}(h) &=\frac{\a}{2}\bar{L}_{R\a}\g^\mu
\left\{\s^3,\tilde{\VL}_\mu\right\} \left(\cF_{1E}(h)\Pd L_R\right)_\b  \\
\cP_{2Q,\a\b}(h) &= i \bar{Q}_{L\a} \g^\mu\VL_\mu
\left(\cF_{2Q}(h)  Q_L\right)_\b \\
\cP_{2L,\a\b}(h) &= i \bar{L}_{L\a} \g^\mu\VL_\mu
\left(\cF_{2L}(h)  L_L\right)_\b \\
\cP_{3Q,\a\b}(h) &= i \bar{Q}_{L\a} \g^\mu\TL\VL_\mu
\TL \left(\cF_{3Q}(h)  Q_L\right)_\b \\
\cP_{3L,\a\b}(h) &= i \bar{L}_{L\a} \g^\mu\TL\VL_\mu
\TL \left(\cF_{3L}(h)  L_L\right)_\b \\
\end{aligned}
\label{OpFermxi}
\eeq
\beq
\begin{aligned}
\cP_{4UU,\a\b\g\d}(h) &=\sum_a \Big[\bar{Q}_{L\a}\s^a \UH 
\left(\cF_{4U}(h)\Pu Q_R\right)_\b-\hc\Big]\Big[\bar{Q}_{L\g}\s^a \UH 
\left(\cF'_{4U}(h) \Pu Q_R\right)_\d-\hc\Big]\\
\cP_{4DD,\a\b\g\d}(h) &=\sum_a\Big[\bar{Q}_{L\a}\s^a \UH 
\left(\cF_{4D}(h)\Pd Q_R\right)_\b-\hc\Big]\Big[\bar{Q}_{L\g}\s^a \UH 
\left(\cF'_{4D}(h) \Pd Q_R\right)_\d-\hc\Big]\\
\cP_{4UD,\a\b\g\d}(h) &=\sum_a\Big[\bar{Q}_{L\a}\s^a \UH 
\left(\cF_{4U}(h)\Pu Q_R\right)_\b-\hc\Big]\Big[\bar{Q}_{L\g}\s^a \UH 
\left(\cF'_{4D}(h) \Pd Q_R\right)_\d-\hc\Big]\\
\cP_{4DU,\a\b\g\d}(h) &=\sum_a\Big[\bar{Q}_{L\a}\s^a \UH 
\left(\cF_{4D}(h)\Pd Q_R\right)_\b-\hc\Big]\Big[\bar{Q}_{L\g}\s^a \UH 
\left(\cF'_{4U}(h) \Pu Q_R\right)_\d-\hc\Big]\\
\cP_{4EE,\a\b\g\d}(h) &=\sum_a\Big[\bar{L}_{L\a}\s^a \UH 
\left(\cF_{4E}(h)\Pd L_R\right)_\b-\hc\Big]\Big[\bar{L}_{L\g}\s^a \UH 
\left(\cF'_{4E}(h) \Pd L_R\right)_\d-\hc\Big]\\
\cP_{4UE,\a\b\g\d}(h) &=\sum_a\Big[\bar{Q}_{L\a}\s^a \UH 
\left(\cF_{4U}(h)\Pu Q_R\right)_\b-\hc\Big]\Big[\bar{L}_{L\g}\s^a \UH 
\left(\cF'_{4E}(h) \Pd L_R\right)_\d-\hc\Big]\\
\cP_{4DE,\a\b\g\d}(h) &=\sum_a\Big[\bar{Q}_{L\a}\s^a \UH 
\left(\cF_{4D}(h)\Pd Q_R\right)_\b-\hc\Big]\Big[\bar{L}_{L\g}\s^a \UH 
\left(\cF'_{4E}(h) \Pd L_R\right)_\d-\hc\Big]\\
\cP_{4EU,\a\b\g\d}(h) &=\sum_a\Big[\bar{L}_{L\a}\s^a \UH 
\left(\cF_{4E}(h)\Pd L_R\right)_\b-\hc\Big]\Big[\bar{Q}_{L\g}\s^a \UH 
\left(\cF'_{4U}(h) \Pu Q_R\right)_\d-\hc\Big]\\
\cP_{4ED,\a\b\g\d}(h) &=\sum_a\Big[\bar{L}_{L\a}\s^a \UH 
\left(\cF_{4E}(h)\Pd L_R\right)_\b-\hc\Big]\Big[\bar{Q}_{L\g}\s^a \UH 
\left(\cF'_{4D}(h) \Pd Q_R\right)_\d-\hc\Big]\\
\cP_{5UU,\a\b\g\d}(h) &=\Big[\bar{Q}_{L\a}\TL \UH 
\left(\cF_{5U}(h) \Pu Q_R\right)_\b-\hc\Big]\Big[\bar{Q}_{L\g}\TL\UH 
\left(\cF'_{5U}(h) \Pu Q_R\right)_\d-\hc\Big]\\
\cP_{5DD,\a\b\g\d}(h) &=\Big[\bar{Q}_{L\a}\TL \UH 
\left(\cF_{5D}(h) \Pd Q_R\right)_\b-\hc\Big]\Big[\bar{Q}_{L\g}\TL\UH 
\left(\cF'_{5D}(h) \Pd Q_R\right)_\d-\hc\Big]\\
\cP_{5UD,\a\b\g\d}(h) &=\Big[\bar{Q}_{L\a}\TL \UH 
\left(\cF_{5U}(h) \Pu Q_R\right)_\b-\hc\Big]\Big[\bar{Q}_{L\g}\TL\UH 
\left(\cF'_{5D}(h) \Pd Q_R\right)_\d-\hc\Big]\\
\cP_{5DU,\a\b\g\d}(h) &=\Big[\bar{Q}_{L\a}\TL \UH 
\left(\cF_{5D}(h) \Pd Q_R\right)_\b-\hc\Big]\Big[\bar{Q}_{L\g}\TL\UH 
\left(\cF'_{5U}(h) \Pu Q_R\right)_\d-\hc\Big]\\
\cP_{5EE,\a\b\g\d}(h) &=\Big[\bar{L}_{L\a}\TL \UH 
\left(\cF_{5E}(h) \Pd L_R\right)_\b-\hc\Big]\Big[\bar{L}_{L\g}\TL\UH 
\left(\cF'_{5E}(h) \Pd L_R\right)_\d-\hc\Big]\\
\cP_{5UE,\a\b\g\d}(h) &=\Big[\bar{Q}_{L\a}\TL \UH 
\left(\cF_{5U}(h) \Pu Q_R\right)_\b-\hc\Big]\Big[\bar{L}_{L\g}\TL\UH 
\left(\cF'_{5E}(h) \Pd L_R\right)_\d-\hc\Big]\\
\cP_{5EU,\a\b\g\d}(h) &=\Big[\bar{L}_{L\a} \TL\UH 
\left(\cF_{5E}(h) \Pd L_R\right)_\d-\hc\Big]\Big[\bar{Q}_{L\g}\TL\UH 
\left(\cF'_{5U}(h) \Pu Q_R\right)_\d-\hc\Big]\\
\cP_{5DE,\a\b\g\d}(h) &=\Big[\bar{Q}_{L\a}\TL \UH 
\left(\cF_{5D}(h) \Pd Q_R\right)_\b-\hc\Big]\Big[\bar{L}_{L\g}\TL\UH 
\left(\cF'_{5E}(h) \Pd L_R\right)_\d-\hc\Big]\\
\cP_{5ED,\a\b\g\d}(h) &=\Big[\bar{L}_{L\a} \TL\UH 
\left(\cF_{5E}(h) \Pd L_R\right)_\d-\hc\Big]\Big[\bar{Q}_{L\g}\TL\UH 
\left(\cF'_{5D}(h) \Pd Q_R\right)_\d-\hc\Big]\,.
\end{aligned} \nn
\eeq
\item[Weighted by $\xi\sqrt\xi$:]
\beq
\begin{aligned}
\cP_{6U,\a\b}(h) &=\bar{Q}_{L\a}\VL_\mu \UH 
\left(\derp^\mu\cF_{6U}(h) \Pu Q_R\right)_\b\\
\cP_{6D,\a\b}(h) &=\bar{Q}_{L\a}\VL_\mu \UH 
\left(\derp^\mu\cF_{6D}(h) \Pd Q_R\right)_\b\\
\cP_{6N,\a\b}(h) &=\bar{L}_{L\a}\VL_\mu \UH 
\left(\derp^\mu\cF_{6N}(h) \Pu L_R\right)_\b\\
\cP_{6E,\a\b}(h) &=\bar{L}_{L\a}\VL_\mu \UH 
\left(\derp^\mu\cF_{6E}(h) \Pd L_R\right)_\b\\
\cP_{7U,\a\b}(h) &=\tr[\TL\VL^\mu]\bar{Q}_{L\a}\TL\UH 
\left(\derp^\mu\cF_{7U}(h) \Pu Q_R\right)_\b\\
\cP_{7D,\a\b}(h) &=\tr[\TL\VL^\mu]\bar{Q}_{L\a}\TL\UH 
\left(\derp^\mu\cF_{7D}(h) \Pd Q_R\right)_\b\\
\cP_{7N,\a\b}(h) &=\tr[\TL\VL^\mu]\bar{L}_{L\a}\TL\UH 
\left(\derp^\mu\cF_{7N}(h) \Pu L_R\right)_\b\\
\cP_{7E,\a\b}(h) &=\tr[\TL\VL^\mu]\bar{L}_{L\a}\TL\UH 
\left(\derp^\mu\cF_{7E}(h) \Pd L_R\right)_\b\\
\cP_{8U,\a\b}(h) &=\tr[\TL\VL^\mu] \bar{Q}_{L\a}[\TL,\VL_\mu]\UH
\left(\cF_{8U}(h) \Pu Q_R\right)_\b \\
\cP_{8D,\a\b}(h) &=\tr[\TL\VL^\mu] \bar{Q}_{L\a}[\TL,\VL_\mu]\UH
\left(\cF_{8D}(h) \Pd Q_R\right)_\b \\
\cP_{8N,\a\b}(h) &=\tr[\TL\VL^\mu] \bar{L}_{L\a}[\TL,\VL_\mu]\UH
\left(\cF_{8N}(h) \Pu L_R\right)_\b \\
\cP_{8E,\a\b}(h) &=\tr[\TL\VL^\mu] \bar{L}_{L\a}[\TL,\VL_\mu]\UH
\left(\cF_{8E}(h) \Pd L_R\right)_\b\,.
\end{aligned}
\label{OpFermxisqrtxi}
\eeq
\end{description}

Rearranging Eqs.~(\ref{EOMW})-(\ref{EOMU}), one can derive the following relations between bosonic and fermionic operators: 
\beq
\begin{aligned}
&2\cP_B(h)+\dfrac{1}{2}\cP_1(h)+\dfrac{1}{2}\cP_2(h)+\cP_4(h) 
- g'^2 \cP_T(h)\left(1+\dfrac{h}{v}\right)^2 = 
\sum_\a\Big\{ \dfrac{1}{3}g'^2\cP_{1Q,\a\a}(h)
+\dfrac{4}{3}g'^2\cP_{1U,\a\a}(h)\\
&\hspace{8cm}-\dfrac{2}{3}g'^2\cP_{1D,\a\a}(h)
-g'^2\cP_{1L,\a\a}(h)-2g'^2\cP_{1E,\a\a}(h)\Big\}\,,\\
&-\cP_W(h)-g^2\cP_C(h)\left(1+\dfrac{h}{v}\right)^2
-\dfrac{1}{4}\cP_1(h)-\dfrac{1}{2}\cP_3(h)+\cP_5(h) = 
  \dfrac{g^2}{2}\sum_\a\Big\{\cP_{2Q,\a\a}(h)+\cP_{2L,\a\a}(h)\Big\}\,,\\
&\cP_H(h)+2 \cP_C(h)\left(1+\dfrac{h}{v}\right)^2 
+(v+h)\cF(h)\dfrac{\delta V}{\delta h}=
 s_Y\dfrac{v+h}{\sqrt2}\sum_{f=U,D,E}
\sum_{\a\b}\Big\{Y_{f,\a\b} \cP_{f,\a\b}(h)+\hc\Big\}\,,\\
&g^2\cP_T(h)\left(1+\dfrac{h}{v}\right)^2
-\dfrac{1}{2}\cP_1(h)-\cP_3(h)+\dfrac{1}{2}\cP_{12}(h)+\cP_{13}(h)+\cP_{17}(h) = \\
&\hspace{4cm} = \dfrac{g^2}{2} 
\sum_\a\Big\{\left(\cP_{3Q,\a\a}(h)+\cP_{2Q,\a\a}(h)\right)+
 \left(\cP_{3L,\a\a}(h)+\cP_{2L,\a\a}(h)\right)\Big\}\,.
\end{aligned}
\label{RelH}
\eeq
The $\cF_i(h)$ functions in all operators in these relations are the same, 
except for $\cP_H$ in the third line of Eq.~(\ref{RelH}), which is related 
to it by 
\beq
\cF_H(h) = \cF_C(h) + \left(1+\dfrac{h}{v}\right)\dfrac{\d\cF_C(h)}{\d h}\,.
\eeq 

Applying the EOM in Eq.~(\ref{EOMh}) to the operators $\cP_{20}(h)$
and $\cP_{21}(h)$ allows us to express them in terms of other operators
in the basis, $h$-gauge boson couplings and Yukawa-like interactions:
\beq
\begin{aligned}
\cP_{20}(h)=&2 \cF(h)\cP_{6}(h)+2\cF(h)\cP_{7}(h)
-\dfrac{16}{v^3}\sqrt{\cF(h)}\cP_C(h)\dfrac{\d V}{\d h}\\
&-\dfrac{8\sqrt2s_y}{v^3}\sqrt{\cF(h)}\cP_C(h)\left(\bar{Q}_L \UH 
\cY_Q Q_{R}+\bar{L}_L \UH \cY_L L_{R}+\hc\right)\,,\\
\cP_{21}(h)=&2 \cF(h)\cP_{23}(h)+2\cF(h)\cP_{25}(h)
+\dfrac{16}{v^3}\sqrt{\cF(h)}\cP_T(h)\dfrac{\d V}{\d h}\\
&+\dfrac{8\sqrt2s_y}{v^3}\sqrt{\cF(h)}\cP_T(h)\left(\bar{Q}_L \UH 
\cY_Q Q_{R}+\bar{L}_L \UH \cY_L L_{R}+\hc\right)\,,
\end{aligned}
\eeq
where all $\cF_i(h)$ appearing explicitly in these expressions and
included in the definition of the operators $\cP_i(h)$ are the same
and defined by
\beq
\cF(h)=\left(1+\dfrac{h}{v}\right)^2\,.
\eeq
From Eqs.~(\ref{EOMW}), (\ref{EOMB}) and (\ref{EOMQ}), it follows that  
\beq
\begin{aligned}
\frac{iv}{\sqrt{2}}\tr(\s^j\DLL_\mu\VL^\mu)\left(1+\dfrac{h}{v}\right)^2 =
&\dfrac{v+s_Y h}{v}\left(i\bar{Q}_L\s^j\UH\cY_Q Q_R +i\bar{L}_L\s^j\UH
\cY_L L_R+\hc\right)\\
&-\dfrac{iv}{\sqrt2}\tr(\s^j\VL_\mu)\derp^\mu\left(1+\dfrac{h}{v}\right)^2\,,\\
 \frac{iv}{\sqrt{2}}\tr(\TL\DLL_\mu\VL^\mu)\left(1+\dfrac{h}{v}\right)^2 =
&\dfrac{v+s_Y h}{v} \left(i\bar{Q}_L\TL\UH\cY_Q Q_R+i\bar{L}_L\TL\UH
\cY_L L_R+\hc\right)\\
 &-\dfrac{iv}{\sqrt2}\tr(\TL\VL_\mu)\derp^\mu\left(1+\dfrac{h}{v}\right)^2\,,
\end{aligned}
\eeq
which allows us to rewrite the pure bosonic operators $\cP_{11-13}(h)$,
$\cP_{10}(h)$ and $\cP_{19}(h)$ as combination of other pure bosonic
ones in Eqs.~(\ref{Opxi})-(\ref{Opxi4}) plus fermionic operators in
Eqs.~(\ref{OpFermxi}) and (\ref{OpFermxisqrtxi}):
\beq
\begin{aligned}
\cP_{9}(h) -\cP_{8}(h)=& \dfrac{1}{v^2}\sum_{f_1,f_2=U,D,E}\sum_{\a\b\g\d} 
Y_{f_1,\a\b}Y_{f_2,\g\d}\cP_{4f_1f_2,\a\b\g\d}(h)\\
&-\dfrac{2\sqrt2}{v}\sum_{f=U,D,N,E}\sum_{\a\b}
(Y_{f,\a\b}\cP_{6f,\a\b}(h)-\hc)\,,\\
\cP_{15}(h) -\cP_{22}(h)=& \dfrac{2}{v^2}\sum_{f_1,f_2=U,D,E}\sum_{\a\b\g\d} 
Y_{f_1,\a\b\g\d}Y_{f_2,\g\d}\cP_{5f_1f_2,\a\b\g\d}(h)\\
&-\dfrac{2\sqrt2}{v\cos\theta_W}\sum_{f=U,D,N,E}
\sum_{\a\b}(Y_{f,\a\b}\cP_{7f,\a\b}(h)-\hc)\,,\\
\cP_{16}(h)+\cP_{18}(h) =& \sum_{f=U,D,N,E}\sum_{\a\b} 
\frac{\sqrt2}{v}\left(Y_{f,\a\b}\cP_{8f,\a\b}(h)-\hc\right)\,,\\
\cP_{10}(h)+\cP_{8}(h) =& \sum_{f=U,D,N,E}\sum_{\a\b} 
\frac{\sqrt2}{v}\left(Y_{f,\a\b}\cP_{6f,\a\b}(h)-\hc\right)\,,\\
\cP_{19}(h)+\cP_{22}(h) =& \sum_{f=U,D,N,E}\sum_{\a\b} 
\frac{\sqrt2}{v}\left(Y_{f,\a\b}\cP_{7f,\a\b}(h)-\hc\right)\,.
\end{aligned}
\label{PDVfermions}
\eeq
A straightforward consequence is that once the $\cF_i(h)$ functions in
the operators on the left-hand side of Eq.~(\ref{PDVfermions}) are
specified, then the $\cF_i(h)$ functions in the operators on the
right-hand side are no longer general, but take the form of specific
expressions.

%
%%%%%%%%%%%%%%%%%%%%%%%%%%   Appendix B       %%%%%%%%%%%%%%%%%%%%%%%%
%
\boldmath
\section{Equivalence of the $d=6$ basis with the SILH Lagrangian}
\label{AppSILH}
\unboldmath

The SILH Lagrangian~\cite{Giudice:2007fh}  is defined by the following 
10 $d=6$ linear operators:
\begin{align}
&
\cO^\text{SILH}_{g}= \Phi^\dagger \Phi \, {G}^{a}_{\mu\nu} {G}^{a\mu\nu}\,,
&&
\cO^\text{SILH}_{\gamma}=\Phi^{\dagger}\hat{B}_{\mu \nu}\hat{B}^{\mu \nu}\Phi\,
,\nn\\
&\cO^\text{SILH}_{B} = \left(\Phi^{\dagger}\Dfb_\mu\Phi\right) 
\derp_\nu \hat{B}^{\mu \nu}\,,
&&\cO^\text{SILH}_{W} = \dfrac{ig}{2}\left(\Phi^{\dagger}\s^i
\Dfb_\mu\Phi\right) \DL_\nu W_i^{\mu \nu}\,,\nn\\
&\cO^\text{SILH}_{HB}  = \left(\DL_{\mu} \Phi\right)^{\dagger} 
(\DL_{\nu} \Phi) \hat{B}^{\mu \nu}\,,
&&\cO^\text{SILH}_{HW}  = (\DL_{\mu} \Phi)^{\dagger} \hat{W}^{\mu \nu}  
(\DL_{\nu} \Phi)\,,
\label{LinearOpsGaugeSILH}  \\
&\cO^\text{SILH}_{T} = \dfrac{1}{2} \left (\Phi^\dag\Dfb_\mu \Phi \right)
\left(\Phi^\dagger\Dfb^\mu\Phi\right)\,,
&&\cO^\text{SILH}_{H} = \frac{1}{2} \partial^\mu
\left ( \Phi^\dagger \Phi \right)\partial_\mu\left 
( \Phi^\dagger \Phi \right)\,,\nn\\
&\cO^\text{SILH}_{6} = \dfrac{1}{3}\left(\Phi^\dag\Phi\right)^3\,,
&&\cO^\text{SILH}_{y} =  \left (\Phi^\dag \Phi \right)f_L\Phi 
\cY f_R+\hc\,,\nn 
\end{align}
where
$\Phi^\dag\Dfb_\mu\Phi\equiv\Phi^\dag\DL_\mu\Phi-\DL_\mu\Phi^\dag\Phi$
and $\Phi^{\dagger}\s^i\Dfb_\mu\Phi\equiv \Phi^\dag\s^i\DL_\mu\Phi
-\DL_\mu\Phi^\dag\s^i\Phi$. They can be related directly to the
operators in Eqs.~(\ref{LinearOpsGauge}) and
(\ref{LinearOpsGaugePhi3}):
\beq
\begin{aligned}
\cO^\text{SILH}_{g} & \equiv\,\, \cO_{GG}\,,  \qquad 
&\cO^\text{SILH}_{\gamma} &\equiv\,\, \cO_{BB}\,,\\
\cO^\text{SILH}_{B} &\equiv\,\, 2\cO_{B}+\cO_{BW}+\cO_{BB}\,,\qquad
&\cO^\text{SILH}_{W} &\equiv\,\, 2\cO_{W}+\cO_{BW}+\cO_{WW}\,,\\
\cO^\text{SILH}_{HW}  &\equiv\,\, \cO_{W}\,,\qquad
&\cO^\text{SILH}_{HB}  &\equiv\,\, \cO_{B}\,, \\
\cO^\text{SILH}_{T} &\equiv\,\, \cO_{\Phi,2}-2\cO_{\Phi,1}\,,\qquad
&\cO^\text{SILH}_{H} &\equiv\,\, \cO_{\Phi,2}\,, \\
\cO^\text{SILH}_{6} &\equiv\,\,\cO_{\Phi,3}\,, \qquad
&\cO^\text{SILH}_{y} &\equiv\,\, 2\cO_{\Phi,2} + 2\cO_{\Phi,4}-\left(\Phi^\dag\Phi\right) \Phi^\dag\dfrac{\d V(h)}{\d\Phi^\dag}\,.
\end{aligned}
\eeq
This shows the equivalence of the two linear expansions.

It can also be interesting to show explicitly the connection between
the SILH operators and those of the chiral basis in
Eqs.~(\ref{Opxi})-(\ref{Opxi4}), which is as follows:

\begin{align}
&\begin{aligned}
\cO^\text{SILH}_{g} &
=\,\, \dfrac{v^2}{2g_s^2}\cP_{G}(h)\,,  \qquad
&\cO^\text{SILH}_{\gamma} &
=\,\, \dfrac{v^2}{2}\cP_{B}(h)\,,\\
\cO^\text{SILH}_{B} &
=\,\,\dfrac{v^2}{8}(\cP_{2}(h)+2\cP_4(h))+\dfrac{v^2}{8}\cP_{1}(h)
+\dfrac{v^2}{2}\cP_{B}(h)\,,\qquad
&\cO^\text{SILH}_{HB}  &
=\,\, \dfrac{v^2}{16}(\cP_{2}(h)+2\cP_4(h))\,,\\
\cO^\text{SILH}_{W} &
=\,\, \dfrac{v^2}{4}(\cP_{3}(h)-2\cP_5(h))+\dfrac{v^2}{8}\cP_{1}(h)
+\dfrac{v^2}{2}\cP_{W}(h)\,,\qquad
&\cO^\text{SILH}_{HW}  &
=\,\,\dfrac{v^2}{8} (\cP_{3}(h)-2\cP_5(h))\,, \\
\cO^\text{SILH}_{T} &
=\,\,\dfrac{v^2}{2} \cF(h)\cP_{T}(h)\,,\qquad
&\cO^\text{SILH}_{H} &
=\,\, v^2\cP_{H}(h)\,, \\
\end{aligned}\\[-5mm]
&\:\cO^\text{SILH}_{y} 
=\,\, 3v^2\cP_{H}(h) + v^2\cF(h)\cP_{C}(h)-\dfrac{(v+h)^3}{2}
\dfrac{\d V(h)}{\d h}\,,\nn
\end{align}
where the $\cF_i(h)$ appearing in these relations and inside the
individual $\cP_i(h)$ operators are all defined by
\beq
\cF(h)=\left(1+\dfrac{h}{v}\right)^2\,.
\eeq

%
%%%%%%%%%%%%%%%%%%%%%%%%%   Appendix C       %%%%%%%%%%%%%%%%%%%%%%%%
%
\section{Relations between chiral and linear operators}
\label{AppSiblings}

In this Appendix, the connections between the operators of the chiral and linear bases is discussed. As the number and nature of the leading order operators in the chiral and linear expansion are not
the same, there are pairs of chiral operators that correspond to the same
lowest dimensional linear one: in order to get then a one-to-one correspondence between these chiral operators and (combinations of) linear ones, operators of higher dimension should be
taken into consideration.  For those weighted by a single
power of $\xi$, the list of the siblings can be read from Eq.~(\ref{LinearChiralCorrelations}). Below, we also indicate which chiral operators, weighted by higher powers of $\xi$, should be combined in order to generate the gauge interactions contained in specific linear ones.
\begin{description}
\item[For operators weighted by \boldmath$\xi$:]
\begin{align}
&\begin{aligned}
\cP_B(h)&\to\cO_{BB}\qquad
&\cP_W(h)&\to\cO_{WW}\qquad
&\cP_G(h)&\to\cO_{GG} \\
\cP_C(h)&\to\cO_{\Phi,4}\qquad
&\cP_T(h)&\to\cO_{\Phi,1}\qquad
&\cP_H(h)&\to\cO_{\Phi,2}\\
\cP_1(h)&\to\cO_{BW}\qquad
&\cP_2(h)\,,\cP_4(h)&\to\cO_{B}\qquad\qquad
&\cP_3(h)\,,\cP_5(h)&\to\cO_{W}
\end{aligned}\\
&\hspace{1.5cm}\cP_{6}(h)\,,\cP_{7}(h)\,,\cP_{8}(h)\,,\cP_{9}(h)\,,\cP_{10}(h)\,,\cP_{\Box H}(h)\to\cO_{\Box\Phi}
\nn
\end{align}

\item[For operators weighted by \boldmath$\xi^2$:]
\beq
\hspace{-0.5cm}
\begin{aligned}
\cP_{DH}(h)\,,\cP_{20}(h)&\to\left[\DL_\mu\Phi^\dag \DL^\mu \Phi\right]^2\\
\cP_{11}(h)\,,\cP_{18}(h)\,,\cP_{21}(h)\,,\cP_{22}(h)\,,\cP_{23}(h)\,,\cP_{24}(h)&\to
\left[\DL^\mu\Phi^\dag \DL^\nu \Phi\right]^2\\
\cP_{12}(h)&\to \left( \Phi^\dag \WWu \Phi\right)^2 \\
\cP_{13}(h)\,,\cP_{17}(h)&\to \left( \Phi^\dag \WWu \Phi\right) 
\DL_\mu\Phi^\dag \DL_\nu \Phi \\
\cP_{14}(h)&\to\e^{\mu\nu\rho\lambda}\left(\Phi^\dag \Dfb_\rho\Phi\right)
\left(\Phi^\dag\s_i\, \Dfb_\lambda\Phi \right)\WWd^i\\
\cP_{15}(h)\,,\cP_{19}(h)&\to\left[\Phi^\dag \DL_\mu \DL^\mu\Phi-
\DL_\mu \DL^\mu\Phi^\dag\Phi\right]^2\\ 
\cP_{16}(h)\,,\cP_{25}(h)&\to\left(\DL^\nu\Phi^\dag\DL_\mu \DL^\mu\Phi-
\DL_\mu \DL^\mu\Phi^\dag\DL^\nu\Phi\right)\left(\Phi^\dag \Dfb_\nu\Phi\right)
\end{aligned}
\label{Siblingsxi2}
\eeq

\item[For operators weighted by \boldmath$\xi^4$:]
\beq
\cP_{26}(h)\to\left[\left(\Phi^\dag \Dfb_\mu\Phi \right)
\left(\Phi^\dag\Dfb_\nu\Phi \right)\right]^2\,.
\eeq
\end{description}

%
%%%%%%%%%%%%%%%%%%%%%%%%%   Appendix D       %%%%%%%%%%%%%%%%%%%%%%%%
%

\section{Feynman rules}
\label{AppFR}
This Appendix provides a complete list of all the Feynman rules
resulting from the operators discussed here in the Lagrangian
$\LL_{chiral}$ of Eq.~(\ref{Lchiral}) (except for the pure Higgs ones
weighted by powers of $\xi$ higher than one). Only diagrams with up to
four legs are shown and the notation $\cF_i(h) = 1+2\tilde
a_i\,h/v+\tilde b_i\,h^2/v^2+\dots$ has been adopted. Moreover, for
brevity, the products $c_i \tilde a_i$ and $c_i \tilde b_i$ have been
redefined as $a_i$ and $b_i$, respectively. For the operators
$\cP_{8}$, and $\cP_{20-22}$, that contain two functions
$\cF_X(h)$ and $\cF'_X(h)$ we redefine $c_X\tilde{a}_X\tilde{a}'_X\to
a_X$.  In all Feynman diagrams the momenta are chosen to be flowing
inwards in the vertices and are computed in the unitary gauge, with
the exception of the propagator of the photon which is written in a
generic gauge.

Finally, the standard (that is SM-like) and non-standard Lorentz
structures are reported in two distinct columns, on the left and on
the right, respectively. Greek indices indicate flavour and are
assumed to be summed over when repeated; whenever they do not appear,
it should be understood that the vertex is flavour diagonal. All the
quantities entering the Feynman diagrams can be expressed in terms of
the parameters of the $Z$-renormalization scheme, as shown in
Eq.~(\ref{param}).

\fancypagestyle{mylandscape}{%
  \fancyhf{}% Clear header/footer
  \fancyfoot{% Footer
    \makebox[\textwidth][r]{% Right
      \rlap{\hspace{\footskip}% Push out of margin by \footskip
        \smash{% Remove vertical height
          \raisebox{\dimexpr.5\baselineskip+\footskip+.8\textheight}{% Raise vertically
            \rotatebox{90}{\thepage}}}}}}% Rotate counter-clockwise
  \renewcommand{\headrulewidth}{0pt}% No header rule
  \renewcommand{\footrulewidth}{0pt}% No footer rule
}

\begin{landscape}
\footnotesize

\addtolength{\hoffset}{-1.5cm}
\addtolength{\voffset}{2cm}

\newcommand{\nr}{\stepcounter{diagram}(FR.\arabic{diagram})}
\newcommand{\Anr}{\stepcounter{diagramDV}(A.\arabic{diagramDV})}
\addtolength{\linewidth}{3cm}
\newcounter{diagram}

\pagestyle{mylandscape}

\begin{center}
\vspace*{0.8cm}
\begin{tabular}{c@{\hspace*{5mm}}>{\centering}p{5cm}l@{\hspace{5mm}}l}

& & \bf standard structure& \bf non-standard structure\\\hline

% propagators and masses
    
  \nr& \parbox{3cm}{\input{Fdiagrams/AA}}   	& $\dfrac{-i}{ p^2}\left[g^{\mu\nu}-(1-\eta)\dfrac{p^\mu p^\nu}{p^2}\right]\quad(\eta \text{ is the gauge fixing parameter})$&\\

\nr&  \parbox{3cm}{\input{Fdiagrams/ZZ}}	& $\dfrac{-i}{p^2-m_Z^2}\left[g^{\mu\nu}-\dfrac{p^\mu p^\nu}{p^2-m_Z^2/X}\left(1-\dfrac{1}{X}\right)\right];\quad X=\dfrac{g^2\xi (c_{9}+2\xi c_{15})}{2\ct^2 }	$&\\

\nr&  \parbox{3cm}{\input{Fdiagrams/WW}}	& $\dfrac{-i}{p^2-m_W^2}\left[g^{\mu\nu}-\dfrac{p^\mu p^\nu}{p^2-m_W^2/X}\left(1-\dfrac{1}{X}\right)\right];\quad 
  \begin{array}{l}X=g^2\xi c_{9}\\
  m_W^2 = m_Z^2\ct^2 \left(1+\dfrac{4e^2\xi c_1}{\ct^2-\st^2}+\dfrac{2\xi  c_T \ct^2 }{\ct^2-\st^2}-4 g^2\xi^2 c_{12}\right)
  \end{array}
  $&\\[1cm]

\nr& \parbox{3cm}{\input{Fdiagrams/GG}}	& $\dfrac{-ig^{\mu\nu}}{ p^2}\left[g^{\mu\nu}-(1-\eta)\dfrac{p^\mu p^\nu}{p^2}\right]$&\\
 
\nr& \parbox{3cm}{\input{Fdiagrams/hh}}	& $\dfrac{-i}{ p^2-m_h^2} $\hspace{2cm}$(*)^{\blue{19}}$&\\  
\nr& \parbox{3cm}{\input{Fdiagrams/qq}}	& $\dfrac{i(\slashed{p}+m_{f})}{p^2-m_{f}^2};\quad m_{f} = -\dfrac{v\mathbf{y}_{f}}{\sqrt2},\qquad f=U,D,E$&
\end{tabular}

\begin{flushleft}
\footnotesize${}^{\blue{19}}$\,\, When considering the operator $\cP_{\square H}(h)$, a $p^4$ contribution to the $h$ propagator arises and it can be written as 
$ {-i}( p^2-m_h^2-p^4 c_{\square H}/v^2)^{-1}$ . 
However, the physical interpretation of this contribution is not straightforward \cite{Jansen:1993jj,Jansen:1993ji,Grinstein:2007mp} and will not be developed here, but in Ref.~\cite{BoxInPreparation}.
\end{flushleft}

\newpage
\renewcommand{\arraystretch}{5}
\begin{tabular}{c@{\hspace*{5mm}}>{\centering}p{5cm}l@{\hspace*{1cm}}l}
& & \bf standard structure& \bf non-standard structure\\\hline
  
  % triple vertices

\nr&  \parbox{3cm}{\input{Fdiagrams/WWA}} &   
$\begin{array}{l}ie[g_{\mu\nu}(p_+-p_-)_\rho-g_{\nu\rho}p_{-\mu}
+g_{\mu\rho}p_{+\nu}]\\[-1.3cm]
-ie[-g_{\mu\rho}p_{A\nu} +g_{\nu\rho} p_{A\mu}]
\left(1-g^2\xi (2c_1-2c_2-c_3)+2g^2\xi^2(c_{13}-2c_{12})\right)
\end{array}
  $&
  $-ieg^2\xi(p_+^\mu g_{\nu\rho}-p_-^\nu g_{\mu\rho}) c_{9}$\\

\nr&  \parbox{3cm}{\input{Fdiagrams/WWZ}} &     $\begin{array}{l}
  ig\ct\Big[g_{\mu\nu}(p_+-p_-)_\rho-g_{\nu\rho}p_{-\mu}+g_{\mu\rho}p_{+\nu}\Big]\left(1+\dfrac{\xi c_T }{\ct^2-\st^2}+\dfrac{g^2\xi c_3}{\ct^2 }+\dfrac{2e^2\xi c_1}{\ct^2 (\ct^2-\st^2)}\right)\\[-1cm]
  -ig\ct[-g_{\mu\rho}p_{Z\nu} +g_{\nu\rho} p_{Z\mu}]\left(1+\dfrac{\xi (c_T +4e^2 c_1)}{\ct^2-\st^2}+\xi g^2 c_3-\dfrac{2e^2\xi c_2}{\ct^2 }+2g^2\xi^2(c_{13}-2c_{12})\right)
  \end{array}$&
  $\begin{array}{l}
  -\e^{\mu\nu\rho\lambda}[p_{+\lambda}-p_{-\lambda}]
\dfrac{g^3\xi^2c_{14}}{\ct}\\[-1cm]
  +\dfrac{ig^3\xi}{\ct^2 }[p_+^\mu g_{\nu\rho}-p_-^\nu g_{\mu\rho}](\st^2  c_{9}-\xi c_{16})
  \end{array}
  $ \\
  
  % quartic vertices

\nr&  \parbox{3cm}{\input{Fdiagrams/WWAA}} &  $-ie^2[2g_{\mu\nu}g_{\lambda\rho}-\left(g_{\lambda\mu}g_{\rho\nu}+g_{\lambda\nu}g_{\rho\mu}\right)\left(1-g^2 \xi c_{9}\right)]$&
  \tabularnewline

\nr&  \parbox{3cm}{\input{Fdiagrams/WWZZ}} & $\begin{array}{l}
    -ig^2\ct^2 \left(1+\dfrac{2\xi c_T }{\ct^2-\st^2}+\dfrac{4e^2\xi c_1}{\ct^2 (\ct^2-\st^2)}+\dfrac{2g^2\xi c_3}{\ct^2 }\right)
    \Big[2g_{\mu\nu}g_{\lambda\rho}\left(1-\dfrac{g^2\xi c_{6}}{\ct^4 }-\dfrac{g^2\xi^2 c_{23}}{\ct^4 }\right)\\[-1cm]
    \qquad-\left(g_{\lambda\mu}g_{\rho\nu}+g_{\lambda\nu}g_{\rho\mu}\right)
    \left(1+\dfrac{\xi(- e^2\st^2  c_{9}+g^2\xi c_{11}+2e^2\xi c_{16})}{\ct^4 }+\dfrac{g^2\xi^2 c_{24}}{\ct^4 }\right)\Big]
  \end{array}$&\tabularnewline

\nr&  \parbox{3cm}{\input{Fdiagrams/WWZA}} &  $\begin{array}{l}-ieg\ct\left(1+\dfrac{\xi c_T }{\ct^2-\st^2}+\dfrac{2\xi e^2 c_1}{\ct^2 (\ct^2-\st^2)}+\dfrac{g^2\xi  c_3}{\ct^2 }\right)\cdot\\[-1cm]
  \cdot\left[2g_{\mu\nu}g_{\lambda\rho}-\left(g_{\lambda\nu}g_{\rho\nu}+g_{\lambda\mu}g_{\rho\nu}\right)\left(1+\dfrac{\xi(e^2 c_{9}-g^2\xi c_{16})}{\ct^2 }\right)\right]
  \end{array}
  $&
  $-\e^{\mu\nu\rho\lambda}\dfrac{2 e g^3 \xi^2c_{14}}{\ct}$
  \tabularnewline
  
\end{tabular}

\begin{tabular}{c@{\hspace*{5mm}}>{\centering}p{4cm}l@{\hspace*{0cm}}l}
& & \bf standard structure& \bf non-standard structure\\\hline
  \nr&  \parbox{3cm}{\input{Fdiagrams/WWWW}} &   $\begin{array}{l}
  ig^2\left(1+\dfrac{2\xi( c_T \ct^2 +2e^2 c_1)}{\ct^2-\st^2}+2g^2\xi c_3\right)\cdot\\[-1.3cm]
  \cdot\Big[-\left(g_{\mu\nu}g_{\lambda\rho}+g_{\lambda\nu}g_{\mu\rho}\right)\left(1+g^2\xi(-2c_{6}-\xi c_{11}-8\xi c_{12}+4\xi c_{13})\right)
\\[-1.5cm]
  \qquad+2g_{\lambda\mu}g_{\nu\rho}\left(1+g^2\xi^2(c_{11}-8c_{12}+4c_{13})\right)\Big]
  \end{array}$&\tabularnewline

\nr& \parbox{3cm}{\input{Fdiagrams/ZZZZ}} & & $\dfrac{i g^4\xi}{\ct^4 }[g_{\mu\nu}g_{\lambda\rho}+g_{\mu\lambda}g_{\nu\rho}+g_{\mu\rho}g_{\nu\lambda}]\left(c_{6}+\xi(c_{11}+2c_{23}+2c_{24})+4\xi^3 c_{26}\right) $\\

 % triple with Higgs
\nr& \parbox{3cm}{\input{Fdiagrams/AAh}} & & $i\dfrac{8 e^2\xi}{v}[g^{\mu\nu}p_{A1}\cdot p_{A2}-p_{A1}^\nu p_{A2}^\mu]\left(\dfrac{ a_B+ a_W}{4}-a_1-\xi  a_{12}\right) $\\
   
\nr& \parbox{3cm}{\input{Fdiagrams/ZZh}} & $2i \dfrac{m_Z^2}{v}g^{\mu\nu}\left(1-\dfrac{\xi c_H}{2}+\dfrac{\xi}{2} (2a_C-c_C)-2\xi(a_T-c_T)\right)$
 & $\begin{array}{l}  
  i\dfrac{2 g^2\ct^2\xi }{v}[g^{\mu\nu}p_{Z1}\cdot p_{Z2}-p_{Z1}^\nu p_{Z2}^\mu]\left( a_B \dfrac{\st^4}{\ct^4} + a_W +4\dfrac{\st^2}{\ct^2}  a_1-4\xi  a_{12}\right) \\[-1cm]
  -\dfrac{ie^2\xi}{v}[p_{Z1}^\nu p_h^\mu+p_{Z2}^\mu p_h^\nu-(p_{Z1}+p_{Z2})\cdot p_h g^{\mu\nu}]\left(\dfrac{2 a_4}{\ct^2 }-\dfrac{a_5+2\xi a_{17}}{\st^2 }\right)\\[-1cm]
  +\dfrac{2ig^2\xi}{v\ct^2 }g_{\mu\nu} p_{h}^2(2a_{7}+4\xi a_{25})
  +\dfrac{2ig^2\xi}{v\ct^2 }p_{Z1}^\mu p_{Z2}^\nu({a_{9}+2\xi a_{15}})\\[-1cm]
  +\dfrac{i g^2\xi}{v\ct^2 }[p_{Z1}^\mu p_h^\nu+p_{Z2}^\nu p_h^\mu]  (a_{10}+2\xi a_{19})
  \end{array}$\\
  
\end{tabular}

\begin{tabular}{c@{\hspace*{3mm}}>{\centering}p{5cm}l@{\hspace*{2cm}}l}
& & \bf standard structure& \bf non-standard structure\\\hline

\nr&  \parbox{3cm}{\input{Fdiagrams/AZh}} & & $\begin{array}{l} 
  \dfrac{-2i g^2\ct\st\xi }{v}[g^{\mu\nu}(p_{A}\cdot p_{Z})-p_{A}^\nu p_{Z}^\mu]\cdot\\[-1.3cm]
  \cdot\left(a_B\dfrac{\st^2}{\ct^2}- a_W+2 a_1\dfrac{\ct^2-\st^2}{\ct^2 }+4\xi  a_{12}\right)\\[-1cm]
  +\dfrac{ieg\xi}{v\ct}[p_A^\nu p_h^\mu-p_A\cdot p_h g^{\mu\nu}](2 a_4+ a_5+2\xi  a_{17})
  \end{array}$\\

\nr&  \parbox{3cm}{\input{Fdiagrams/WWh}} & 
  $\begin{array}{l}
  i\dfrac{2m_Z^2\ct^2 }{v}g_{\mu\nu}\Big(1-\dfrac{\xi c_H}{2}+\dfrac{\xi}{2}  (2a_C-c_C)+\dfrac{4e^2\xi c_1 }{(\ct^2-\st^2) }\\[-1cm] 
  \quad +\dfrac{2 \ct^2\xi   c_T }{(\ct^2-\st^2) }-4 g^2\xi^2  c_{12}\Big)  
   \end{array}$& 
  $\begin{array}{l} 
  \dfrac{i g^2\xi}{v}[2g^{\mu\nu}(p_{+}\cdot p_{-})-2p_{+}^\nu p_{-}^\mu] a_W
\\[-1.3cm]
  -\dfrac{ig^2\xi}{v}[(p_+^\nu p_h^\mu+p_-^\mu p_h^\nu)-(p_++p_-)\cdot p_h g^{\mu\nu}] a_5\\[-1cm]
  +\dfrac{2ig^2\xi}{v}g_{\mu\nu} p_h^2  a_{7}
  +\dfrac{2ig^2\xi}{v}p_+^\mu p_-^\nu a_{9}
  +\dfrac{ig^2\xi}{v}(p_+^\mu p_h^\nu+p_-^\nu p_h^\mu)c_{10}
  \end{array}
  $\\

\nr&  \parbox{3cm}{\input{Fdiagrams/GGh}} & & $ i\dfrac{2g_s^2\xi }{v} [g^{\mu\nu}p_{G1}\cdot p_{G2}-p_{G1}^\nu p_{G2}^\mu] a_G$ \\
 
\nr&  \parbox{3cm}{\input{Fdiagrams/WWAh}} & & $\begin{array}{l} 
  -\dfrac{2ie g^2 \xi }{v}[g_{\mu\nu}(p_+-p_-)_\rho+g_{\mu\rho}p_{-\nu}+g_{\nu\rho}p_{+\mu}] a_W\\[-1.3cm]
  -\dfrac{2ieg^2 \xi}{v}[-g_{\mu\rho}p_{A\nu} +g_{\nu\rho} p_{A\mu}](a_W-2 a_1+2  a_2+ a_3+2\xi( a_{13}-2 a_{12}))\\[-1cm]
  -\dfrac{ieg^2\xi}{v}[g_{\mu\rho}p_{h\nu}-g_{\nu\rho}p_{h\mu}]( a_5-a_{10})
  -\dfrac{2ieg^2\xi}{v}(p_+^\mu g_{\nu\rho}-p_-^\nu g_{\mu\rho}) a_{9}
  \end{array}$\\

\end{tabular}
 
\begin{tabular}{c@{\hspace*{5mm}}>{\centering}p{3cm}l@{\hspace*{1cm}}l}
& & \bf standard structure& \bf non-standard structure\\\hline

\nr&  \parbox{3cm}{\input{Fdiagrams/WWZh}} & &  $\begin{array}{l} 
  -\dfrac{2i g^3  \ct\xi }{v}[g_{\mu\nu}(p_+-p_-)_\rho+g_{\mu\rho}p_{-\nu}+g_{\nu\rho}p_{+\mu}]\left( a_W+\dfrac{ a_3}{\ct}\right)\\[-1.3cm]
  -\dfrac{2ig^3 \ct\xi}{v}[-g_{\mu\rho}p_{Z\nu} +g_{\nu\rho} p_{Z\mu}]\left( a_W+ a_3+2\dfrac{\st^2}{\ct^2}( a_1- a_2)+2\xi( a_{13}-2 a_{12})\right)\\[-1cm]
  +\dfrac{ige^2\xi}{v\ct}[g_{\mu\rho}p_{h\nu}-g_{\nu\rho}p_{h\mu}]\left(a_5-a_{10}+\dfrac{2\xi(a_{17}+a_{18})}{\st^2 }\right)\\[-1cm]
  +\dfrac{i2g^3\xi}{v\ct^2 }[p_+^\mu g_{\nu\rho}-p_-^\nu g_{\mu\rho}](\st^2  a_{9}-\xi a_{16})
  -\dfrac{2g^3\xi^2}{v\ct}\e^{\mu\nu\rho\lambda}[p_{+\lambda}-p_{-\lambda}]a_{14}
  \end{array}$\\

\nr&  \parbox{3cm}{\input{Fdiagrams/AAhh}} & &$i\dfrac{8e^2\xi }{v^2}[g^{\mu\nu}p_{A1}\cdot p_{A2}-p_{A1}^\nu p_{A2}^\mu]\left(\dfrac{ b_B+ b_W}{4}- b_1-\xi  b_{12}\right) $ \\
    
\nr&   \parbox{2.5cm}{\input{Fdiagrams/ZZhh}} & $i \dfrac{2m_Z^2}{v^2}g^{\mu\nu}\left(1-\xi c_H+\xi  b_C-2\xi(b_T-c_T)\right)$ 
   & $\begin{array}{l}
   i\dfrac{2 g^2\ct^2\xi }{v^2}[g^{\mu\nu}p_{Z1}\cdot p_{Z2}-p_{Z1}^\nu p_{Z2}^\mu]
   \left(b_B \dfrac{\st^4}{\ct^4} + b_W +4\dfrac{\st^2}{\ct^2}  b_1-\xi  b_{12}\right)\\[-1cm]
  -\dfrac{ie^2\xi}{v^2}[p_{Z1}^\nu (p_{h1}+p_{h2})^\mu+p_{Z2}^\mu (p_{h1}+p_{h2})^\nu-(p_{Z1}+p_{Z2})\cdot (p_{h1}+p_{h2})g^{\mu\nu}]\left(\dfrac{2 b_4}{\ct^2 }-\dfrac{ b_5+2\xi b_{17}}{\st^2 }\right)\\[-1cm]
  + \dfrac{2ig^2\xi}{v^2\ct^2 }g_{\mu\nu} \Big[2p_{h1}\cdot p_{h2} (b_{7}+2\xi a_{20}+4 \xi a_{21}+2\xi b_{25}) +(p_{h1}^2+p_{h2}^2)(b_{7}+2\xi b_{25})\Big]\\[-1cm]
  +\dfrac{2ig^2\xi}{v^2\ct^2 }p_{Z1}^\mu p_{Z2}^\nu (b_{9}+2\xi b_{15})
  +\dfrac{4ig^2\xi}{v^2\ct^2 }[p_{h1}^\mu p_{h2}^\nu+p_{h1}^\nu p_{h2}^\mu](a_{8}+2\xi a_{22})\\[-1cm]
  +\dfrac{ig^2 \xi}{v^2\ct^2 }\left[p_{Z1}^\mu (p_{h1}+p_{h2})^\nu+p_{Z2}^\nu (p_{h1}+p_{h2})^\mu\right] (b_{10}+2\xi b_{19})
  \end{array}$\\
 
\end{tabular}
 
\begin{tabular}{c@{\hspace*{4mm}}>{\centering}p{4cm}l@{\hspace*{2cm}}l}
& & \bf standard structure& \bf non-standard structure\\\hline

\nr&  \parbox{2.5cm}{\input{Fdiagrams/AZhh}} & &$\begin{array}{l} 
  \dfrac{-2i g^2\ct\st\xi }{v^2}[g^{\mu\nu}(p_{A}\cdot p_{Z})-p_{A}^\nu p_{Z}^\mu]\cdot\\[-1.3cm]
  \cdot\left(b_B\dfrac{\st^2}{\ct^2}- b_W+2 b_1\dfrac{\ct^2-\st^2}{\ct^2 }+4\xi  b_{12}\right)\\[-1cm]
  +\dfrac{ieg\xi}{v^2\ct}[p_A^\nu (p_{h1}+p_{h2})^\mu-p_A\cdot (p_{h1}+p_{h2})g^{\mu\nu}](2 b_4+ b_5+2\xi  b_{17})
  \end{array}$\\
  
\nr&  \parbox{2.5cm}{\input{Fdiagrams/WWhh}} & $\begin{array}{l}
  i\dfrac{2m_Z^2\ct^2 }{v^2}g_{\mu\nu}\Big(1-\xi c_H+\xi b_C+\dfrac{4e^2\xi  c_1}{(\ct^2-\st^2) }\\[-1cm]
  \quad +\dfrac{2\ct^2\xi   c_T  }{(\ct^2-\st^2) }-4 g^2\xi^2  c_{12}\Big)
  \end{array}$
  & $\begin{array}{l}
  \dfrac{i\xi}{v^2}[2g^{\mu\nu}(p_{+}\cdot p_{-})-2p_{+}^\nu p_{-}^\mu]g^2  b_W+
  \dfrac{2ig^2\xi}{v^2}p_+^\mu p_-^\nu b_{9}\\[-1cm]
   +\dfrac{ig^2\xi}{v^2}[p_+^\nu (p_{h1}+p_{h2})^\mu+p_-^\mu (p_{h1}+p_{h2})^\nu-(p_++p_-)\cdot (p_{h1}+p_{h2})g^{\mu\nu}] b_5\\[-1cm]
  +\dfrac{2ig^2\xi}{v^2}g_{\mu\nu} \left[2p_{h1}\cdot p_{h2} (b_{7}+2\xi a_{20})+(p_{h1}^2+p_{h2}^2)b_{7}\right]\\[-1.3cm]
  +\dfrac{4ig^2\xi}{v^2}[p_{h1}^\mu p_{h2}^\nu+p_{h1}^\nu p_{h2}^\mu]a_{8}
\\[-1cm]
  +\dfrac{ig^2 \xi}{v^2}\left[p_+^\mu (p_{h1}+p_{h2})^\nu+p_-^\nu (p_{h1}+p_{h2})^\mu\right] b_{10}
  \end{array}$\\
  
\nr&   \parbox{2.5cm}{\input{Fdiagrams/GGhh}} & & $ i\dfrac{2 g_s^2 \xi}{v^2} [g^{\mu\nu}p_{G1}\cdot p_{G2}-p_{G1}^\nu p_{G2}^\mu] b_G$

\end{tabular}

\newpage
\thispagestyle{empty}
\begin{tabular}{c@{\hspace*{5mm}}>{\centering}p{4cm}ll}
& & \bf standard structure& \bf non-standard structure\\\hline

\nr&  \parbox{2.5cm}{\input{Fdiagrams/Aqq}} & $-ie q_{f} \g_\mu;\quad q_U=\dfrac{2}{3},\:q_D=-\dfrac{1}{3},\:q_N=0,\:q_E=-1$& \\
  
\nr&  \parbox{2.5cm}{\input{Fdiagrams/Wud}} & $\dfrac{ig}{\sqrt2}\dfrac{\g_\mu(1-\g_5)}{2}(V^\dag)_{\a\b}
  \left(1+\dfrac{\xi (c_T  \ct^2 +2e^2c_1)}{\ct^2-\st^2}-2g^2\xi^2 c_{12}\right)$& \\
  
\nr&  \parbox{2.5cm}{\input{Fdiagrams/Zuu}} & $\begin{array}{l}
                                        \dfrac{ig\ct}{2}\left(\dfrac{\st^2}{3\ct^2}-1\right) \dfrac{\g_\mu(1-\g_5)}{2}\left[1+\dfrac{8\xi e^2 c_1}{(\ct^2-\st^2)(1+2(\ct^2-\st^2))}+\dfrac{\xi c_T}{\ct^2-\st^2}\dfrac{1+2\ct^2 }{1+2(\ct^2-\st^2)}\right]\\[-1cm]
                                        \dfrac{2}{3}ie\dfrac{\st}{\ct} \dfrac{\g_\mu(1+\g_5)}{2}\left[1-\dfrac{\xi (c_T+2g^2 c_1)}{\ct^2-\st^2}\right]
                                          \end{array}$ & \\
                                          
\nr&  \parbox{2.5cm}{\input{Fdiagrams/Zdd}} & $\begin{array}{l}
				      \dfrac{ig\ct}{2}\left(\dfrac{\st^2}{3\ct^2}+1\right)\dfrac{\g_\mu(1-\g_5)}{2}\left[1+\dfrac{\xi 4e^2 c_1}{(\ct^2-\st^2)(2+(\ct^2-\st^2))}+\dfrac{\xi c_T}{\ct^2-\st^2}\dfrac{ 4\ct^2 -1}{2+(\ct^2-\st^2)}\right]\\[-1cm]
				        -\dfrac{1}{3}ie\dfrac{\st}{\ct}\dfrac{\g_\mu(1+\g_5)}{2}\left[1-\dfrac{\xi (c_T +2g^2 c_1)}{\ct^2-\st^2}\right]
					\end{array}$ & \\
					
\nr&  \parbox{2.5cm}{\input{Fdiagrams/Znn}} & $\begin{array}{l}
				      -\dfrac{ig}{2\ct}\dfrac{\g_\mu(1-\g_5)}{2}\left[1+\dfrac{\xi 4e^2 c_1}{(\ct^2-\st^2)(2+(\ct^2-\st^2))}+\dfrac{\xi c_T}{\ct^2-\st^2}\dfrac{ 4\ct^2 -1}{2+(\ct^2-\st^2)}\right]
					\end{array}$ & \\					
					
\nr&  \parbox{2.5cm}{\input{Fdiagrams/Zee}} & $\begin{array}{l}
				      \dfrac{ig(\ct^2-\st^2)}{2\ct}\dfrac{\g_\mu(1-\g_5)}{2}\left[1+\dfrac{\xi 4e^2 c_1}{(\ct^2-\st^2)(2+(\ct^2-\st^2))}+\dfrac{\xi c_T}{\ct^2-\st^2}\dfrac{ 4\ct^2 -1}{2+(\ct^2-\st^2)}\right]\\[-1cm]
				        -ie\dfrac{\st}{\ct}\dfrac{\g_\mu(1+\g_5)}{2}\left[1-\dfrac{\xi (c_T +2g^2 c_1)}{\ct^2-\st^2}\right]
					\end{array}$& \\					

\end{tabular}

\begin{tabular}{c@{\hspace*{5mm}}>{\centering}p{4cm}l@{\hspace*{2cm}}l}
& & \bf standard structure& \bf non-standard structure\\\hline  

\nr&  \parbox{2.5cm}{\input{Fdiagrams/hqq}} & $ -i\dfrac{s_Y}{\sqrt2}Y_{f,\a\b}\left(1-\dfrac{\xi c_H}{2}\right)\,,\quad  f=U,D,E$ & \\

\nr&  \parbox{2.5cm}{\input{Fdiagrams/Wudh}} & & $\dfrac{\sqrt2ig\xi}{v}\dfrac{\g_\mu(1-\g_5)}{2}(V^\dag)_{\a\b} \left(\dfrac{ a_T  \ct^2 +2e^2 a_1}{\ct^2-\st^2}-2g^2\xi  a_{12}\right)$ \\
  
\nr&  \parbox{2.5cm}{\input{Fdiagrams/Zuuh}} & & $\begin{array}{l}
                                        \dfrac{ig\xi\ct}{v}\left(\dfrac{\st^2}{3\ct^2}-1\right) \dfrac{\g_\mu(1-\g_5)}{2}\left[\dfrac{8 e^2 a_1}{(\ct^2-\st^2)(1+2(\ct^2-\st^2))}+\dfrac{ a_T}{\ct^2-\st^2}\dfrac{1+2\ct^2 }{1+2(\ct^2-\st^2)}\right]\\[-1cm]
                                        +\dfrac{4}{3}\dfrac{ie\xi}{v}\dfrac{\st}{\ct} \dfrac{\g_\mu(1+\g_5)}{2}\left[\dfrac{ a_T+2g^2  a_1}{\ct^2-\st^2}\right]
                                          \end{array}$  \\
                                          
\nr&  \parbox{2.5cm}{\input{Fdiagrams/Zddh}} & &$\begin{array}{l}
				      \dfrac{i\xi g\ct}{v}\left(\dfrac{\st^2}{3\ct^2}+1\right)\dfrac{\g_\mu(1-\g_5)}{2}\left[\dfrac{ 4e^2  a_1}{(\ct^2-\st^2)(2+(\ct^2-\st^2))}+\dfrac{ a_T}{\ct^2-\st^2}\dfrac{ 4\ct^2 -1}{2+(\ct^2-\st^2)}\right]\\[-1cm]
				        -\dfrac{2}{3}\dfrac{i\xi e}{v}\dfrac{\st}{\ct}\dfrac{\g_\mu(1+\g_5)}{2}\left[\dfrac{ a_T +2g^2  a_1}{\ct^2-\st^2}\right]
					\end{array}$   \\
\end{tabular}

\begin{tabular}{c@{\hspace*{5mm}}>{\centering}p{4cm}l@{\hspace*{1cm}}l}
& & \bf standard structure& \bf non-standard structure\\\hline  

\nr&  \parbox{2.5cm}{\input{Fdiagrams/Wneh}} & & $\dfrac{\sqrt2ig\xi}{v}\dfrac{\g_\mu(1-\g_5)}{2}
  \left(\dfrac{ a_T  \ct^2 +2e^2 a_1}{\ct^2-\st^2}-2g^2\xi  a_{12}\right)$ \\
  
\nr&  \parbox{2.5cm}{\input{Fdiagrams/Znnh}} & & $\begin{array}{l}
                                        -\dfrac{ig\xi}{v\ct} \dfrac{\g_\mu(1-\g_5)}{2}\left[\dfrac{8 e^2 a_1}{(\ct^2-\st^2)(1+2(\ct^2-\st^2))}+\dfrac{ a_T}{\ct^2-\st^2}\dfrac{1+2\ct^2 }{1+2(\ct^2-\st^2)}\right]
                                          \end{array}$  \\
                                          
\nr&  \parbox{2.5cm}{\input{Fdiagrams/Zeeh}} & &$\begin{array}{l}
				      \dfrac{i\xi g(\ct^2-\st^2)}{v\ct}\dfrac{\g_\mu(1-\g_5)}{2}\left[\dfrac{ 4e^2  a_1}{(\ct^2-\st^2)(2+(\ct^2-\st^2))}+\dfrac{ a_T}{\ct^2-\st^2}\dfrac{ 4\ct^2 -1}{2+(\ct^2-\st^2)}\right]\\[-1cm]
				        -\dfrac{i\xi e}{v}\dfrac{\st}{\ct}\dfrac{\g_\mu(1+\g_5)}{2}\left[\dfrac{ a_T +2g^2  a_1}{\ct^2-\st^2}\right]
					\end{array}$  
							
\end{tabular}

\end{center}

\end{landscape}

%%%%%%%%%%%%%%%%%%%%%%%%%%%%%%%%%%%%%%%%%%%%%%%%%%%%%%%%%%%%
% Bibliography     
%%%%%%%%%%%%%%%%%%%%%%%%%%%%%%%%%%%%%%%%%%%%%%%%%%%%%%%%%%%%
\normalsize
\bibliographystyle{BiblioStyle}
\providecommand{\href}[2]{#2}\begingroup\raggedright\endgroup
\end{document}

%% file: Fdiagrams/WWA.tex
\begin{fmffile}{wwacoupl}
\begin{fmfgraph*}(50,50)

\fmfleft{i1,i2,i3}
\fmfright{o3,o2,o1}
  \fmf{boson}{i2,v1}
  \fmf{boson}{v1,o1}
  \fmf{boson}{v1,o3}
  
\fmfv{lab=$W^+_\mu$,l.angle=0}{o1}
\fmfv{lab=$W^-_\nu$,l.angle=0}{o3}
\fmfv{lab=$A_\rho$}{i2}
\end{fmfgraph*}
\end{fmffile}

%% file: Fdiagrams/WWZ.tex
\begin{fmffile}{wwzcoupl}
\begin{fmfgraph*}(50,50)

\fmfleft{i1,i2,i3}
\fmfright{o3,o2,o1}
  \fmf{boson}{i2,v1}
  \fmf{boson}{v1,o1}
  \fmf{boson}{v1,o3}
  
\fmfv{lab=$W^+_\mu$,l.angle=0}{o1}
\fmfv{lab=$W^-_\nu$,l.angle=0}{o3}
\fmfv{lab=$Z_\rho$}{i2}
\end{fmfgraph*}
\end{fmffile}

%% file: Fdiagrams/WWZA.tex
\begin{fmffile}{wwzacoupl}
\begin{fmfgraph*}(50,50)

\fmfleft{i1,i2}
\fmfright{o2,o1}
  \fmf{boson}{i2,v1}
  \fmf{boson}{i1,v1}
  \fmf{boson}{v1,o1}
  \fmf{boson}{v1,o2}
  
\fmfv{lab=$W^+_\lambda$,l.angle=0}{o1}
\fmfv{lab=$W^-_\rho$,l.angle=0}{o2}
\fmfv{lab=$A_\nu$,l.angle=180}{i1}
\fmfv{lab=$Z_\mu$,l.angle=180}{i2}
\end{fmfgraph*}
\end{fmffile}

%% file: Fdiagrams/AA.tex
\begin{fmffile}{Aprop}
\begin{fmfgraph*}(80,50)

\fmfleft{i1}
\fmfright{o1}
  \fmf{boson,label=$\gamma$}{i1,o1}
  
\fmfv{lab=$\nu$,l.angle=0}{o1}
\fmfv{lab=$\mu$,l.angle=180}{i1}
\end{fmfgraph*}
\end{fmffile}

%% file: Fdiagrams/ZZ.tex
\begin{fmffile}{Zprop}
\begin{fmfgraph*}(80,50)

\fmfleft{i1}
\fmfright{o1}
  \fmf{boson,label=$Z$}{i1,o1}
  
\fmfv{lab=$\nu$,l.angle=0}{o1}
\fmfv{lab=$\mu$,l.angle=180}{i1}
\end{fmfgraph*}
\end{fmffile}

%% file: Fdiagrams/WW.tex
\begin{fmffile}{Wprop}
\begin{fmfgraph*}(80,50)

\fmfleft{i1}
\fmfright{o1}
  \fmf{boson,label=$W$}{i1,o1}
  
\fmfv{lab=$\nu$,l.angle=0}{o1}
\fmfv{lab=$\mu$,l.angle=180}{i1}
\end{fmfgraph*}
\end{fmffile}

%% file: Fdiagrams/GG.tex
\begin{fmffile}{Gprop}
\begin{fmfgraph*}(80,50)

\fmfleft{i1}
\fmfright{o1}
  \fmf{gluon,label=$G$,l.side=left}{i1,o1}
  
\fmfv{lab=$\nu$,l.angle=0}{o1}
\fmfv{lab=$\mu$,l.angle=180}{i1}
\end{fmfgraph*}
\end{fmffile}

%% file: Fdiagrams/hh.tex
\begin{fmffile}{hprop}
\begin{fmfgraph*}(80,50)

\fmfleft{i1}
\fmfright{o1}
  \fmf{dashes,label=$h$}{i1,o1}
  
\end{fmfgraph*}
\end{fmffile}

%% file: Fdiagrams/qq.tex
\begin{fmffile}{qprop}
\begin{fmfgraph*}(80,50)

\fmfleft{i1}
\fmfright{o1}
  \fmf{fermion,label=$f$}{i1,o1}
  
\end{fmfgraph*}
\end{fmffile}

%% file: Fdiagrams/WWAA.tex
\begin{fmffile}{wwaacoupl}
\begin{fmfgraph*}(50,50)

\fmfleft{i1,i2}
\fmfright{o2,o1}
  \fmf{boson}{i2,v1}
  \fmf{boson}{i1,v1}
  \fmf{boson}{v1,o1}
  \fmf{boson}{v1,o2}
  
\fmfv{lab=$W^+_\lambda$,l.angle=0}{o1}
\fmfv{lab=$W^-_\rho$,l.angle=0}{o2}
\fmfv{lab=$A_\nu$,l.angle=180}{i1}
\fmfv{lab=$A_\mu$,l.angle=180}{i2}
\end{fmfgraph*}
\end{fmffile}

%% file: Fdiagrams/WWZZ.tex
\begin{fmffile}{wwzzcoupl}
\begin{fmfgraph*}(50,50)

\fmfleft{i1,i2}
\fmfright{o2,o1}
  \fmf{boson}{i2,v1}
  \fmf{boson}{i1,v1}
  \fmf{boson}{v1,o1}
  \fmf{boson}{v1,o2}
  
\fmfv{lab=$W^+_\lambda$,l.angle=0}{o1}
\fmfv{lab=$W^-_\rho$,l.angle=0}{o2}
\fmfv{lab=$Z_\nu$,l.angle=180}{i1}
\fmfv{lab=$Z_\mu$,l.angle=180}{i2}
\end{fmfgraph*}
\end{fmffile}

%% file: Fdiagrams/WWWW.tex
\begin{fmffile}{wwwwcoupl}
\begin{fmfgraph*}(50,50)

\fmfleft{i1,i2}
\fmfright{o2,o1}
  \fmf{boson}{i2,v1}
  \fmf{boson}{i1,v1}
  \fmf{boson}{v1,o1}
  \fmf{boson}{v1,o2}
  
\fmfv{lab=$W^+_\lambda$,l.angle=0}{o1}
\fmfv{lab=$W^-_\rho$,l.angle=0}{o2}
\fmfv{lab=$W^+_\mu$,l.angle=180}{i1}
\fmfv{lab=$W^-_\nu$,l.angle=180}{i2}
\end{fmfgraph*}
\end{fmffile}

%% file: Fdiagrams/ZZZZ.tex
\begin{fmffile}{zzzzcoupl}
\begin{fmfgraph*}(50,50)

\fmfleft{i1,i2}
\fmfright{o2,o1}
  \fmf{boson}{i2,v1}
  \fmf{boson}{i1,v1}
  \fmf{boson}{v1,o1}
  \fmf{boson}{v1,o2}
  
\fmfv{lab=$Z_\lambda$,l.angle=0}{o1}
\fmfv{lab=$Z_\rho$,l.angle=0}{o2}
\fmfv{lab=$Z_\mu$,l.angle=180}{i1}
\fmfv{lab=$Z_\nu$,l.angle=180}{i2}
\end{fmfgraph*}
\end{fmffile}

%% file: Fdiagrams/AAh.tex
\begin{fmffile}{aahcoupl}
\begin{fmfgraph*}(50,50)

\fmfleft{i1,i2,i3}
\fmfright{o3,o2,o1}
  \fmf{dashes}{i2,v1}
  \fmf{boson,label=$p_{A1}$}{v1,o1}
  \fmf{boson,label=$p_{A2}$}{v1,o3}
  
\fmfv{lab=$A_\mu$,l.angle=0}{o1}
\fmfv{lab=$A_\nu$,l.angle=0}{o3}
\fmfv{lab=$h$}{i2}
\end{fmfgraph*}
\end{fmffile}

%% file: Fdiagrams/ZZh.tex
\begin{fmffile}{zzhcoupl}
\begin{fmfgraph*}(50,50)

\fmfleft{i1,i2,i3}
\fmfright{o3,o2,o1}
  \fmf{dashes}{i2,v1}
  \fmf{boson,label=$p_{Z1}$}{v1,o1}
  \fmf{boson,label=$p_{Z2}$,l.side=right }{v1,o3}
  
\fmfv{lab=$Z_\mu$,l.angle=0}{o1}
\fmfv{lab=$Z_\nu$,l.angle=0}{o3}
\fmfv{lab=$h$}{i2}
\end{fmfgraph*}
\end{fmffile}

%% file: Fdiagrams/AZh.tex
\begin{fmffile}{azhcoupl}
\begin{fmfgraph*}(50,50)

\fmfleft{i1,i2,i3}
\fmfright{o3,o2,o1}
  \fmf{dashes}{i2,v1}
  \fmf{boson}{v1,o1}
  \fmf{boson}{v1,o3}
  
\fmfv{lab=$A_\mu$,l.angle=0}{o1}
\fmfv{lab=$Z_\nu$,l.angle=0}{o3}
\fmfv{lab=$h$}{i2}
\end{fmfgraph*}
\end{fmffile}

%% file: Fdiagrams/WWh.tex
\begin{fmffile}{wwhcoupl}
\begin{fmfgraph*}(50,50)

\fmfleft{i1,i2,i3}
\fmfright{o3,o2,o1}
  \fmf{dashes}{i2,v1}
  \fmf{boson}{v1,o1}
  \fmf{boson}{v1,o3}
  
\fmfv{lab=$W^+_\mu$,l.angle=0}{o1}
\fmfv{lab=$W^-_\nu$,l.angle=0}{o3}
\fmfv{lab=$h$}{i2}
\end{fmfgraph*}
\end{fmffile}

%% file: Fdiagrams/GGh.tex
\begin{fmffile}{gghcoupl}
\begin{fmfgraph*}(50,50)

\fmfleft{i1,i2,i3}
\fmfright{o3,o2,o1}
  \fmf{dashes}{i2,v1}
  \fmf{gluon,label=$p_{G1}$}{v1,o1}
  \fmf{gluon,label=$p_{G2}$}{v1,o3}
  
\fmfv{lab=$G_\mu$,l.angle=0}{o1}
\fmfv{lab=$G_\nu$,l.angle=0}{o3}
\fmfv{lab=$h$}{i2}
\end{fmfgraph*}
\end{fmffile}

%% file: Fdiagrams/WWAh.tex
\begin{fmffile}{wwahcoupl}
\begin{fmfgraph*}(50,50)

\fmfleft{i1,i2}
\fmfright{o2,o1}
  \fmf{dashes}{i2,v1}
  \fmf{boson}{i1,v1}
  \fmf{boson}{v1,o1}
  \fmf{boson}{v1,o2}
  
\fmfv{lab=$W^+_\mu$,l.angle=0}{o1}
\fmfv{lab=$W^-_\nu$,l.angle=0}{o2}
\fmfv{lab=$A_\rho$,l.angle=180}{i1}
\fmfv{lab=$h$,l.angle=180}{i2}
\end{fmfgraph*}
\end{fmffile}

%% file: Fdiagrams/WWZh.tex
\begin{fmffile}{wwzhcoupl}
\begin{fmfgraph*}(50,50)

\fmfleft{i1,i2}
\fmfright{o2,o1}
  \fmf{dashes}{i2,v1}
  \fmf{boson}{i1,v1}
  \fmf{boson}{v1,o1}
  \fmf{boson}{v1,o2}
  
\fmfv{lab=$W^+_\mu$,l.angle=0}{o1}
\fmfv{lab=$W^-_\nu$,l.angle=0}{o2}
\fmfv{lab=$Z_\rho$,l.angle=180}{i1}
\fmfv{lab=$h$,l.angle=180}{i2}
\end{fmfgraph*}
\end{fmffile}

%% file: Fdiagrams/AAhh.tex
\begin{fmffile}{aahhcoupl}
\begin{fmfgraph*}(50,50)

\fmfleft{i1,i2}
\fmfright{o2,o1}
  \fmf{dashes}{i2,v1}
  \fmf{dashes}{i1,v1}
  \fmf{boson,label=$p_{A1}$,l.side=right}{v1,o1}
  \fmf{boson,label=$p_{A2}$,l.side=right}{v1,o2}
  
\fmfv{lab=$A_\mu$,l.angle=0}{o1}
\fmfv{lab=$A_\nu$,l.angle=0}{o2}
\fmfv{lab=$h$,l.angle=180}{i1}
\fmfv{lab=$h$,l.angle=180}{i2}
\end{fmfgraph*}
\end{fmffile}

%% file: Fdiagrams/ZZhh.tex
\begin{fmffile}{zzhhcoupl}
\begin{fmfgraph*}(50,50)

\fmfleft{i1,i2}
\fmfright{o2,o1}
  \fmf{dashes}{i2,v1}
  \fmf{dashes}{i1,v1}
  \fmf{boson,label=$p_{Z1}$,l.side=right}{v1,o1}
  \fmf{boson,label=$p_{Z2}$,l.side=right}{v1,o2}
  
\fmfv{lab=$Z_\mu$,l.angle=0}{o1}
\fmfv{lab=$Z_\nu$,l.angle=0}{o2}
\fmfv{lab=$h$,l.angle=180}{i1}
\fmfv{lab=$h$,l.angle=180}{i2}
\end{fmfgraph*}
\end{fmffile}

%% file: Fdiagrams/AZhh.tex
\begin{fmffile}{azhhcoupl}
\begin{fmfgraph*}(50,50)

\fmfleft{i1,i2}
\fmfright{o2,o1}
  \fmf{dashes}{i2,v1}
  \fmf{dashes}{i1,v1}
  \fmf{boson}{v1,o1}
  \fmf{boson}{v1,o2}
  
\fmfv{lab=$A_\mu$,l.angle=0}{o1}
\fmfv{lab=$Z_\nu$,l.angle=0}{o2}
\fmfv{lab=$h$,l.angle=180}{i1}
\fmfv{lab=$h$,l.angle=180}{i2}
\end{fmfgraph*}
\end{fmffile}

%% file: Fdiagrams/WWhh.tex
\begin{fmffile}{wwhhcoupl}
\begin{fmfgraph*}(50,50)

\fmfleft{i1,i2}
\fmfright{o2,o1}
  \fmf{dashes}{i2,v1}
  \fmf{dashes}{i1,v1}
  \fmf{boson}{v1,o1}
  \fmf{boson}{v1,o2}
  
\fmfv{lab=$W^+_\mu$,l.angle=0}{o1}
\fmfv{lab=$W^-_\nu$,l.angle=0}{o2}
\fmfv{lab=$h$,l.angle=180}{i1}
\fmfv{lab=$h$,l.angle=180}{i2}
\end{fmfgraph*}
\end{fmffile}

%% file: Fdiagrams/GGhh.tex
\begin{fmffile}{gghhcoupl}
\begin{fmfgraph*}(50,50)

\fmfleft{i1,i2}
\fmfright{o2,o1}
  \fmf{dashes}{i2,v1}
  \fmf{dashes}{i1,v1}
  \fmf{gluon,label=$p_{G1}$,l.side=right}{v1,o1}
  \fmf{gluon,label=$p_{G2}$,l.side=right}{v1,o2}
  
\fmfv{lab=$G_\mu$,l.angle=0}{o1}
\fmfv{lab=$G_\nu$,l.angle=0}{o2}
\fmfv{lab=$h$,l.angle=180}{i1}
\fmfv{lab=$h$,l.angle=180}{i2}
\end{fmfgraph*}
\end{fmffile}

%% file: Fdiagrams/Aqq.tex
\begin{fmffile}{aqqcoupl}
\begin{fmfgraph*}(50,50)

\fmfleft{i1,i2,i3}
\fmfright{o3,o2,o1}
  \fmf{boson}{i2,v1}
  \fmf{fermion,label=$f$}{o1,v1,o3}
  
\fmfv{lab=$A_\mu$}{i2}
\end{fmfgraph*}
\end{fmffile}

%% file: Fdiagrams/Wud.tex
\begin{fmffile}{wudcoupl}
\begin{fmfgraph*}(50,50)

\fmfleft{i1,i2,i3}
\fmfright{o3,o2,o1}
  \fmf{boson}{i2,v1}
  \fmf{fermion}{o1,v1,o3}
  
\fmfv{lab=$W^-_\mu$}{i2}
\fmfv{lab=$U_\beta$,l.angle=0}{o1}
\fmfv{lab=$D_\alpha$,l.angle=0}{o3}
\end{fmfgraph*}
\end{fmffile}

%% file: Fdiagrams/Zuu.tex
\begin{fmffile}{zuucoupl}
\begin{fmfgraph*}(50,50)

\fmfleft{i1,i2,i3}
\fmfright{o3,o2,o1}
  \fmf{boson}{i2,v1}
  \fmf{fermion,label=$U$}{o1,v1,o3}
  
\fmfv{lab=$Z_\mu$}{i2}

\end{fmfgraph*}
\end{fmffile}

%% file: Fdiagrams/Zdd.tex
\begin{fmffile}{zddcoupl}
\begin{fmfgraph*}(50,50)

\fmfleft{i1,i2,i3}
\fmfright{o3,o2,o1}
  \fmf{boson}{i2,v1}
  \fmf{fermion,label=$D$}{o1,v1,o3}
  
\fmfv{lab=$Z_\mu$}{i2}

\end{fmfgraph*}
\end{fmffile}

%% file: Fdiagrams/Znn.tex
\begin{fmffile}{znncoupl}
\begin{fmfgraph*}(50,50)

\fmfleft{i1,i2,i3}
\fmfright{o3,o2,o1}
  \fmf{boson}{i2,v1}
  \fmf{fermion,label=$N$}{o1,v1,o3}
  
\fmfv{lab=$Z_\mu$}{i2}

\end{fmfgraph*}
\end{fmffile}

%% file: Fdiagrams/Zee.tex
\begin{fmffile}{zeecoupl}
\begin{fmfgraph*}(50,50)

\fmfleft{i1,i2,i3}
\fmfright{o3,o2,o1}
  \fmf{boson}{i2,v1}
  \fmf{fermion,label=$E$}{o1,v1,o3}
  
\fmfv{lab=$Z_\mu$}{i2}

\end{fmfgraph*}
\end{fmffile}

%% file: Fdiagrams/hqq.tex
\begin{fmffile}{hqqcoupl}
\begin{fmfgraph*}(50,50)

\fmfleft{i1,i2,i3}
\fmfright{o3,o2,o1}
  \fmf{dashes}{i2,v1}
  \fmf{fermion,label=$f_\beta$}{o1,v1}
  \fmf{fermion,label=$f_\alpha$}{v1,o3}
  
\fmfv{lab=$h$}{i2}
\end{fmfgraph*}
\end{fmffile}

%% file: Fdiagrams/Wudh.tex
\begin{fmffile}{wudhcoupl}
\begin{fmfgraph*}(50,50)

\fmfleft{i1,i2,i3}
\fmfright{o3,o2,o1}
  \fmf{boson}{i1,v1}
  \fmf{dashes}{i3,v1}
  \fmf{fermion}{o1,v1,o3}
  
\fmfv{lab=$W^-_\mu$,l.angle=180}{i1}
\fmfv{lab=$h$,l.angle=180}{i3}
\fmfv{lab=$U_\beta$,l.angle=0}{o1}
\fmfv{lab=$D_\alpha$,l.angle=0}{o3}
\end{fmfgraph*}
\end{fmffile}

%% file: Fdiagrams/Zuuh.tex
\begin{fmffile}{zuuhcoupl}
\begin{fmfgraph*}(50,50)

\fmfleft{i1,i2,i3}
\fmfright{o3,o2,o1}
  \fmf{boson}{i1,v1}
  \fmf{dashes}{i3,v1}
  \fmf{fermion}{o1,v1,o3}
  
\fmfv{lab=$Z_\mu$,l.angle=180}{i1}
\fmfv{lab=$h$,l.angle=180}{i3}
\fmfv{lab=$U$,l.angle=0}{o1}
\fmfv{lab=$U$,l.angle=0}{o3}

\end{fmfgraph*}
\end{fmffile}

%% file: Fdiagrams/Zddh.tex
\begin{fmffile}{zddhcoupl}
\begin{fmfgraph*}(50,50)

\fmfleft{i1,i2,i3}
\fmfright{o3,o2,o1}
  \fmf{boson}{i1,v1}
  \fmf{dashes}{i3,v1}
  \fmf{fermion}{o1,v1,o3}
  
\fmfv{lab=$Z_\mu$,l.angle=180}{i1}
\fmfv{lab=$h$,l.angle=180}{i3}
\fmfv{lab=$D$,l.angle=0}{o1}
\fmfv{lab=$D$,l.angle=0}{o3}
\end{fmfgraph*}
\end{fmffile}

%% file: Fdiagrams/Wneh.tex
\begin{fmffile}{wnehcoupl}
\begin{fmfgraph*}(50,50)

\fmfleft{i1,i2,i3}
\fmfright{o3,o2,o1}
  \fmf{boson}{i1,v1}
  \fmf{dashes}{i3,v1}
  \fmf{fermion}{o1,v1,o3}
  
\fmfv{lab=$W^-_\mu$,l.angle=180}{i1}
\fmfv{lab=$h$,l.angle=180}{i3}
\fmfv{lab=$N$,l.angle=0}{o1}
\fmfv{lab=$E$,l.angle=0}{o3}
\end{fmfgraph*}
\end{fmffile}

%% file: Fdiagrams/Znnh.tex
\begin{fmffile}{znnhcoupl}
\begin{fmfgraph*}(50,50)

\fmfleft{i1,i2,i3}
\fmfright{o3,o2,o1}
  \fmf{boson}{i1,v1}
  \fmf{dashes}{i3,v1}
  \fmf{fermion}{o1,v1,o3}
  
\fmfv{lab=$Z_\mu$,l.angle=180}{i1}
\fmfv{lab=$h$,l.angle=180}{i3}
\fmfv{lab=$N$,l.angle=0}{o1}
\fmfv{lab=$N$,l.angle=0}{o3}
\end{fmfgraph*}
\end{fmffile}

%% file: Fdiagrams/Zeeh.tex
\begin{fmffile}{zeehcoupl}
\begin{fmfgraph*}(50,50)

\fmfleft{i1,i2,i3}
\fmfright{o3,o2,o1}
  \fmf{boson}{i1,v1}
  \fmf{dashes}{i3,v1}
  \fmf{fermion}{o1,v1,o3}
  
\fmfv{lab=$Z_\mu$,l.angle=180}{i1}
\fmfv{lab=$h$,l.angle=180}{i3}
\fmfv{lab=$E$,l.angle=0}{o1}
\fmfv{lab=$E$,l.angle=0}{o3}
\end{fmfgraph*}
\end{fmffile}